# Distribution of $CO_2$ ice on the large moons of Uranus and evidence for compositional stratification of their near-surfaces

R. J. Cartwright[1], J. P. Emery[1], A. S. Rivkin[2], D. E. Trilling[3,4,5], N. Pinilla-Alonso[1]

[1]Department of Earth and Planetary Sciences, 1412 Circle Drive, University of Tennessee, Knoxville, TN 37916, rcartwri@utk.edu; [2]John Hopkins University Applied Physics Laboratory 11100 Johns Hopkins Road, Laurel, Maryland 20723; [3]Northern Arizona University, Department of Physics and Astronomy, Northern Arizona University, PO Box 6010 Flagstaff, AZ 86011. [4]Visiting Scientist, South African Astronomical Observatory, [5]Visiting Professor, University of the Western Cape

## Abstract

The surfaces of the large Uranian satellites are characterized by a mixture of $H_2O$ ice and a dark, potentially carbon-rich, constituent, along with $CO_2$ ice. At the mean heliocentric distance of the Uranian system, native $CO_2$ ice should be removed on timescales shorter than the age of the Solar System. Consequently, the detected $CO_2$ ice might be actively produced. Analogous to irradiation of icy moons in the Jupiter and Saturn systems, we hypothesize that charged particles caught in Uranus' magnetic field bombard the surfaces of the Uranian satellites, driving a radiolytic $CO_2$ production cycle. To test this hypothesis, we investigated the distribution of $CO_2$ ice by analyzing near-infrared (NIR) spectra of these moons, gathered using the SpeX spectrograph at NASA's Infrared Telescope Facility (IRTF) (2000 – 2013). Additionally, we made spectrophotometric measurements using images gathered by the Infrared Array Camera (IRAC) onboard the Spitzer Space Telescope (2003 – 2005). We find that the detected $CO_2$ ice is primarily on the trailing hemispheres of the satellites closest to Uranus, consistent with other observations of these moons. Our band parameter analysis indicates that the detected $CO_2$ ice is pure and segregated from other constituents. Our spectrophotometric analysis indicates that IRAC is not sensitive to the $CO_2$ ice detected by SpeX, potentially because $CO_2$ is retained beneath a thin surface layer dominated by $H_2O$ ice that is opaque to photons over IRAC wavelengths. Thus, our combined SpeX and IRAC analyses suggest that the near-surfaces (*i.e.*, top few 100 microns) of the Uranian satellites are compositionally stratified. We briefly compare the spectral characteristics of the $CO_2$ ice detected on the Uranian moons to icy satellites elsewhere, and we also consider the most likely drivers of the observed distribution of $CO_2$ ice.

## 1. Introduction

Spatially resolved images gathered by the Imaging Science System (ISS) onboard Voyager 2 revealed that the surfaces of the classical (*i.e.*, large and tidally-locked) Uranian satellites Miranda, Ariel, Umbriel, Titania, and Oberon (Table 1) are grayish in tone, with bright patches generally associated with impact features and tectonized terrain (*e.g.*, Smith et al., 1986). Analysis of ISS images demonstrated that the Uranian satellites' leading hemispheres are spectrally redder than their trailing hemispheres, and the amount of reddening appears to increase with increasing distance from Uranus (Bell and McCord, 1991; Buratti and Mosher, 1991; Helfenstein et al., 1991). These satellites display abundant evidence of tectonic resurfacing, ranging from tectonized coronae and chasmata on Miranda (*e.g.*, Pappalardo et al., 1997; Beddingfield et al., 2015) to subtle polygonal basins observed on Umbriel (Helfenstein et al., 1989). Numerous studies have presented evidence for potential cryovolcanic landforms on each of these moons, from flow bands with medial grooves on Ariel, smooth patches with convex edges on Titania, and smooth, low and high albedo deposits on crater floors on Umbriel and Oberon (*e.g.*, Jankowski and Squyres, 1989; Schenk, 1991; Croft and Soderblom, 1991; Kargel 1994). Surface age estimates range from a few 100 Ma for younger terrains on Ariel to > 4 Ga for most of the ancient surface of Umbriel (Zahnle et al., 2003). Geologic interpretation of these satellites' surfaces, however, is limited by the low spatial resolution of the ISS dataset, ranging from a few 100 m/pixel on Miranda to ~6 km/pixel on Oberon.

While Voyager 2's flyby of Uranus returned a wealth of information, (*e.g.*, Stone and Miller, 1986) it is the only spacecraft to visit the Uranian system. Consequently, compositional analysis of Oberon, Titania, Umbriel, Ariel, and Miranda is much less-well developed than that of their Jovian and Saturnian counterparts, which have been imaged extensively by the Galileo and Cassini spacecraft, respectively. Ground-based observations of the large Uranian satellites indicate that their surfaces are dominated by $H_2O$ ice mixed with a low-albedo constituent, which is spectrally neutral over NIR wavelengths (*e.g.*, Soifer et al., 1981; Brown and Cruikshank, 1983; Brown and Clark, 1984). Although the low albedo constituent detected on the surfaces of the Uranian satellites has yet to be uniquely identified, spectral modeling suggests that it is likely carbon-rich (*e.g.*, Clark and Lucey, 1984).

More recent NIR observations of these moons led to the detection of $CO_2$ ice on Ariel, Umbriel, and Titania, principally on their trailing hemispheres (Grundy et al., 2003, 2006). Additionally, Grundy et al. (2006) found that the abundance of $CO_2$ ice decreases with increasing orbital radius, with no detection on the furthest classical satellite, Oberon. A host of loss mechanisms (*e.g.*, sublimation, UV photolysis, micrometeorite bombardment, and charged particle sputtering) should effectively remove $CO_2$ from their surfaces over timescales shorter than the age of the Solar System (*e.g.*, Grundy et al., 2006). Consequently, the detection of $CO_2$ on the Uranian moons suggests that it is actively produced by non-native processes. Grundy et al. (2006) suggest that bombardment of native $H_2O$ ice and C-rich constituents on the Uranian satellite surfaces by magnetospherically-bound charged particles could drive a radiolytic production cycle of $CO_2$ ice.

Magnetic field interactions with icy satellite surfaces are well documented in the Jupiter and Saturn systems. UV spectra of Europa's trailing hemisphere display an enhanced absorption feature near 280 nm (attributed to $SO_2$) that likely originated from magnetospherically-embedded sulfur ions irradiating $H_2O$ ice on Europa's surface (Lane et al., 1981; Ockert et al., 1987; Noll et al., 1995). An albedo minimum near 260 nm in UV spectra of Rhea and Dione has been attributed to magnetospherically-generated $O_3$ trapped in the $H_2O$ ice matrix on these moons' surfaces (Noll et al., 1997). The magnetic fields of Jupiter, Saturn, and Uranus all co-rotate with the planets, and at a faster rate than the orbital periods of their regular moons. Consequently, charged particles caught in these planet's magnetic fields preferentially interact with the trailing hemispheres of their classical satellite systems. Unlike the magnetic fields of Jupiter and Saturn, Uranus' magnetic field is substantially offset from its rotational axis (~58.6°) and from its center of mass (~0.3 Uranian radii); modeling interactions between Uranus' moons and its magnetic field is therefore difficult.

In order to test the hypothesis that $CO_2$ ice on the surfaces of Ariel, Umbriel, Titania, and Oberon is generated by magnetospherically-trapped charged particle irradiation, we have collected new SpeX spectra of these moons' previously unobserved northern hemispheres. Using these spectra, along with two other SpeX datasets collected over their southern hemispheres, and spectrophotometry measured by IRAC, we investigate the spatial distribution and mixing regime of $CO_2$ ice on these satellites. We also explore the distribution of $H_2O$ ice on these moons. Additionally, the broad wavelength range of the SpeX (~0.81 – 2.42 μm) and IRAC (~3.1 – 9.5 μm) datasets enables us to characterize vertical layering in these moons' near-surfaces.

## 2. Observations and data reduction

### 2.1 IRTF/SpeX

Observations of the Uranian satellites reported in this work were gathered between 2000 and 2013 using the short wavelength cross-dispersed mode (SXD) of the NIR spectrograph/imager SpeX at the IRTF on Mauna Kea, Hawaii (Rayner et al., 1998, 2003). These observations were made by three different teams (summarized in Table 2, mid-observation 'sub-observer' latitudes and longitudes displayed in Figure 1). The observations by Cartwright and those by Rivkin are presented here for the first time, whereas the data from Grundy were published in Grundy et al. (2003, 2006), and we refer the reader to these publications for details regarding their data reduction procedures. SpeX includes two detectors: a 1024 x 1024 InSb array for the spectrograph (0.15 arcsec/pixel), and a 512 x 512 InSb array that images the slit (0.12 arcsec/pixel). All Uranian satellite spectra gathered between 2000 and 2012 have a spectral range of ~0.81 – 2.42 μm. Data gathered by Cartwright in 2013 utilized the visible/near-infrared (VNIR) MORIS camera co-mounted with SpeX, resulting in a slightly reduced spectral range for those SpeX spectra (covering ~0.94 – 2.42 μm). Observations made by Rivkin and Cartwright used a 0.8 x 15 arcsecond slit, providing spectral resolution (R = $\lambda/\Delta\lambda$) of ~650 – 750. Grundy et al. (2003, 2006) used two different slit width settings (0.3 x 15 arcseconds and 0.5 x 15 arcseconds), resulting in spectral resolutions of ~1600 – 1700 and ~1300 – 1400, respectively.

For all three SpeX datasets, spectra were gathered as pairs, with the object imaged in two positions (referred to as A and B beams) separated by 7.5 arcsec along the 15-arcsec slit. Subtraction of these A-B image pairs provides first order removal of sky emission. Maximum exposure time per frame was limited to 120 s to minimize sky emission variability. In order to improve the signal to noise ratio (S/N), object frames from each night were co-added during data reduction. Nearby solar analog stars were observed by each team to ensure good correction of atmospheric absorption (summarized in Table 3). These solar analogs were observed repeatedly throughout the observations at multiple airmasses. Flat field images and wavelength calibration files were obtained using observations of SpeX's internal integrating sphere illuminated by a quartz lamp and an argon lamp, respectively.

Background subtraction and extraction of spectra gathered by Rivkin and those gathered by Cartwright were conducted using custom programs and the Spextool data reduction package (Cushing et al., 2004; Vacca et al., 2003). Extracted satellite spectra were divided by solar analog spectra from the same night, at similar air masses. Solar-analog-divided spectra were then combined using custom programs and the Spextool program suite (Cushing et al., 2004). We applied several additional corrections to the spectra gathered by Cartwright in 2012 and 2013 to remove residual telluric contributions, including: sub-pixel shifting of object and solar analog spectra, interpolation of solar analog airmasses to better match object airmasses, and dividing spectra by an appropriately scaled atmospheric transmission spectra (gathered at Mauna Kea). A scaled spectrum of Uranus (Rayner et al., 2009) was used to correct for scattered light in the Ariel and Umbriel spectra collected in 2013. Finally, we photometrically scaled all three SpeX datasets to I-band geometric albedos (~0.957 μm, albedos listed in Table 1), using values presented in Karkoschka (2001).

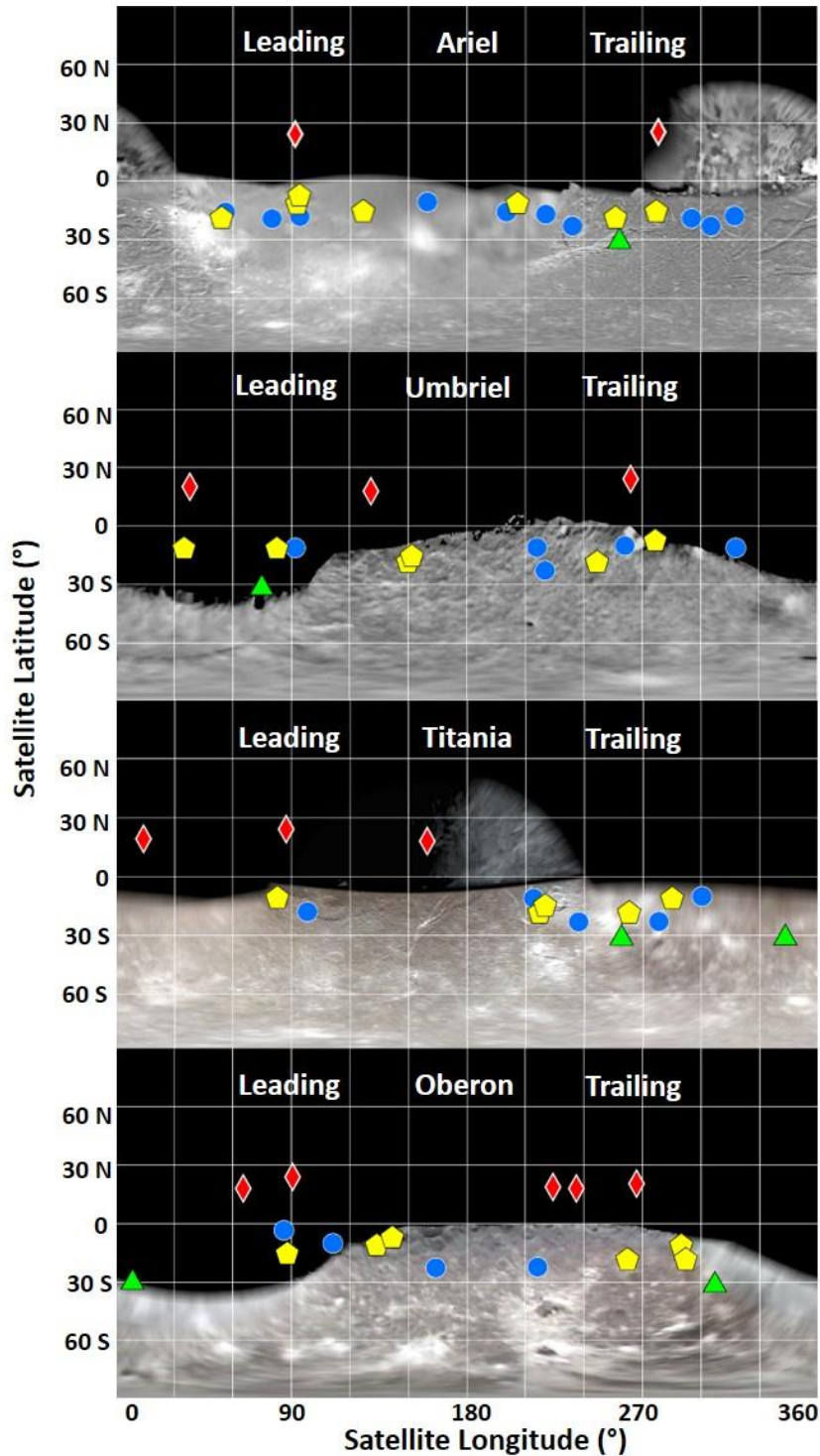

**Figure 1:** Position of the mid-observation satellite latitudes and longitudes (*i.e.,* center of the target disc) observed by three different teams using SpeX: Cartwright (red diamonds), Rivkin (green triangles), and Grundy et al., 2006 (blue circles). Position of the mid-observation satellite latitudes and longitudes observed by IRAC (Program 71) are also indicated (yellow pentagons). Base maps for each moon are Voyager 2 ISS mosaics (Courtesy NASA/JPL-Caltech/USGS, http://maps.jpl.nasa.gov/uranus.html), including night-side regions of Ariel and Titania illuminated by scattered light from Uranus (Stryk and Stooke, 2008).

*2.2 Spitzer/IRAC*

The Spitzer Space Telescope was launched in 2003 into an Earth-trailing, heliocentric orbit to observe the universe at infrared wavelengths (~3.1 – 160 μm) (Werner et al., 2004). IRAC is a NIR imager onboard Spitzer that gathers data in four broadband channels, centered near 3.6, 4.5, 5.8, and 8.0 μm with pass band widths of 0.68, 0.87, 1.25, 2.52 μm, respectively (Fazio et al., 2004). Each channel has a field of view (FOV) of 5.8 x 5.8 arcmin, with ~1.2 arcsec/pixel spatial resolution. IRAC exposures are collected as simultaneous image pairs; channel 1 and 3 share the same FOV, as do channel 2 and 4. While one pair of channels observes the target, the other pair records data from the adjacent sky (with no overlap in FOV).

Near-infrared photometry of the large Uranian satellites was obtained by IRAC on eight nights between 2003 and 2005 (Program 71). The IRAC dither pattern was set to cycling, gathering three images per channel pair (six total frames per observation) with an exposure time of 26.8 seconds per frame. IRAC observed each moon twice, with one targeted observation of their leading hemispheres and one of their trailing hemispheres (close to 'sub-observer' satellite longitudes of 90° and 270°, respectively). In each of these leading/trailing Uranian satellite image pairs, the other three moons were also imaged in the frames; thus, IRAC effectively imaged each moon a total of eight times. Some of the IRAC observations of the four largest Uranian moons are clearly contaminated by scattered light from Uranus and other background sources, making reliable aperture photometry much more difficult on these frames. Here, we only report our aperture photometry results for the IRAC frames devoid of significant flux contamination (summarized in Table 4, mid-observation 'sub-observer' latitudes and longitudes displayed in Figure 1).

IRAC frames have been processed by the standard Spitzer Science Center (SSC) data reduction pipeline, where dark subtraction, flat fielding, and flux calibration procedures are conducted (reduction procedures are described in greater detail in the IRAC instrument handbook[1]). The SSC reduction pipeline generates corrected basic calibrated data (CBCD) products, which were further processed to remove common IRAC artifacts like mux-stripe, column pulldown, banding, saturation, and stray light contributions (see the instrument handbook for details of each of these potential artifacts and methods used to mitigate them). All IRAC images analyzed in this study are CBCD products. Using corrections supplied by the SSC, we corrected for variations in the pixel solid angle and for photometric variations in different parts of the array before converting the images to units of mJy/pixel.

We conducted aperture photometry using aperture sizes of 2 and 3 pixels, with corresponding background annulus radii of 2 – 6 and 3 – 7 pixels, respectively. We also performed aperture photometry using a background annulus with a radius of 10 – 20 pixels for each aperture, making a total of four flux estimates per moon. Our chosen aperture sizes allow us to measure the maximum possible source flux while minimizing sky background, cosmic ray hits, and other non-source contributions to flux within the aperture. In order to account for the finite size of our apertures compared to the SSC-calibrated aperture sizes (10 pixel radius), we multiplied our flux estimates by the channel-dependent aperture corrections listed in the IRAC instrument handbook. Because we are measuring reflected flux from Solar System objects, we assumed that the primary source of the measured fluxes matches a solar spectral slope through each IRAC channel's pass band. Consequently, we divided our flux estimates by color

---

corrections calculated for the spectrum of the Sun (from Smith and Gottlieb, 1974). The absolute flux calibration of IRAC is accurate to 2% (Reach et al. 2005).

After calibration of the data was completed, we calculated an average flux using the 2 and 3 pixel flux measurements with the small background annuli and another average flux using the two apertures and the larger annuli. We then combined these two averages into one final flux estimate and propagated errors (Table 5). The listed uncertainties in Table 5 account for photon counting statistics, variation among the individual CBCD frames, and uncertainty from our chosen aperture/annulus size combinations.

After completing photometry on the IRAC frames, we converted the object fluxes in each channel into geometric albedos using $p_\lambda = F_\lambda r^2_{AU} \Delta^2 / F_{\odot,\lambda} \Phi R^2$, where $p_\lambda$ is the broadband geometric albedo for each IRAC channel, $F_\lambda$ is the measured flux, $r_{AU}$ is the target's heliocentric distance in AU, $\Delta$ is the distance between the observer and the target, $F_{\odot,\lambda}$ is the solar flux at 1 AU for each channel, $\Phi$ is a phase correction, and $R$ is the target's radius. We used Uranian satellite photometric parameters provided in Karkoschka (2001) for the phase correction: $\Phi = 10exp(- (((\beta * \alpha) + (0.5 * \alpha)/(\alpha_0 + \alpha)))/2.5)$, where $\alpha$ is the phase angle of the observation, $\beta$ is the phase coefficient for each moon (Ariel 0.025, Umbriel 0.027, Titania 0.023, and Oberon 0.023) and $\alpha_0$ is the width of the opposition surge (Ariel 0.2, Umbriel 0.7, Titania 0.5, and Oberon 0.5) ($\beta$ and $\alpha_0$ values provided in Table 7 of Karkoschka, 2001).

## 3. Results

### 3.1 SpeX spectra

Figure 2 displays examples of leading and trailing observations of each moon gathered by Cartwright and Rivkin. The rest of the spectra we collected are displayed in Appendix A, along with spectra previously presented in Grundy et al. (2003, 2006). Observations made by Rivkin and Grundy et al. (2003, 2006) were centered on the southern hemispheres of the Uranian moons (sub-solar latitudes of $\sim 10 - 30°$ S). Observations made by Cartwright were centered on these satellites' northern hemispheres (sub-solar latitudes of $\sim 17 - 24°$ N). Throughout the following sections, we refer to all SpeX and IRAC data points gathered between satellite longitudes of 1 and 180° as *leading hemisphere* observations, and all data points gathered between 181 and 360° are referred to as *trailing hemisphere* observations.

All 43 of the spectra analyzed in this study show clear evidence for overlapping combination and overtone $H_2O$ ice bands centered near 1.52 μm and 2.02 μm, along with another feature centered near 1.65 μm, which is indicative of crystalline $H_2O$ ice (*e.g.*, Mastrapa et al., 2008). Visual comparison with icy satellites in the Saturnian system reveals that $H_2O$ bands on the Uranian moons are weaker (*i.e.*, smaller band areas and depths) than those on Saturn's moons (see Emery et al. (2005) for examples of icy Saturnian moon spectra gathered using SpeX). All 43 Uranian satellite spectra are relatively neutral or display reddish slopes from $\sim 0.8$ to 1.3 μm. From $\sim 1.4$ to 2.5 μm, the spectra are slightly blueish, characteristic of $H_2O$ ice-rich objects (*e.g.*, Clark et al., 2013 and references therein). Additionally, the 'arch-like' shape of the $H_2O$ ice continuum between $\sim 2.15$ and 2.3 μm is noticeably flattened in our spectra as compared to pure $H_2O$ ice. The subtle slopes of these spectra's continua, along with the absence of the 1.05 and 1.25 μm $H_2O$ ice bands, and relatively weak 1.52 and 2.02 μm $H_2O$ ice bands, is consistent with surfaces dominated by small $H_2O$ ice grains mixed with a spectrally featureless material.

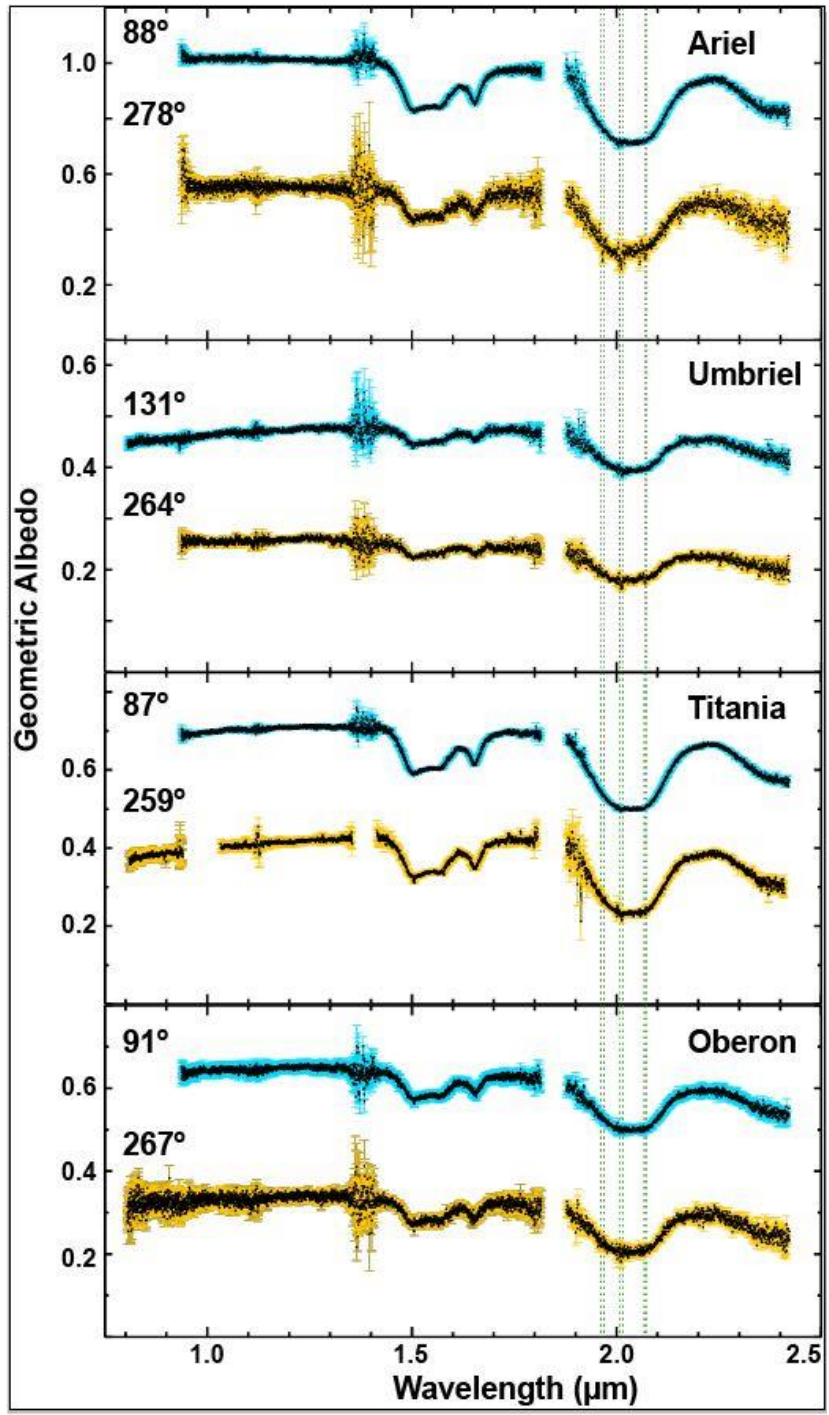

**Figure 2:** SpeX spectra of the leading (blue) and trailing (orange) hemispheres of Ariel, Umbriel, Titania, and Oberon, scaled to I band geometric albedos (Karkoschka 2001). Mid-observation satellite longitudes are listed above each spectrum in the top left-hand corner of each plot. Position and width of the 1.966, 2.012, and 2.070 μm $CO_2$ ice bands are highlighted by the green dashed lines. For clarity, leading hemisphere spectra are offset upwards in each plot (Ariel +0.45, Umbriel +0.2, Titania +0.3, Oberon +0.3). Using a moving window procedure (30 pixels wide), we removed outlier data points that were beyond three standard deviations from the median value. All of the spectra were gathered by Cartwright, except for the trailing hemisphere spectrum of Titania (259°), which was collected by Rivkin.

The spectra gathered over the trailing hemispheres of Ariel and Umbriel display visibly apparent $CO_2$ ice combination and overtone bands between 1.9 and 2.2 μm. Weak $CO_2$ bands between 1.9 and 2.2 μm are also present in some of the spectra collected over the trailing hemisphere of Titania. We do not detect visibly apparent $CO_2$ bands in the Oberon spectra. We discuss our analysis of the $CO_2$ and $H_2O$ bands in greater details in sections 4.1 and 4.3, respectively.

*3.2 IRAC photometry*

We report our IRAC flux and geometric albedo results for all four channels in Table 5 and present lightcurves in Figure 3. While the S/N ratios are > 100 for channels 1 and 2 (3.6 and 4.5 μm), they are as low as ~12 and ~6 for channels 3 and 4 (5.8 and 8.0 μm), respectively. In some cases the channel 3 and 4 fluxes are below the $3\sigma$ threshold of the background flux uncertainties (indicated with italicized $3\sigma$ upper limits in Table 5). We exclude all IRAC observations that are below the $3\sigma$ threshold from our subsequent analyses. We also report mean IRAC geometric albedos for the leading and trailing hemispheres of each moon (Table 6).

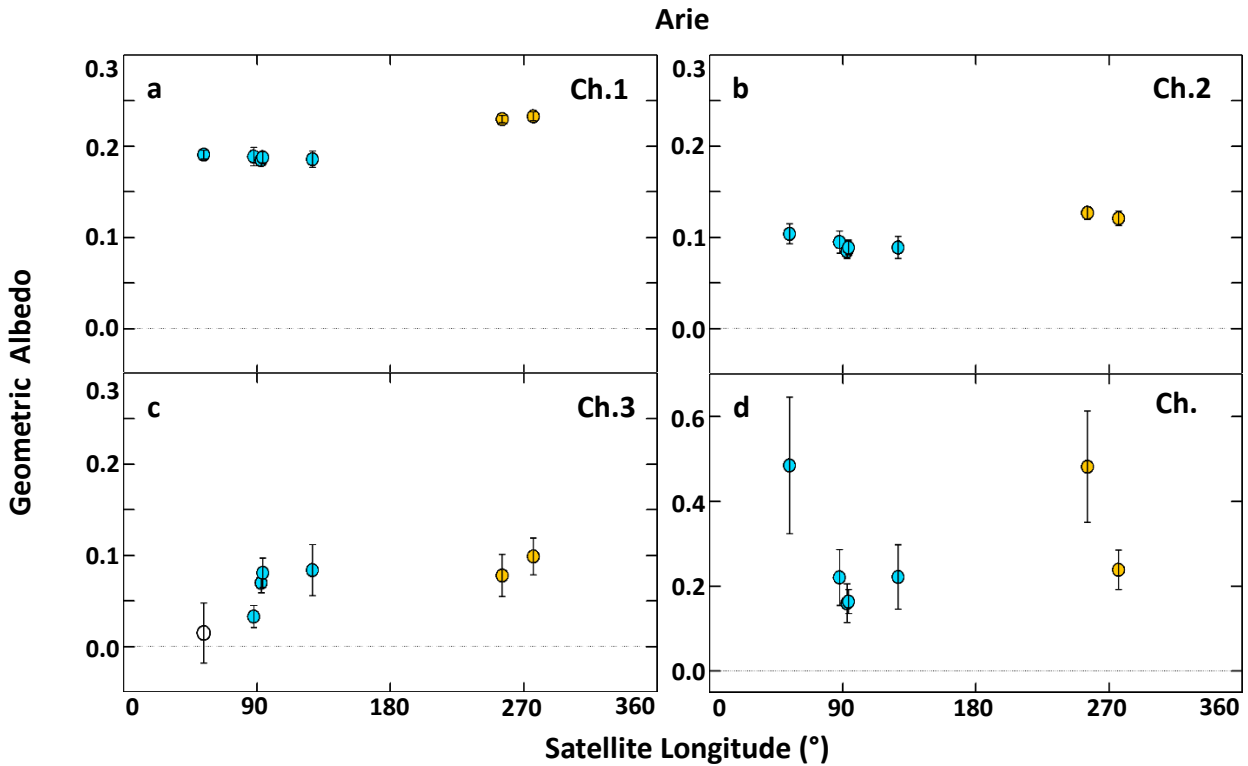

**Figure 3:** Geometric albedos as a function of satellite longitude in each IRAC channel for Ariel (a – d) , Umbriel (e – h) , Titania (i – l) , and Oberon (m – p). Filled light blue circles are leading hemisphere observations, filled orange circles are trailing hemisphere observations, and unfilled circles are data points not included in the analysis (but presented here for completeness). The mean geometric albedo of each satellite decreases from Ch.1 to Ch.2 and decreases again from Ch.2 to Ch.3. The Ch.4 geometric albedos display more variation than the other three channels, which may in part be due to their low S/N. We do not detect any discernable trends in the satellites' Ch.4 albedos.

**Umbriel**

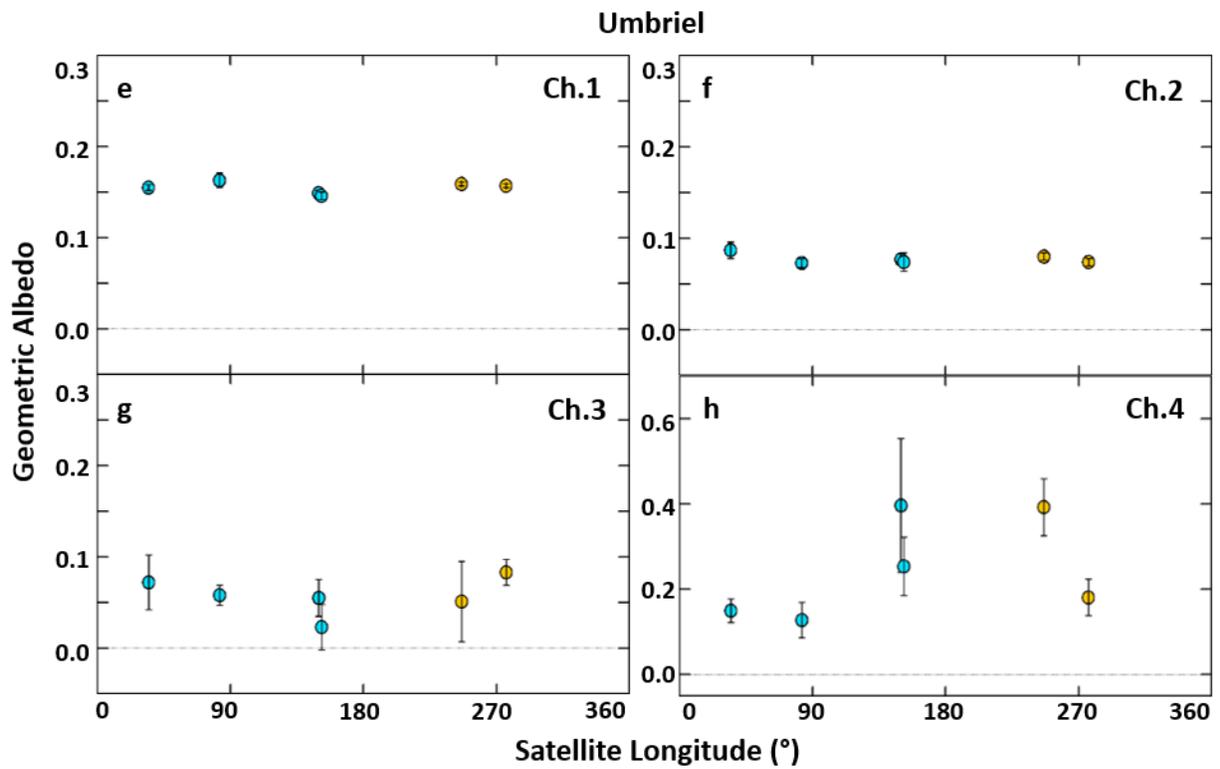

**Titania**

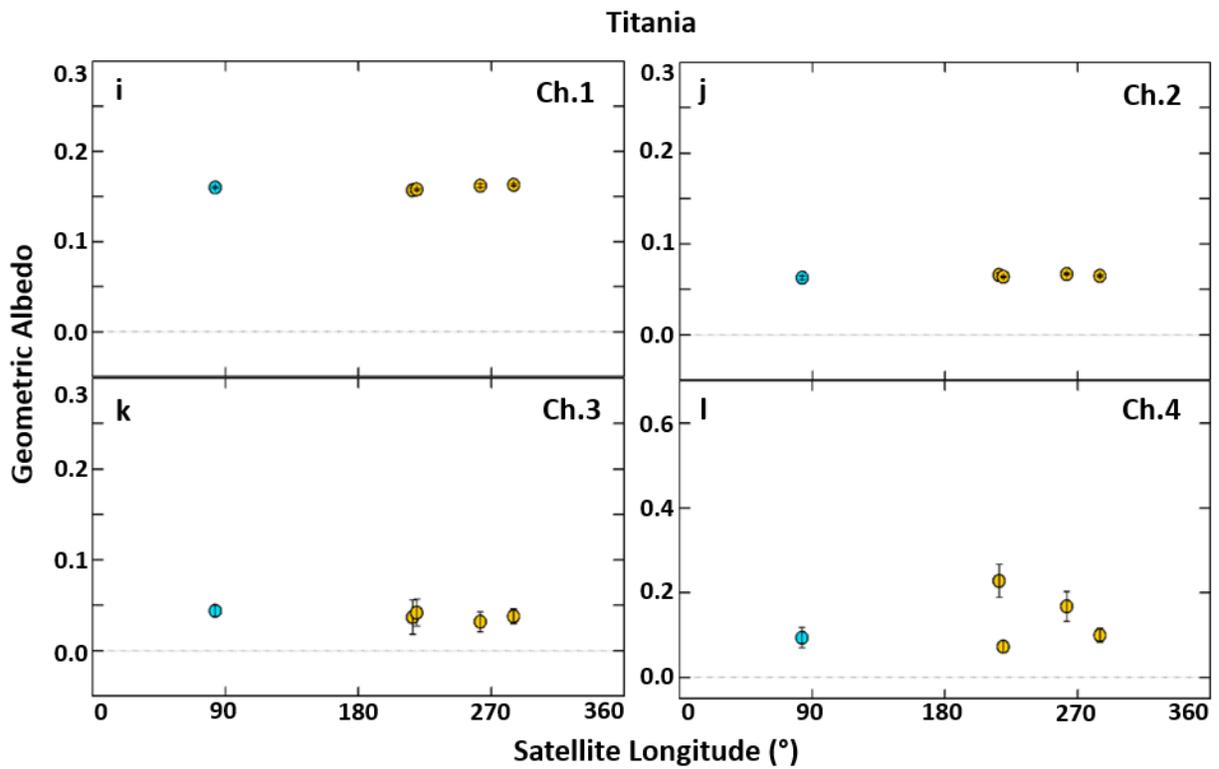

**Figure 3 (continued)**

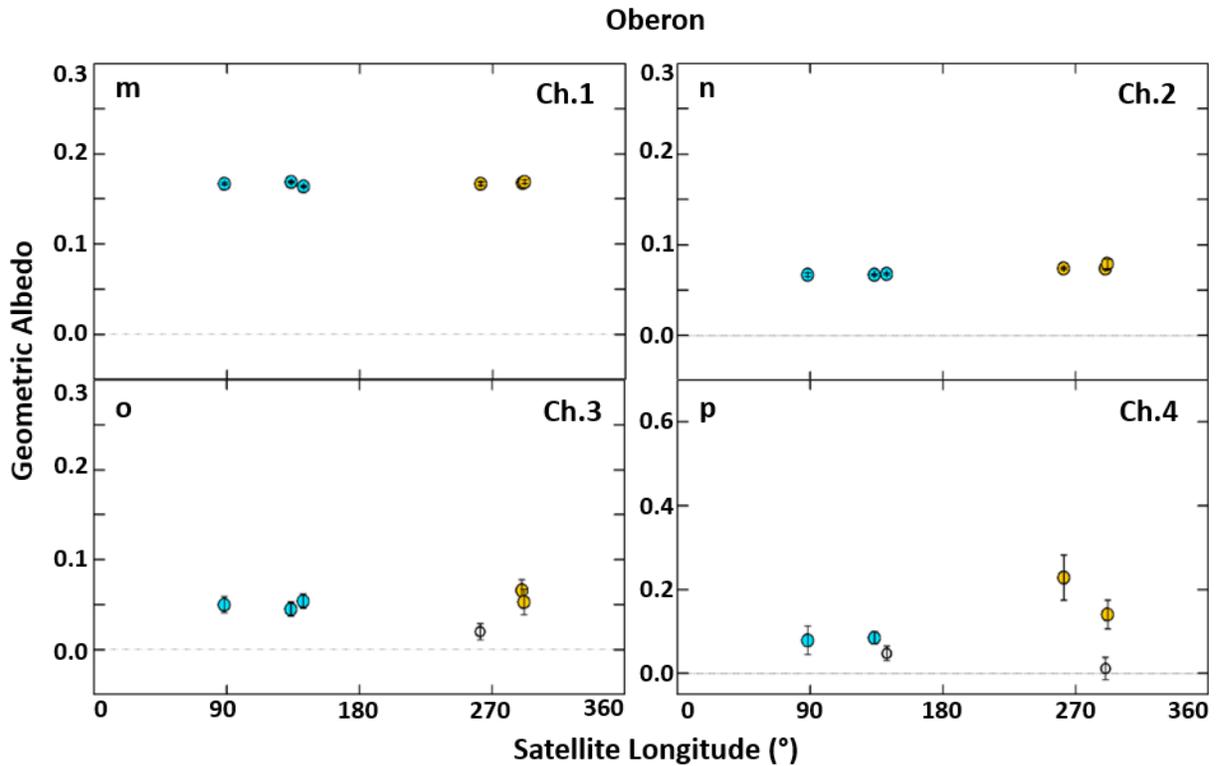

**Figure 3 (continued)**

Ariel displays a clear leading/trailing hemispherical asymmetry in its mean albedos for both channels 1 and 2, with $> 3\sigma$ higher albedos on its trailing hemisphere. Umbriel displays a subtle hemispherical asymmetry, with a slightly brighter mean albedo on its trailing hemisphere in channel 1 ($> 1\sigma$). We see no evidence for brighter trailing hemispheres on the other satellites. For all four moons, we see a clear drop in geometric albedo from channel 1 to 2, and then another reduction in albedo from channel 2 to 3. Channel 4 geometric albedos vary widely on each moon, and we see no discernable trends within the results from this channel, which may in part be due to their low S/N. Of note, the IRAC images utilized by this study represent a previously unexplored wavelength range for these satellites.

## 4. Analysis

In the following sections we analyze the distribution of $CO_2$ and $H_2O$ on the Uranian moons and provide compositional analysis of our IRAC results. We also use synthetic spectra, generated using numerical models, to further test the robustness of our $CO_2$ ice detections (described in Appendix B). As defined above, we refer to all SpeX and IRAC data points gathered between satellite longitudes of 1 and 180° as *leading hemisphere* observations, and all data points gathered between 181 and 360° are referred to as *trailing hemisphere* observations. In the following sections, when our analysis deals specifically with data gathered within 45° longitude of the apex (46° − 135°) and antapex (226 − 315°), we refer to these data as *leading quadrant* and *trailing quadrant* observations, respectively.

*4.1 CO₂ ice band parameter analysis*

In order to investigate the distribution of $CO_2$ on the large moons of Uranus, we measured the band areas of the three strongest $CO_2$ bands between 1.9 and 2.2 μm in each spectrum. We then used an F-test (*e.g.*, Spiegel, 1992) to determine if there are statistically significant differences between the measured $CO_2$ band areas on the leading and trailing hemispheres of these satellites. For spectra with visibly apparent $CO_2$ bands, we fit their band centers using Gaussian curves in order to estimate deviations between the position of the detected $CO_2$ bands and pure $CO_2$ ice (*e.g.*, Gerakines et al., 2005). Band parameter analyses were conducted using a modified version of the Spectral Analysis Routine for Asteroids program (SARA, Lindsay et al., 2015), which we repurposed for the analysis of $CO_2$ and $H_2O$ ice bands. Each of these steps is described in greater detailed in sub-sections 4.1.1 – 4.1.5.

Pure $CO_2$ ice displays many IR-active features, including the strong asymmetric stretch fundamental mode ($v_3$) centered near 4.27 μm, two degenerate bending modes centered near 15.2 μm ($v_2$), and numerous combination and overtone bands between ~1.0 and 3.5 μm (the symmetric stretch mode ($v_1$) of $CO_2$ is not IR-active). The three strongest $CO_2$ ice combination and overtone bands that are readily observable from Earth's surface occur between 1.9 and 2.1 μm, which we refer to as $CO_2$ band 1, $CO_2$ band 2, and $CO_2$ band 3 (Table 7). Figure 4 displays a SpeX spectrum collected over the trailing hemisphere of Ariel with visibly apparent $CO_2$ bands.

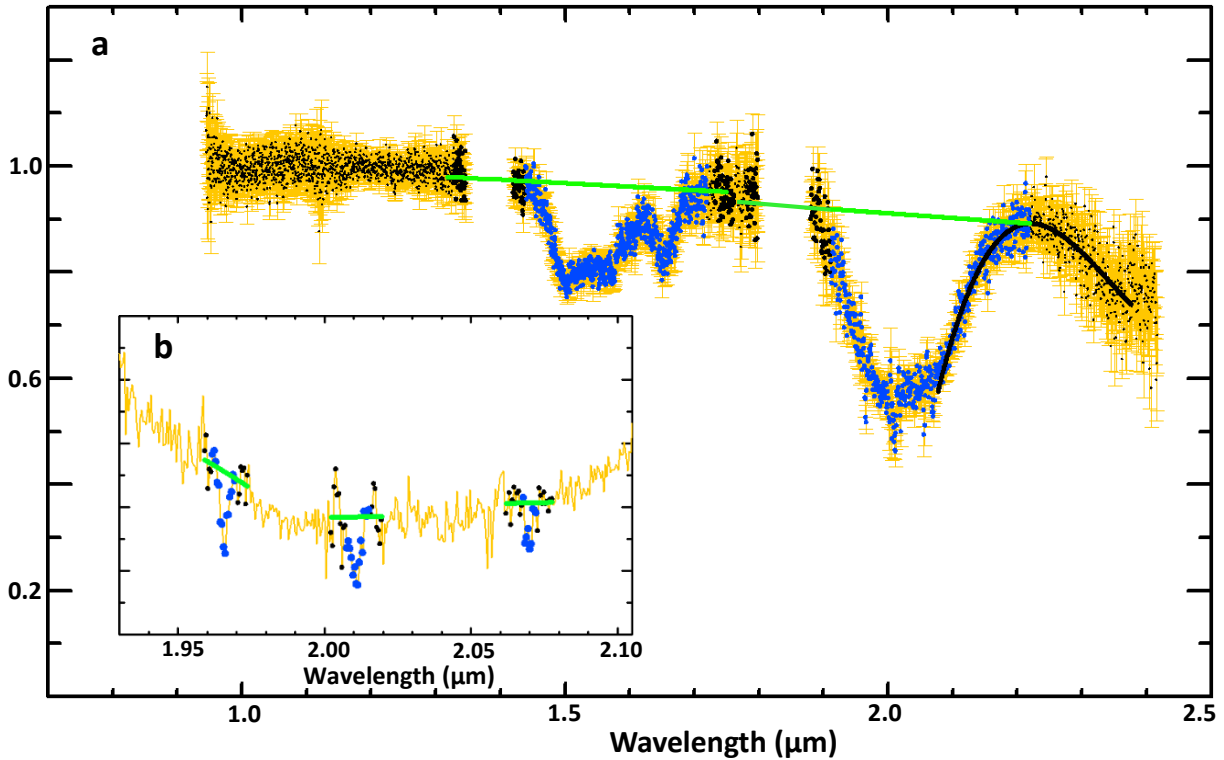

**Figure 4:** Example of our $H_2O$ and $CO_2$ ice band analyses performed on a spectrum of Ariel gathered by our team in 2013 (mid-observation satellite longitude, 278°). (**a**) Shows 1.52 and 2.02 μm $H_2O$ ice bands (blue points) and their associated continua (larger black data points and peak of arch) connected with green lines. (**b**) Close up of same image shown in (a), focused on the positions of the detected $CO_2$ ice bands (blue points) and their associated continua (black points) connected with green lines. Error bars have been suppressed in (b) for clarity.

*4.1.1 CO$_2$ band area measurements*

We measured the integrated areas of CO$_2$ bands 1, 2, and 3 in all 43 SpeX spectra (Table 8). To do this, we first defined the width of these bands and adjacent continua using the same wavelength ranges as Grundy et al. (2006, 2010) (summarized in Table 7). Using a line, we connected the two continua pieces on either side of each band, and divided each band by its continuum. Using the trapezoidal rule, we computed the integrated area of each continuum-divided band, and then added these three bands together to get the total CO$_2$ area in each spectrum. We estimated errors for our band area measurements by running Monte Carlo simulations that randomly sampled our spectra's uncertainty over 20,000 iterations, generating a probabilistic range of uncertainty. The summed CO$_2$ band areas for each spectrum, along with their $1\sigma$ errors, are shown in Figure 5 as a function of satellite longitude.

Our band area results are consistent with two trends previously observed by Grundy et al. (2006): (1) peak abundances in CO$_2$ ice tend to be in spectra gathered near the antapex (270° longitude) of each moon, while the spectra with the lowest CO$_2$ abundances were collected near their apexes (90° longitude). (2) The overall abundance in CO$_2$ appears to decrease with increasing distance from Uranus, with the most CO$_2$ on Ariel and the least on Oberon. A computational error in the band measurement procedure used by Grundy et al. (2006) led to systematic under-estimation of their CO$_2$ ice band areas (Grundy, personal communication). Consequently, our band parameter code is able to reproduce the relative relationships amongst CO$_2$ band areas presented in Grundy et al. (2006), but the CO$_2$ integrated band areas reported here are ~20 – 40% higher.

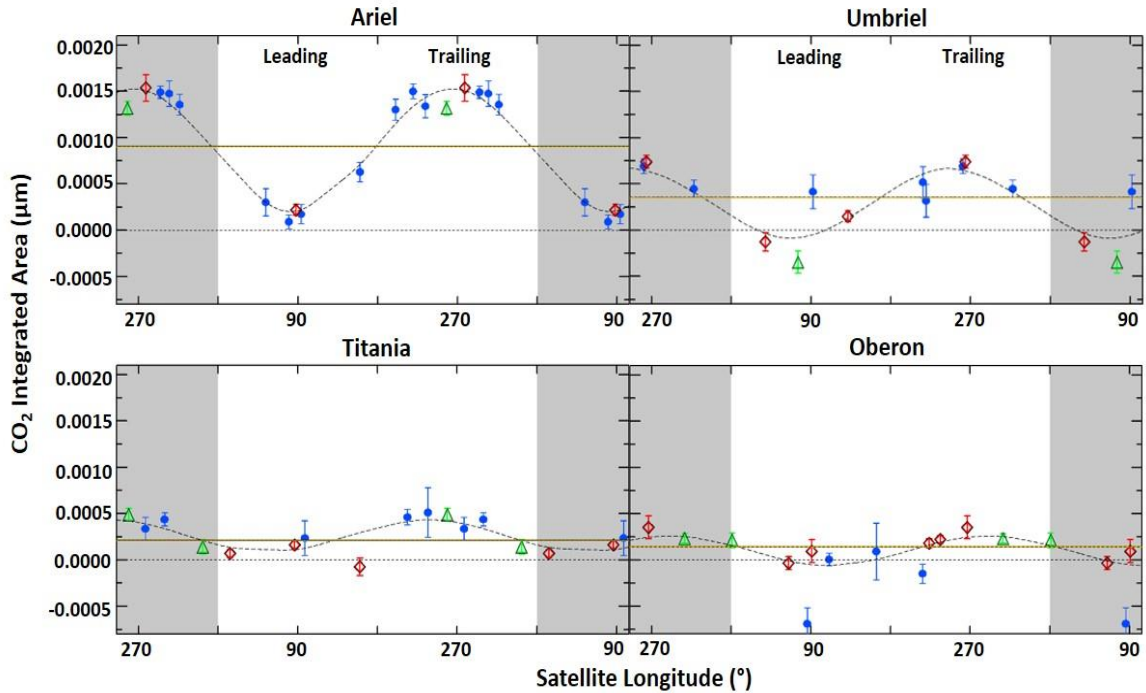

**Figure 5:** CO$_2$ integrated band area as a function of satellite longitude for the three SpeX datasets presented in Figure 1: green triangles (Rivkin), blue filled circles (Grundy et al., 2003, 2006), and red diamonds (Cartwright). Duplicate longitudes are shown to highlight periodic trends in CO$_2$ ice abundance on each satellite (gray-toned regions). Black dashed lines in each plot represent sinusoidal model fits to the data, and yellow lines represent the mean integrated areas for each moon. Ariel displays a clear hemispherical dichotomy in CO$_2$ ice band areas with much stronger CO$_2$ bands on its trailing hemisphere. A similar trend is apparent on the other three moons as well, with larger CO$_2$ band areas on their trailing hemispheres, peaking near 270° longitude.

*4.1.2 Testing the robustness of the $CO_2$ band area measurements*

It is possible that our measured $CO_2$ band areas are enhanced by random noise spikes that coincide with the wavelength regions of the $CO_2$ bands. We see clear evidence of incorporated noise in those cases where our band parameter code returned negative band areas (21 out 129 measured $CO_2$ bands), leading to negative summed $CO_2$ band areas for two spectra gathered over Umbriel (mid-observation longitudes of 38.4° and 75.2°), one spectrum over Titania (160° longitude), and two spectra over Oberon (85.9° and 216.2° longitude) (Figure 5 and Table 8). Importantly, of the 21 cases where our band analysis program generated negative band area measurements, 17 are in spectra gathered over their leading hemispheres, with one trailing hemisphere Oberon spectrum accounting for 3 of the remaining 4 negative band area measurements (summarized in Table 8). Enhancement of both positive and negative band areas with random noise might be particularly important for spectra with limited evidence of visibly apparent $CO_2$ bands (*i.e.*, Titania and Oberon).

In order to further test the robustness of our $CO_2$ ice detections, we investigated the relative band area contributions of $CO_2$ bands 1, 2, and 3 in the SpeX spectra, along with a wide range of noise-free and noise-added synthetic spectra generated using numerical models, including pure $H_2O$ ice, pure $CO_2$ ice, and various mixtures of $H_2O$, $CO_2$, and amorphous C (described in Appendix B). For the mean Oberon spectrum, (*i.e.*, the one with the lowest summed $CO_2$ band areas), none of the pure $H_2O$ ice models we generated match its measured $CO_2$ levels, nor the relative contributions of Oberon's $CO_2$ bands.

In summary, all of the spectral models we tested (both noise-free and noise-added) require $CO_2$ to be present at least at the 1% level, and in many cases at the 5% or higher level, in order to match the measured $CO_2$ band areas and relative band ratios of the mean Uranian satellite spectra and their best fit models. It is therefore unlikely that the low levels of $CO_2$ ice detected in SpeX spectra of Oberon can be explained as the result of integrated noise. Additionally, the relative band area contributions of the models that include areally mixed $CO_2$ ice are more consistent with the relative band area contributions of the mean trailing hemisphere spectra than particulate mixtures of $CO_2$, providing further support for the presence of segregated $CO_2$ on the Uranian moons.

Along with our analysis of the relative band area ratios of synthetic spectra, we also considered how contributions from telluric $CO_2$ might alter the relative band ratios in the mean spectra. Telluric $CO_2$ has several prominent absorption features between 1.9 and 2.1 μm, coincident with solid state $CO_2$. However, the narrow 'peaked' profiles of the detected $CO_2$ bands clearly contrast with the broad 'double lobe' profiles of telluric $CO_2$ bands (Figure B2). Consequently, it appears that telluric contamination of the measured $CO_2$ bands is minimal.

*4.1.3 Statistical analysis of the distribution of $CO_2$*

In order to investigate whether the observed preferential accumulation of $CO_2$ ice on the trailing hemispheres of the Uranian moons is statistically significant, we compared the fits provided by two different models to the measured band areas (weighted by their uncertainties). We used a sinusoidal model with three coefficients (amplitude, phase shift, vertical offset) to represent hemispherical asymmetries in the distribution of $CO_2$ (dashed black lines in Figure 5), and a simple average of the measured band areas on each moon (mean model) to represent a globally homogenous distribution of $CO_2$ (yellow lines in Figure 5). We then used an F-test to determine the level of correlation between these two models. Our null hypothesis is that the sinusoidal and mean models are statistically indistinguishable, and by extension, that there is no statistical significance to the observed asymmetry in the distribution of $CO_2$. We rejected this null hypothesis for the cases where the probability ($p$) of a significant difference between the models was $\leq 0.05$.

The F-test results (Table 9) demonstrate that a sinusoidal fit is significantly better at describing the distribution of $CO_2$ ice on all four moons (Ariel, Umbriel, Titania, and Oberon with $p < 0.001$, $< 0.03$, $< 0.009$, $< 0.006$, respectively), and we reject the null hypothesis for all four moons. Therefore, our F-test analysis indicates that there are statistically significant accumulations of $CO_2$ ice on the trailing hemispheres of these moons.

Previous work has demonstrated that hemispherical asymmetries in cratering rates (*e.g.*, Zahnle et al., 2003) and ion sputtering rates (*e.g.*, Cassidy et al., 2013) are greatest between the leading and trailing *quadrants* ($46° - 135°$ and $226° - 315°$ longitude, respectively) of icy satellites in the Jovian system. If micrometeorite and/or magnetospheric charged particle bombardment dominate the production of $CO_2$ ice in the Uranian system, and, by extension, dominate the observed hemispherical asymmetries in the distribution of $CO_2$ ice, then perhaps these asymmetries are strongest between the leading and trailing quadrants of these moons. Therefore, we also used F-tests to compare sinusoidal and mean model fits to the data points within the leading and trailing quadrants of these moons (Table 10). We find that a sinusoidal fit is significantly better at describing the distribution of the $CO_2$ ice on Ariel and Titania ($P < 0.001$, $< 0.01$, respectively). However, the sinusoidal model is not significantly better than the mean model at describing the distribution of $CO_2$ on Umbriel ($P = 0.45$) and is slightly below the significance threshold for Oberon ($P = 0.06$).

Thus, the full-hemisphere and quadrant-limited F-tests demonstrate that the distribution of $CO_2$ ice on Ariel and Titania is statistically significant, but they generate dissimilar probabilities for the distribution of $CO_2$ on Umbriel and Oberon. Given the lower number of data points in the quadrant-limited F-tests for these two moons (five and seven for Umbriel and Oberon, respectively) compared to the full-hemisphere F-tests (nine and eleven for Umbriel and Oberon, respectively), we favor our full-hemisphere results for the distribution of $CO_2$ ice. Furthermore, the summed $CO_2$ bands in one of the spectra gathered over the leading quadrant of Umbriel (mid-observation longitude of $92.1°$) are substantially larger than the other two leading quadrant spectra for this moon. This spectrum does not have visibly apparent $CO_2$ bands, unlike the spectra collected over Umbriel's trailing quadrant. If this leading quadrant spectrum is removed from the analysis, then the sinusoidal model provides a significant fit ($P = 0.05$) to the data points gathered over the leading and trailing quadrants of Umbriel.

### 4.1.4 $CO_2$ band center modeling

For spectra with visibly apparent $CO_2$ bands (*i.e.*, trailing hemisphere spectra on Ariel and Umbriel), we estimate the position of their band centers (Table 8) by fitting Gaussian curves to the bands. We used an iterative Monte Carlo procedure (described in section 4.1.1) to estimate the uncertainties of these band center fits. The Gauss fits to each band have centers that are close (within $\pm 0.003$ μm) to the band centers of pure $CO_2$ ice (1.966 μm, 2.012 μm, and 2.070 μm), as determined by laboratory experiments (*e.g.*, Gerakines et al., 2005). Laboratory spectra of intimate mixtures of $H_2O$ and $CO_2$ ice also lead to subtle $CO_2$ band position shifts to longer wavelengths ($0.002 - 0.004$ μm) (Bernstein et al., 2005), unlike the $CO_2$ band Gauss fits presented here, which are shifted to both shorter and longer wavelengths – presumably because of the point-to-point variation within each $CO_2$ band. The relatively close matches in wavelength position, along with the strength and narrow profiles of the $CO_2$ ice bands, suggests that our detections are dominated by pure $CO_2$.

*4.1.5 $CO_2$ band parameter analysis summary*

We measured the areas of $CO_2$ bands 1, 2, and 3 in all 43 SpeX spectra. Our band parameter codes measured non-zero $CO_2$ ice band areas on the trailing hemispheres of all four moons, peaking near their antapexes (270° longitude). Our analysis indicates that the abundance of $CO_2$ ice is greatest on Ariel and decreases with distance from Uranus, which is consistent with band area results presented in Grundy et al. (2006). As a check on our band area results, we investigated the relative band area contributions for $CO_2$ bands 1, 2, and 3 in our spectra, and compared them to the relative $CO_2$ band areas in a wide range of synthetic spectra and best fit models. The results of this analysis indicate that the detected $CO_2$ is most consistent with pure $CO_2$ ice that is segregated from other species, and telluric contributions to the measured bands are likely negligible.

Previously, $CO_2$ ice had not been detected on the trailing hemisphere of Oberon (albeit, based on the analysis of only one SpeX spectrum in Grundy et al. (2006)). Our team has increased the number of SpeX observations of Oberon's trailing hemisphere to five (Table 2). While we find no compelling evidence for *visibly apparent* $CO_2$ bands, our F-test analysis indicates that there is a statistically significant accumulation of $CO_2$ on the trailing hemispheres of all four moons, including Oberon.

For the spectra with visibly apparent $CO_2$ bands, we modeled the band centers of the three $CO_2$ bands using Gaussian curves. We detected small shifts to both shorter and longer wavelengths in the center position of some of the $CO_2$ features, which we attribute to noise within the bands. The relatively small shifts in the modeled band centers suggests that the detected $CO_2$ is dominantly pure and segregated. Furthermore, the absence of evidence for the $CO_2$ 'forbidden transition' ($2v_3$) near 2.134 μm in our spectra, which is ~100 times stronger than $CO_2$ bands 1, 2, and 3 in intimate mixtures of $CO_2$ and $H_2O$ (Bernstein et al., 2005), also suggests that the detected $CO_2$ is pure and isolated from other surface species.

*4.2 Investigating the distribution of $CO_2$ using IRAC*

NIR ground-based observations are mostly limited to windows defined by strong telluric bands, which make observations of the Uranian system at wavelengths longer than ~2.5 μm difficult and completely block incoming light near the $CO_2$ asymmetric stretch fundamental ($v_3$). The wavelength range covered by IRAC channel 2 (~4.0 – 5.0 μm) includes the wide $CO_2$ $v_3$ feature. This absorption feature is roughly a factor of 1000 stronger than the shorter wavelength combination and overtone bands (*e.g.*, Hansen, 1997, 2005) detected in SpeX spectra of these satellites. Thus, the wavelength range covered by IRAC is ideal for investigating potentially low levels of $CO_2$ on the Uranian moons.

The $CO_2$ $v_3$ feature (centered near 4.27 μm in pure $CO_2$) is co-incident with numerous overlapping $H_2O$ absorption bands that absorb strongly across the entire IRAC wavelength range (Figure 6). This $CO_2$ absorption feature is flanked by bright 'wings' that cause the geometric albedos of IRAC channels 1 and 3 to be higher than channel 2. This arrangement of channel 1 through 3 geometric albedos leads to a low spectral resolution absorption feature centered in IRAC channel 2. Although the Uranian satellite channel 1 geometric albedos are higher than channel 2 in all cases, only in one trailing hemisphere observation of Umbriel is the channel 3 albedo slightly higher than channel 2 (Table 5). Given the presence of visibly apparent combination and overtone $CO_2$ bands on Ariel and Umbriel over SpeX wavelengths, we expected to find clear evidence for the much stronger $CO_2$ $v_3$ fundamental in our IRAC results, at least on these two moons.

### 4.2.1 Compositional analysis of IRAC photometry

In order to characterize the composition of the Uranian satellites using our IRAC results, we converted their geometric albedos into spectral colors and plotted them in magnitude space. We also extrapolated the best fit models for the mean trailing hemisphere SpeX spectra (described in Appendix B) over IRAC wavelengths, and plotted them in magnitude space as well. $Mag_1 - Mag_2 = 2.5 * \log p_2/p_1$, where $Mag_1$ and $Mag_2$ represent the magnitudes of two IRAC channels, and $p_1$ and $p_2$ represent the geometric albedos of the same two IRAC channels. After converting each moon's mean leading and trailing geometric albedos into magnitudes, we generated a color-color plot with IRAC channel 1 – channel 2 on the y-axis and IRAC channel 2 – channel 3 on the x-axis (Figure 7).

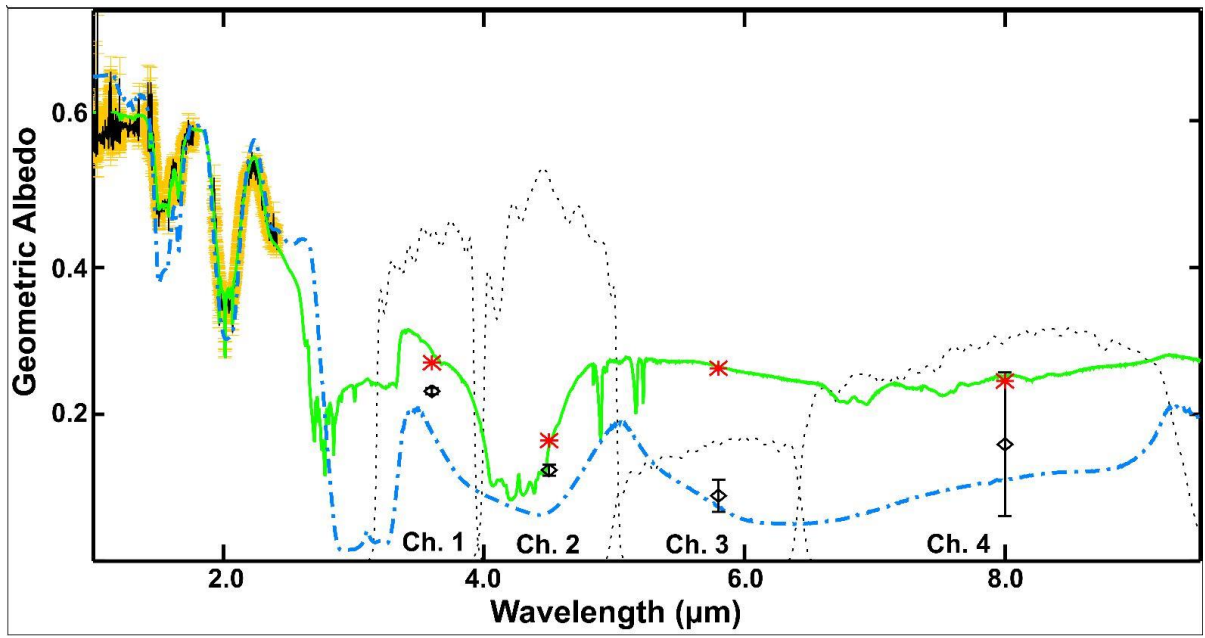

**Figure 6:** Best fit SpeX model for Ariel (described in Table B1, green spectrum) overprinting the mean Ariel trailing hemisphere spectrum (black with orange error bars), extrapolated over IRAC wavelengths (~3.1 – 9.5 µm). A synthetic spectrum of pure $H_2O$ ice (blue dash-dot spectrum) is shown to demonstrate the strong $H_2O$ ice absorption bands that stretch over all four IRAC channels. The mean IRAC geometric albedos for the trailing hemisphere of Ariel (black diamonds with error bars, Table 6) and the geometric albedos for the best fit SpeX model for Ariel extrapolated over IRAC wavelengths (red asterisks) are plotted along with these synthetic spectra. The widths of the IRAC pass bands are indicated by the channel spectral response curves (black dotted lines). Although the best fit SpeX model is a good match for the mean trailing hemisphere SpeX spectrum of Ariel (~1.2 – 2.4 µm), it is unable to adequately fit the mean IRAC geometric albedos for the trailing hemisphere of Ariel.

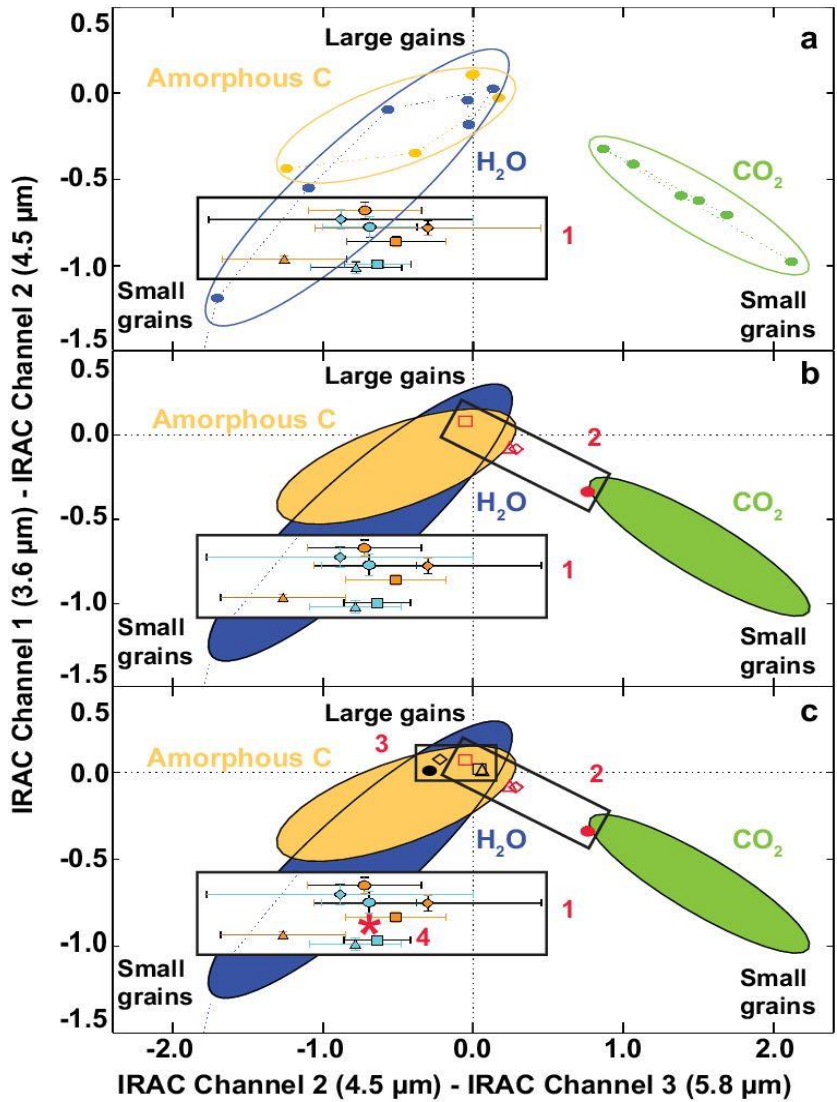

**Figure 7:** Color-color plots displaying IRAC data and compositional models. Positions of the Uranian moon mean geometric albedos, converted into magnitudes, are contained within box 1 in (a – c). Circles represent Ariel, diamonds Umbriel, triangles Titania, and squares Oberon in all three plots. The light blue symbols in box 1 represent the leading hemisphere albedos, and the orange symbols in box 1 represent the trailing hemisphere albedos for these moons. (**a**) This plot includes magnitudes for pure species extracted from synthetic spectra: $H_2O$ ice – blue circles, $CO_2$ ice – green circles, amorphous C – gold circles. Larger grain sizes (5 – 100 μm) for all four compositional groups plot toward the center of the parameter space, and small grains (0.5 – 2 μm) plot toward the edges. (**b**) Same as (a) with best fit models for the mean trailing hemisphere SpeX spectra (shown in Appendix B) contained within box 2. These best fit SpeX models are a mixture of $H_2O$ ice, amorphous carbon, and $CO_2$ ice. (**c**) Same as (b) with a different set of best fit SpeX models composed of only $H_2O$ ice and amorphous carbon (black colored shapes), contained in box 3. Both sets of best fit SpeX models plot in distinct zones that do not overlap with the IRAC geometric albedos. The IRAC albedos are best fit by synthetic spectra dominated by pure $H_2O$ ice, with only minor amounts of $CO_2$ and/or amorphous C. For example, the red asterisk in box 1 (labeled 4 in 7c) represents an IRAC best fit model composed of 80% 1 μm $H_2O$ grains, 15% 2 μm $H_2O$ grains, and 5% 5 μm $CO_2$ grains.

We computed IRAC colors from spectral models of pure $H_2O$, $CO_2$, and amorphous carbon (grain sizes $0.5 - 100$ μm) (Figure 7a), and then plotted these synthetic spectra in the same color-color space in order to compare them to the Uranian satellite IRAC colors. The Uranian satellite IRAC data plot adjacent to the pure $H_2O$ ice zone, which we expected for these $H_2O$-rich icy moons. Surprisingly, the best fit models for the mean SpeX spectra (extended to cover IRAC wavelengths) do not plot near the Uranian moon IRAC data (Figure 7b, box 2). We also generated a set of best fit models for the mean SpeX spectra, without $CO_2$ (*i.e.*, models that only fit the $H_2O$ ice bands and continua), and plotted them in the same magnitude space (Figure 7c, box 3). This set of models also plot in a zone distinctly different from the Uranian satellite IRAC data.

We generated new sets of synthetic spectra using much smaller grain sizes ($0.2 - 5$ μm) than the dominant grain sizes of our SpeX best fit models ($10 - 50$ μm). By including substantial abundances (~$20 - 90$ %) of $0.5 - 2$ μm $H_2O$ ice in our models, we were able to replicate the spectral colors of the Uranian satellites. Additionally, these best fit IRAC models can also include up to ~5 % of larger $H_2O$ ice grains (5 and 10 μm grain sizes), $1 - 5$ % $CO_2$ ice ($1 - 2$ μm grain sizes), and $1 - 20$% amorphous C ($1 - 5$ μm grain sizes) and still plot within the Uranian satellite zone (Figure 7c, asterisk 4).

*4.2.2 IRAC photometry summary*

We analyzed the geometric albedos for IRAC channels 1, 2, and 3, finding that they are inconsistent with the detection of $CO_2$. We converted the geometric albedos of these three channels into spectral colors, and plotted them in magnitude space along with a range of synthetic spectra (including pure end-member compositional models and intimate mixtures). Our color-color analysis suggests that the surfaces of these moons are composed primarily of segregated $H_2O$ ice grains that tend to be small (~$20 - 90$% with $\leq 2$ μm diameters). At the same time, the best fit spectral models for the SpeX spectra suggest that these moons are dominated by intimate mixtures of $H_2O$ ice (mostly $10 - 50$ μm grain sizes) with substantial amounts of amorphous carbon ($2 - 40$%) and $CO_2$ ice ($3 - 27$%). We discuss the apparent mismatch between our best fit SpeX and IRAC models further in section 5.1.

*4.3 $H_2O$ ice band parameter analysis*

Along with characterizing the distribution and near-surface mixing regime of $CO_2$ ice, we analyzed the $H_2O$ ice bands present in the Uranian satellite spectra, measuring their areas and depths. We used an F-test to statistically compare the $H_2O$ band areas and depths on their leading and trailing hemispheres. Additionally, we generated leading/trailing hemisphere ratios of $H_2O$ band areas (as well as band depths) for each moon in order to investigate the distribution of $H_2O$ ice. Each of these steps are described in greater detailed in sub-sections $4.3.1 - 4.3.4$.

Within the spectral region covered by the SXD mode of the SpeX spectrograph (~$0.81 - 2.42$ μm), $H_2O$ ice has multiple band complexes composed of overlapping combination and overtone bands centered near 0.9, 1.04, 1.25, 1.52, and 2.02 μm. These bands increase in depth with increasing wavelength, and the 1.04, 1.25, 1.52, and 2.02 μm bands are readily detected by SpeX on icy Saturnian moons (*e.g.*, Emery et al., 2005). The presence of intimately mixed, low albedo material on the Uranian satellites weakens the $H_2O$ ice bands and effectively masks the 1.04 and 1.25 μm bands in almost all of the SpeX spectra we analyzed. Hence, we focus our analysis on the visibly apparent 1.52 and 2.02 μm bands.

### 4.3.1 $H_2O$ band area measurements

Similar to our $CO_2$ ice band area procedure, we defined two continua on either side of the $H_2O$ ice bands: $1.318 - 1.440$ µm and $1.720 - 1.750$ µm for the 1.52 µm $H_2O$ ice band complex ($1.44 - 1.72$ µm band width), and $1.769 - 1.924$ µm and $2.215 - 2.230$ µm for the 2.02 µm $H_2O$ band ($1.924 - 2.215$ µm band width) (see Figure 4). We used large continua on the short wavelength end of each band in order to ensure robust area measurements for our spectra with significant atmospheric contributions from telluric bands centered near 1.4 and 1.9 µm. We used a third order polynomial to fit the arch-shaped, long wavelength end of the 2.02 µm band. After anchoring the continua for both $H_2O$ ice bands, we connected them with a line, and divided each band by its continuum. We then measured the integrated areas and depths of both bands. To estimate the uncertainty of our band area measurements, we used the same iterative Monte Carlo procedure described in section 4.1.1.

The $H_2O$ ice integrated areas and $1\sigma$ uncertainties are presented in Figure 8 and summarized, along with $H_2O$ ice band depths, in Table 11. A hemispherical asymmetry in Ariel's $H_2O$ ice band areas is visibly apparent, with larger band areas near the apex of its leading hemisphere. The relative difference between $H_2O$ band areas on the leading and trailing hemispheres of Umbriel, Titania, and Oberon are much less pronounced, but the peak in $H_2O$ ice band area is near the apexes of Umbriel and Titania for both the 1.52 and 2.02 µm bands (but only the 2.02 µm band for Oberon).

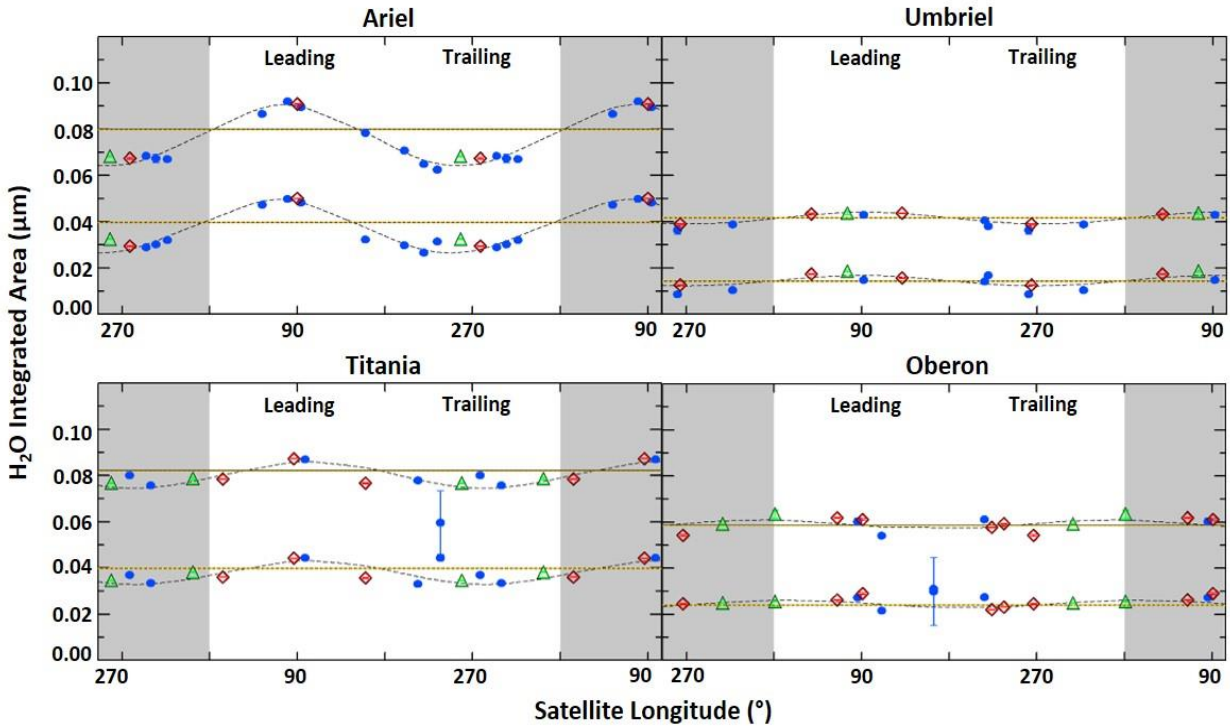

**Figure 8:** Integrated band areas for the 1.52 µm (bottom set of data points in each plot) and the 2.02 µm (top set of data points) $H_2O$ ice bands, as a function of satellite longitude for the three SpeX datasets presented in Figure 1: green triangles (Rivkin), blue filled circles (Grundy et al., 2003, 2006), and red diamonds (Cartwright). Duplicate longitudes shown to highlight periodic trends in $H_2O$ ice abundance (gray-toned regions). Black dashed lines represent sinusoidal model fits, and yellow lines represent the mean band area for each moon. A clear hemispherical asymmetry in $H_2O$ ice band areas is apparent on Ariel, with much stronger $H_2O$ bands on its leading hemisphere. A similar trend is present on the other three moons as well, with slightly larger $H_2O$ band areas on their leading hemispheres, peaking near 90° longitude.

Previous studies (*e.g.*, Brown and Cruikshank, 1983) have noted a trend between Uranian satellite $H_2O$ ice band depths and their visible wavelength (VIS) albedos, with the brightest moon (Ariel) displaying the deepest $H_2O$ ice bands and the darkest moon (Umbriel) displaying the weakest. Our analysis supports this trend on the leading hemispheres of these moons, with both geometric albedo and the mean 1.52 and 2.02 μm $H_2O$ ice band depths and areas decreasing from Ariel, Titania, Oberon, to Umbriel. However, this trend does not hold for their trailing hemispheres, with the mean band depths and areas decreasing from Titania, Ariel, Oberon, to Umbriel (Figure 9, Table 12).

### 4.3.2 Statistical analysis of the distribution of $H_2O$

The same F-test procedure described in section 4.1.3 was applied to the measured $H_2O$ ice band areas in order to quantify the distribution of $H_2O$ ice on these moons. The F-test results (Table 9) demonstrate that a sinusoidal fit is significantly better at describing the distribution of $H_2O$ ice on all four moons (for both the 1.52 and 2.02 μm bands). We calculated $P$ values of $< 0.001$, $< 0.04$, $< 0.01$, and $< 0.04$ for Ariel, Umbriel, Titania, and Oberon, respectively, for both $H_2O$ ice bands. Therefore, our F-test analysis indicates that $H_2O$ ice band areas are significantly larger on the leading hemispheres of these moons compared to their trailing hemispheres.

We also conducted F-tests using data points limited to the leading and trailing *quadrants* (Table 10). F-test results for Ariel and Titania ($P < 0.001$ and $< 0.01$, respectively, for both the 1.52 and 2.02 μm bands) and Umbriel's 2.02 μm $H_2O$ band ($P = 0.01$) indicate that a sinusoidal model is significantly better at describing the distribution of $H_2O$ than a mean model. Asymmetries in the $H_2O$ band areas on Oberon's leading and trailing quadrants are not statistically significant for either band. Similar to the $CO_2$ ice spatial analysis, we favor our full-hemisphere F-test results for the $H_2O$ ice bands because of the lower number of data points in the quadrant-only analysis; there are only 5 and 7 $H_2O$ band area measurements for Umbriel and Oberon, respectively, compared to 9 and 11 data points for the full-hemisphere analysis for Umbriel and Oberon, respectively.

### 4.3.3 $H_2O$ band area and depth ratios

We also present leading/trailing ratios for the mean areas and depths of the 1.52 and 2.02 μm bands on each moon (for example, <band area$_{1.52 μm}$>$_{Ariel-leading}$ / <band area$_{1.52 μm}$>$_{Ariel-trailing}$) (Table 12). When plotted as a function of orbital radius, it is clear that the 1.52 and 2.02 μm leading/trailing band area and depth ratios decrease with distance from Uranus (Figures 10a and 10b). We find no such trend when these same ratios are plotted as a function of satellite geometric albedo (Figures 10c and 10d). Additionally, the 1.52 μm band ratios are greater than the 2.02 μm band ratios, suggesting that the 1.52 μm band is more sensitive to changes in $H_2O$ ice abundance and grain size. The clear trend of decreasing band area and depth ratios with distance from Uranus suggests that the controlling mechanism operates at a system-wide level, and its efficiency either falls off with distance from Uranus (for example, magnetospherically-driven radiolysis), or increases with distance from Uranus (for example, accumulation of in-falling dust from irregular satellites).

We also made mean 1.52 μm / mean 2.02 μm band area ratios for the leading and trailing hemispheres of each moon (Table 12), which increase with increasing geometric albedo (Figure 11). Additionally, the difference between the leading and trailing 1.52 μm/2.02 μm band area ratios for each moon becomes progressively smaller with increasing orbital radius (Figure 11).

### 4.3.4 H₂O band parameter analysis summary

We measured the areas and depths of the 1.52 and 2.02 μm $H_2O$ ice bands in all 43 SpeX spectra. Our $H_2O$ band parameter and F-test results indicate that $H_2O$ bands are stronger on the leading hemispheres of these satellites. Furthermore, there is a clear trend between albedo and $H_2O$ band strength, with the strongest $H_2O$ bands on the leading hemisphere of Ariel and the weakest on the trailing hemisphere of Umbriel. One notable exception to this trend is the $H_2O$ bands on the trailing hemisphere of Ariel, which are weaker than the $H_2O$ bands on the trailing hemisphere of Titania.

Our analysis of the $H_2O$ ice bands demonstrates that hemispherical asymmetries are more pronounced for the 1.52 μm band, compared to the 2.02 μm band for all four moons. Both grain size and contaminants can reduce $H_2O$ ice band depths, and our analysis indicates that the 2.02 μm band is more resistant to these effects than the 1.52 μm band on the Uranian satellites. Determining the relative importance of grain size and contaminant abundance in the reduction of $H_2O$ band depths is beyond the scope of this study, and we will explore the Uranian satellites' $H_2O$ ice bands in greater detail in future work (Cartwright et al. [In prep]). Thus, both the abundance of $CO_2$ ice and hemispherical asymmetries in $H_2O$ band areas decrease with distance from Uranus (Figure 12), suggesting that the process (or processes) modifying the surfaces of these moons are a strong function of orbital radius.

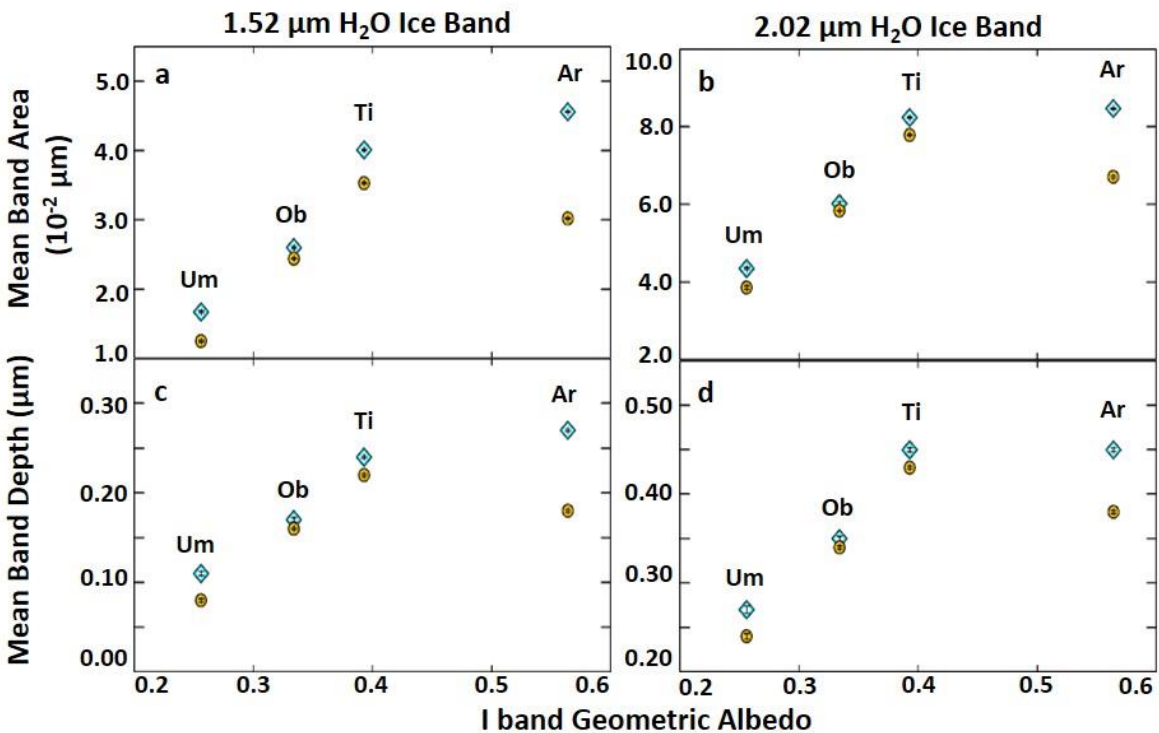

**Figure 9:** Mean leading (blue diamonds) and trailing (orange circles) $H_2O$ band areas (**a, b**) and $H_2O$ band depths (**c, d**) as a function of I band geometric albedos. In each plot, Ar = Ariel, Um = Umbriel, Ti = Titania, and Ob = Oberon. Both the 1.52 and 2.02 μm $H_2O$ bands clearly decrease with decreasing satellite albedo, with the strongest bands on the leading hemisphere of Ariel and the weakest on Umbriel. One notable exception to this trend is the $H_2O$ bands on the trailing hemisphere of Ariel, which are weaker than the $H_2O$ bands on the trailing hemisphere of Titania.

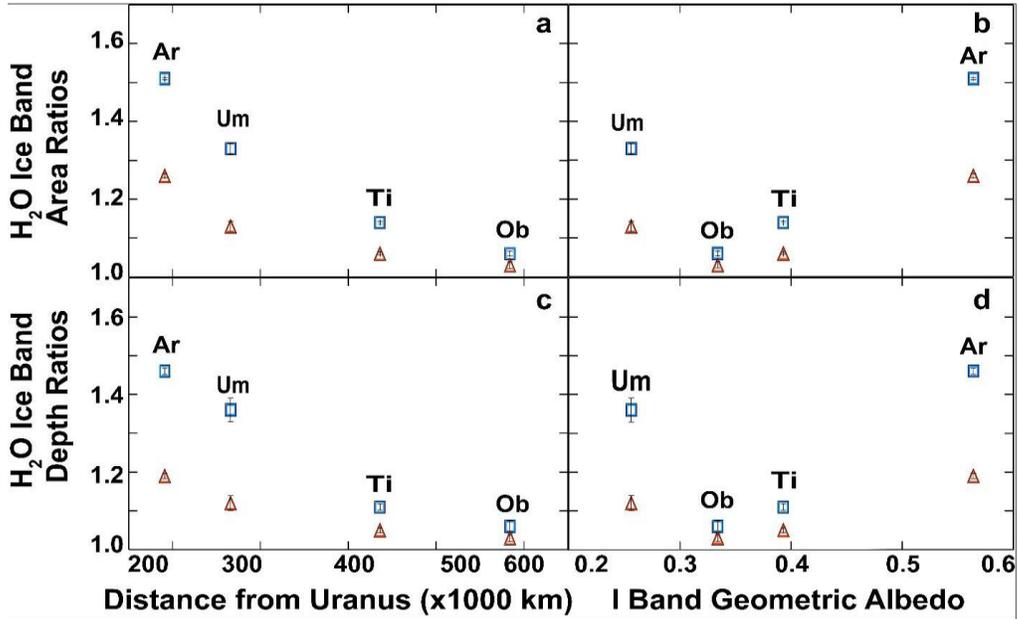

**Figure 10:** (**a**) Ratios between the mean leading and trailing $H_2O$ ice band areas for the 1.52 μm (blue squares) and 2.02 μm (red triangles) bands, as a function of orbital radius. (**b**) Ratios between the mean leading and trailing $H_2O$ ice band depths for the 1.52 (blue squares) and 2.02 μm (red triangles) bands, as a function of orbital radius. (**c**) Same $H_2O$ band area data points shown in (**a**), but as a function of I band geometric albedo. (**d**) Same $H_2O$ band depths shown in (**b**), but as a function of I band geometric albedo. In each plot, Ar = Ariel, Um = Umbriel, Ti = Titania, and Ob = Oberon. Both the mean band area and band depth leading/trailing ratios decrease with increasing distance from Uranus, with the largest ratios on Ariel and the smallest on Oberon. No such trend is present between albedo and either the mean band area or mean band depth leading/trailing ratios. Additionally, the mean 1.52 μm leading/trailing ratios decrease more rapidly than the mean 2.02 μm hemispherical ratios, with only minor differences between these two $H_2O$ band ratios on Oberon.

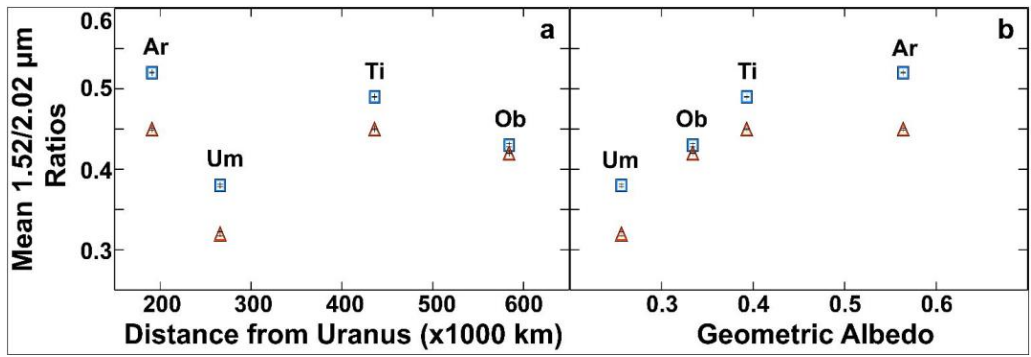

**Figure 11:** (**a**) Ratios of the mean 1.52 μm and 2.02 μm band areas on the leading (blue squares) and trailing (red triangles) hemispheres of the Uranian satellites as a function of orbital radius. (**b**) Mean 1.52 μm / mean 2.02 μm band area ratios for the leading (blue squares) and trailing (red triangles) hemispheres of the Uranian satellites as a function of geometric albedo. In both plots, Ar = Ariel, Um = Umbriel, Ti = Titania, and Ob = Oberon. The 1.52 μm / 2.02 μm band area ratios display a clear trend with albedo, with the largest ratios on the leading hemisphere of Ariel and the smallest on Umbriel. The difference between the 1.52 μm /2.02 μm ratios on the leading and trailing hemispheres of these moons clearly decreases with orbital radius, with the largest difference on Ariel and the smallest on Oberon.

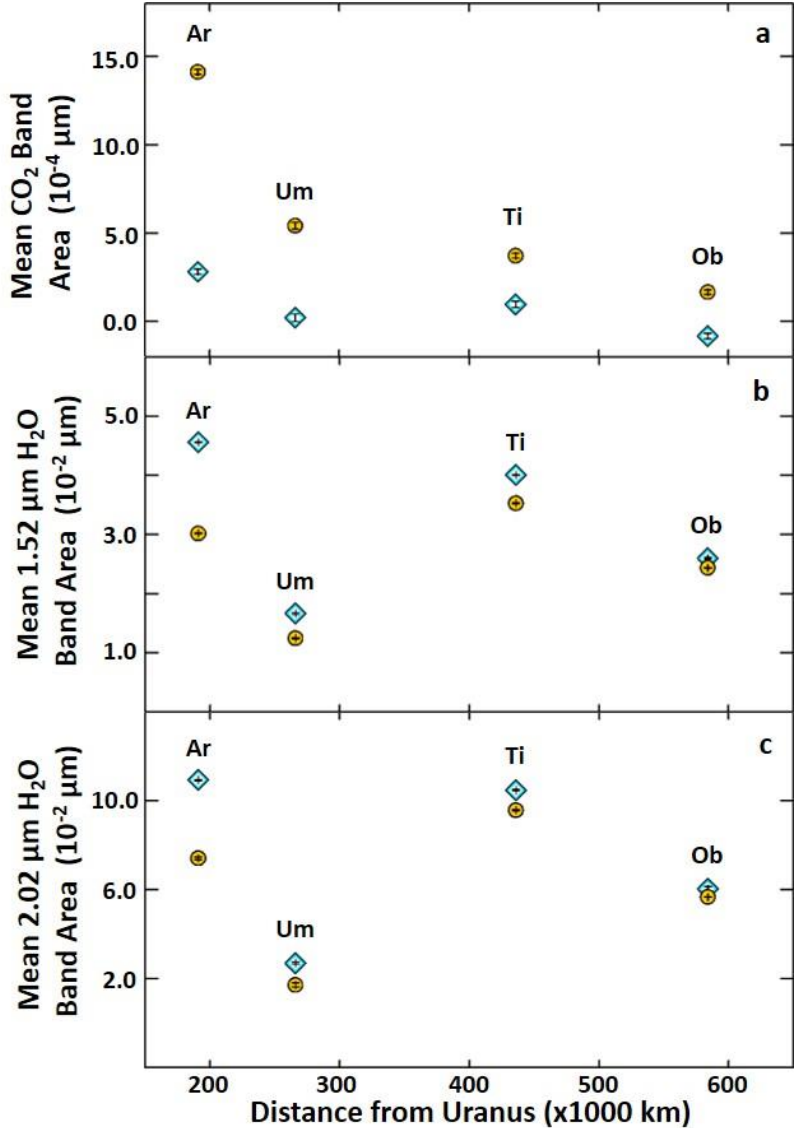

**Figure 12:** (**a**) Mean $CO_2$ band areas, (**b**) mean 1.52 μm $H_2O$ band areas, and (**c**) mean 2.02 μm $H_2O$ band areas as a function of orbital radius (blue diamonds – leading hemisphere, filled orange circles – trailing hemisphere). In each plot, Ar = Ariel, Um = Umbriel, Ti = Titania, and Ob = Oberon. The enhanced $CO_2$ band areas on the trailing hemispheres of these moons, and the reduction in $CO_2$ band areas with distance from Uranus, are apparent in (a). $CO_2$ band areas on these moons' leading hemispheres show no discernable trend, with slightly higher $CO_2$ levels measured on the brighter moons (Ariel and Titania), close to zero $CO_2$ measured on Umbriel, and a slightly negative mean $CO_2$ band area on Oberon. The relative differences between leading/trailing $H_2O$ band areas decreases with distance from Uranus for both the 1.52 and 2.02 μm $H_2O$ bands (b and c), with only subtle hemispherical asymmetries in $H_2O$ band strengths on Oberon.

## 5. Discussion

In the following sections we consider the implications of our analyses, focusing on the differences between the SpeX and IRAC best fit models. We also briefly compare the $CO_2$ ice detected on the classical Uranian moons to $CO_2$ detected on icy satellites elsewhere in the Solar System. Finally, we include a short discussion of the possible sources of $CO_2$ ice, and likely controls on its distribution, in the Uranian system.

### 5.1 Near-surface layering

As described in section 4.2.1, our best fit SpeX models, extrapolated over IRAC wavelengths, are poor matches to the IRAC geometric albedos for these moons (Figures 6 and 7). One possible explanation for this disparity between our SpeX and IRAC results is that these two instruments probe distinct layers in the near-surfaces of the Uranian satellites. To investigate this possibility, we converted the Mastrapa et al. (2008, 2009) crystalline $H_2O$ ice extinction coefficients ($\kappa$) into absorption coefficients ($\alpha$): $\alpha = 4\pi\kappa / \lambda$. The inverse of $\alpha$ gives the e-folding penetration depth of photons, which represents the depth at which there is a ~63% probability of an incident photon (of a given wavelength) being absorbed. The e-folding penetration depth has been used in other studies as an indicator of sensible depths for reflectance spectroscopy (*e.g.*, Palmer and Brown, 2011).

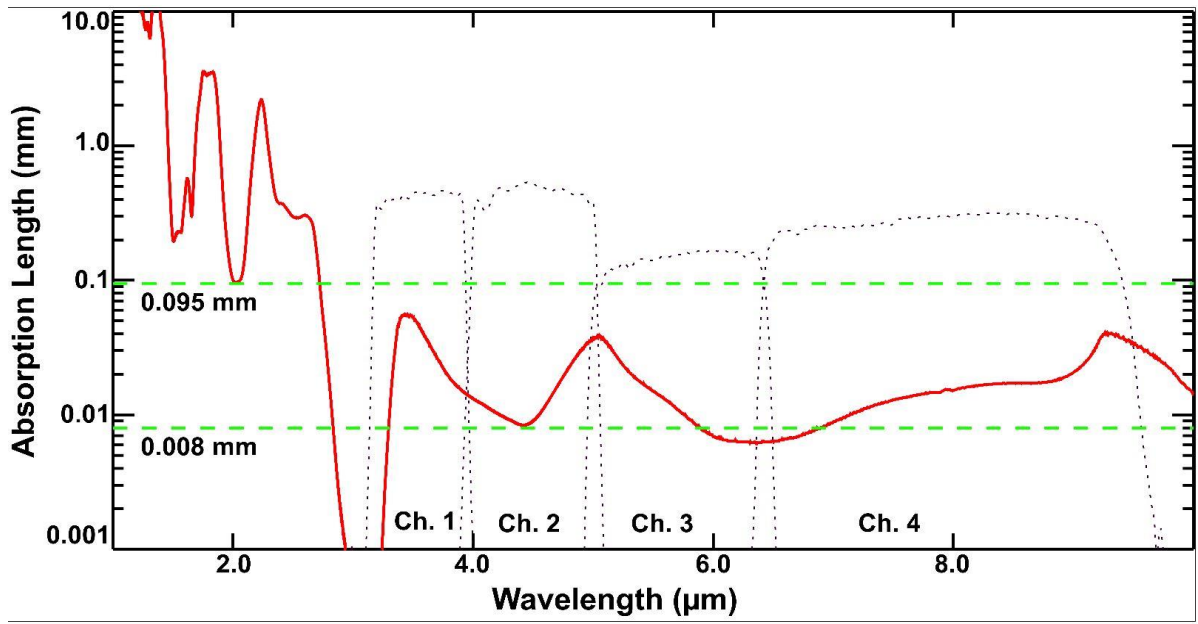

**Figure 13:** Red curve represents the e-folding absorption length for crystalline $H_2O$ ice (80 K) as a function of wavelength. Green dashed lines highlight the $H_2O$ penetration depths in the wavelength regions where $CO_2$ combination and overtone bands (~0.1 mm) and the $CO_2$ asymmetric stretch fundamental (~0.01 mm) are present. The widths of the IRAC pass bands are indicated by the channel spectral response curves (black dotted lines).

Over SpeX wavelengths (~0.81 − 2.42 μm), NIR photons penetrate to greater depths (~0.1 mm to 100's of mm depths) than over IRAC wavelengths (~3.1 − 9.5 μm) into crystalline

H$_2$O ice (~0.0003 to 0.05 mm depths) (Figure 13). Specifically, over the wavelength range of CO$_2$ bands 1, 2, and 3 (~1.9 to 2.1 μm), photons have a high probability of traveling ~100 μm into a pure H$_2$O ice surface, but over the wavelength range of the CO$_2$ $v_3$ fundamental (~$4.0 - 4.5$ μm), photons only travel ~10 μm into H$_2$O ice (see Figure 6 for width and position of CO$_2$ bands over these wavelength ranges). Consequently, SpeX and IRAC are sensitive to NIR photons scattered from different depths, and these instruments may be probing two compositionally distinct layers in the Uranian satellite near-surfaces. Therefore, the CO$_2$ ice detected by SpeX could plausibly be retained beneath a thin veneer of mostly pure H$_2$O frost that strongly absorbs photons over IRAC wavelengths, and alters the expected relationships between the geometric albedos of IRAC channels 1, 2, and 3.

The Hapke-Mie hybrid modeling program utilized by this study (described in Appendix B) generates synthetic spectra composed of areal and intimate mixtures of H$_2$O, CO$_2$, and amorphous C. This program does not include parameters for modeling stratified layers with heterogeneous compositions and/or grain sizes. Subsequent spectral analysis efforts will need to utilize codes that include components for modeling discrete compositional layers, along with multiple scattering of photons from more than one layer, in order to further investigate whether the Uranian satellites' near-surfaces are compositionally stratified. Additionally, the penetration depths discussed above do not take amorphous C into account, which strongly absorbs photons over the wavelength ranges of both SpeX and IRAC. Given the fairly flat absorption of amorphous C over the wavelength ranges covered by SpeX and IRAC, the depths probed by SpeX into H$_2$O-amorphous C particulate mixtures should still be greater than those probed by IRAC (assuming amorphous C is intimately mixed to depths of at least ~100 μm). Subsequent numerical modeling of the penetration potential of photons into intimate mixtures of H$_2$O ice and low albedo materials like amorphous C is also needed to further investigate near-surface layering on these satellites.

*5.2 Jupiter to Neptune: CO$_2$ on icy satellites*

To help provide context, we briefly compare the CO$_2$ ice detected on icy satellites elsewhere in the Solar System to the CO$_2$ detected on the classical Uranian moons. CO$_2$ ice has been detected on icy moons in all four giant planet systems (*e.g.*, McCord et al., 1997; Clark et al., 2005; Grundy et al, 2003; Cruikshank et al., 1993). The nature of CO$_2$ in each of these systems is a strong function of mean surface temperature; complexed and/or trapped CO$_2$ molecules in refractory matrices dominate the spectral signature of CO$_2$ at Jupiter and Saturn, while pure CO$_2$ frost dominates at Uranus and Neptune (*e.g.*, Clark et al., 2013 and references therein). On airless icy satellites with minimal surface pressures, CO$_2$ is susceptible to sublimation, except on distant Triton where surface temperatures are too low to drive CO$_2$ removal via sublimation (Lebofsky, 1975). Therefore, pure CO$_2$ ice is unstable over the age of the Solar System in the Jovian, Saturnian, and Uranian systems, and deposits of pure CO$_2$ must be actively synthesized and/or represent native deposits recently exposed by impacts, tectonism, or emplaced by cryovolcanism.

CO$_2$ deposits in the Jovian and Saturnian systems are associated with low albedo, and likely organic rich regions (*e.g.*, McCord et al., 1998; Hansen and McCord, 2008; Clark et al., 2005, 2008; Cruikshank et al., 2010). Aside from the CO$_2$ ice associated with some fresh impact craters on Callisto (Hibbitts et al., 2000, 2002), and the CO$_2$ detected in Enceladus' plumes (Waite et al., 2006), there is no clear correlation of CO$_2$ with geologic features. The spatial correlation between dirty $H_2O$ ice and $CO_2$ band strength on Jovian and Saturnian icy satellites supports active synthesis mechanisms, driven by UV photolysis (*e.g.*, Palmer and Brown, 2011) and charged particle radiolysis (*e.g.*, Cassidy et al., 2010).

Europa, Callisto, Dione, and Rhea are all embedded within their primary's magnetosphere, and they all have deeper $CO_2$ bands on their trailing hemispheres, which is consistent with $CO_2$ synthesis via charged particle radiolysis on these bodies. While Ganymede is also embedded within the magnetosphere of Jupiter, its intrinsic magnetic field deflects charged particles, which might explain why its surface does not display a clear leading/trailing asymmetry in $CO_2$ band strength (Hibbitts et al., 2003). Hyperion (chaotic rotation) and Phoebe (non-synchronous rotation) also do not display global asymmetries in $CO_2$ distribution. Iapetus does display hemispherical asymmetries in $CO_2$ abundance, but unlike the inner satellites Dione and Rhea, its deepest $CO_2$ bands are found in association with transitional longitudes between its low albedo leading hemisphere and $H_2O$ ice-rich trailing hemisphere (Pinilla-Alonso et al., 2011). Palmer and Brown (2011) estimated that $CO_2$ synthesis via UV photolysis on Iapetus should be rapid enough to roughly balance out sublimation and other loss mechanisms. Photolysis of localized materials (*i.e.*, $H_2O$ and organics) can neatly explain the deep $CO_2$ bands at leading and transitional longitudes on Iapetus, as well as the weak $CO_2$ bands on the trailing hemisphere, which is dominated by relatively pure $H_2O$, devoid of organic materials.

Similar to spectra of the large Uranian satellites, NIR observations have detected $CO_2$ combination and overtone bands between 1.9 and 2.2 μm on Triton (Cruikshank et al., 1993), suggestive of pure $CO_2$ ice. However, these $CO_2$ bands display no longitudinal variation (Grundy et al., 2010), unlike the clear hemispherical asymmetries in $CO_2$ band areas observed on the Uranian moons. At Triton's estimated mean surface temperature (~38 K), $CO_2$ ice is very stable and is unlikely to participate in volatile migration cycles (Grundy et al., 2010). $CO_2$ ice on the relatively warmer Uranian satellite surfaces (~70 – 80 K, Grundy et al., 2006), is unstable and should be effectively removed by sublimation over the age of the Solar System.

Our band parameter and statistical analyses demonstrate that $CO_2$ bands are strongest on the trailing hemispheres of the satellites closest to Uranus. Similar hemispherical asymmetries in $CO_2$ band strength are seen in spectra of Europa, Callisto, Dione and Rhea, which is consistent with $CO_2$ generation via charged particle radiolysis on these magnetospherically-embedded satellites. Furthermore, the low geometric albedos of Ariel, Umbriel, Titania, and Oberon (~0.25 – 0.55), along with Voyager 2 images of their gray-toned surfaces, indicates that they are coated in a mixture of $H_2O$ ice and low albedo, presumably C-rich, materials. Whereas UV photolysis likely generates some amount of $CO_2$ in all four planets' satellite systems, this mechanism should lead to a fairly uniform distribution, without leading/trailing asymmetries in $CO_2$ band strength. Iapetus is a notable exception, with a non-homogenous distribution in $CO_2$ likely driven by UV photolysis involving non-homogeneously distributed carbon-bearing low-albedo material (Palmer and Brown, 2011). Iapetus, however, is unique in its dramatic hemispheric dichotomy due to the moon sweeping up dust from Phoebe.

Thus, comparisons between the distribution of $CO_2$ on the Uranian satellites with the distribution on the Jovian and Saturnian satellites tends to support active production of $CO_2$ in the Uranian system, most likely driven by magnetospheric charged particle radiolysis. The detected spectral features suggest that $CO_2$ is pure and segregated from other surface species on the classical Uranian moons, similar to the $CO_2$ detected on Triton.

*5.3 Possible sources of CO₂ and relevant removal mechanisms*

The statistically significant accumulation of $CO_2$ on the trailing hemispheres of the Uranian satellites suggests that a system-wide mechanism is controlling the observed $CO_2$ distribution – most likely, magnetospherically-embedded charged particles. However, a wide range of processes can account for the presence of $CO_2$ on icy satellites. We explore each of these potential sources of $CO_2$ in Appendix C and only summarize our conclusions here. Potential sources of $CO_2$ fall into three broad categories: (1) native $CO_2$ that was accreted along with other primordial materials as they were forming (impact/tectonically exposed and cryovolcanically emplaced $CO_2$), (2) deposition of non-native $CO_2$ molecules sourced from a different body (intraplanetary, interplanetary, and interstellar dust), or re-deposition of sublimated native $CO_2$ in cold traps, and (3) non-native $CO_2$ synthesized by ion, electron, and UV irradiation of $H_2O$ ice and carbon-rich species.

It seems unlikely that native $CO_2$ exposed by impacts and tectonism, or emplaced by cryovolcanism, could generate the observed preferential accumulation of $CO_2$ on the trailing hemispheres of all four moons, along with the observed planetocentric trend (reduction in $CO_2$ abundance with distance from Uranus). Although the impact energy from heliocentric and planetocentric dust could drive $CO_2$ synthesis, this mechanism would likely lead to preferential accumulation of $CO_2$ on the leading hemispheres of these satellites, counter to the observed distribution. Sublimation of $CO_2$ driven by subsolar heating is unlikely to influence the observed distribution given that this mechanism should operate evenly on both hemispheres of all four satellites, which would tend to homogenize the distribution of $CO_2$ and mask hemispherical and planetocentric trends. Similarly, $CO_2$ synthesis and destruction via UV photolysis should also operate evenly throughout the Uranian system and is an unlikely driver of hemispherical and planetocentric asymmetries.

Laboratory experiments demonstrate that radiolysis of $H_2O$ ice and C-rich substrates, driven by charged particle irradiation, can readily generate $CO_2$ (*e.g.*, Spinks and Wood, 1990; Mennella et al., 2004; Gomis and Strazzulla, 2005; Sedlacko et al., 2005; Jamieson et al., 2006; Kim and Kaiser, 2012; Raut et al., 2012) (Appendix C). Charged particles embedded in Uranus' magnetosphere should preferentially bombard the trailing hemispheres of these satellites, which is consistent with the observed hemispherical trend. Additionally, magnetospheric particle densities fall off rapidly with distance from the primary, which is consistent with the observed planetocentric trend in $CO_2$. The Uranian magnetosphere is dominated by protons and electrons, with no evidence for $C^+$ or other heavy ions (see Appendix D for a brief review of Uranus' magnetic field and plasma environment). However, the surfaces of these moons are dominated by intimately mixed $H_2O$ ice and a low albedo, presumably C-rich, constituent, which represents an ideal substrate for $CO_2$ production via irradiation. Consequently, $C^+$ ion implantation is not required to explain the presence of radiolytically-generated $CO_2$ on the Uranian satellites. Additionally, experiments investigating the irradiation potential of the Uranian satellites demonstrated that the electron and proton fluxes measured by Voyager 2 are sufficient to chemically modify their surfaces (Lanzerotti et al., 1987). Thus, we conclude that irradiation by magnetospherically-trapped charged particles is the most likely mechanism for explaining the observed distribution in $CO_2$. Detailed numerical modeling of Uranian satellite-magnetospheric interactions, which is beyond the scope of this study, would help constrain the regions on these satellites' surfaces that experience the largest charged particle fluxes.

## 6. Summary and Conclusions

We have used multiple methodologies to quantify the distribution of $CO_2$ ice on Ariel, Umbriel, Titania, and Oberon, including: (1) visual identification of bands, (2) band area measurements, (3) band center modeling, (4) statistical analysis of the distribution of $CO_2$, (5) spectral modeling of mean trailing hemisphere spectra, and (6) relative $CO_2$ band area ratios for the mean spectra and a wide range of synthetic spectra. Combining the results of these procedures, we find clear evidence for significant amounts of $CO_2$ ice on the trailing hemispheres of Ariel, Umbriel, Titania, and Oberon. While Oberon lacks visibly apparent $CO_2$ bands, our statistical analysis of the distribution of $CO_2$, as well as our extensive spectral modeling and relative band area ratio analysis (Appendix B), supports the presence of $CO_2$ on this satellite. Additionally, the narrow profiles of the $CO_2$ bands and the lack of the $CO_2$ $2\nu_3$ feature near 2.134 μm, along with the results of our $CO_2$ relative band area ratio analysis, suggests that the detected $CO_2$ is dominantly pure and segregated from other constituents.

The mismatch between our best fit SpeX and IRAC numerical models suggests that we are probing distinct layers in the near-surfaces of these moons. Our SpeX models include intimate mixtures of $H_2O$ and amorphous C, mixed areally with $CO_2$ ice (5 – 50 μm grain sizes for all three components), whereas our IRAC models are dominated by small $H_2O$ ice grains with only minor amounts of $CO_2$ ice and/or amorphous C (0.5 – 2 μm grain sizes for all three components). Thus, the results of our spectral color analysis, along with our photon penetration depth estimates, suggests that the $CO_2$ ice detected by SpeX is present at greater depths (~100 μm) than those probed by IRAC (~10 μm) into pure crystalline $H_2O$ ice. Spectral models that can account for multiple heterogeneous layers are needed to further investigate whether the near-surfaces of these moons are compositionally stratified.

We also investigated the distribution of $H_2O$ ice on these moons using band parameter and statistical analyses on the 1.52 and 2.02 μm $H_2O$ bands. Our results indicate that $H_2O$ ice bands are stronger on the leading hemispheres of all four moons. Ratios between the mean $H_2O$ band areas on the leading and trailing hemispheres of these satellites demonstrate that hemispherical asymmetries in $H_2O$ band strength decrease with distance from Uranus, with only a slight asymmetry on the furthest classical moon, Oberon.

Our analyses demonstrate that these moons display clear hemispherical asymmetries in composition. Both the abundance of $CO_2$ ice and hemispherical asymmetries in $H_2O$ ice band strengths decrease with increasing orbital radius. We have identified magnetospheric charged particle bombardment as the most likely driver of the observed distribution of $CO_2$. Radiolytic production of $CO_2$, and other oxidized carbonaceous species, could lead to the removal and/or masking of native $H_2O$ ice on the trailing hemispheres of these moons, which in turn, would enhance leading/trailing asymmetries in $H_2O$ band strengths on the moons closest to Uranus. Thus, charged particle bombardment might account for the hemispherical and planetocentric trends in the distribution of both $CO_2$ and $H_2O$ ices on the classical Uranian satellites. Numerical models that investigate interactions between Uranus' magnetosphere and the Uranian satellites are needed to further test whether magnetospheric charged particle bombardment represents the primary control on the distribution of $CO_2$ and $H_2O$. Additionally, future NIR observations of the Uranian satellites will continue to improve our understanding of their surface composition and further constrain the distribution of $CO_2$ and $H_2O$ on these moons.

## Acknowledgments

This study was funded by a NASA Earth and Space Science Fellowship (grant number NNX14AP16H), as well as NASA Planetary Astronomy grant NNX10AB23G. Additional funding was provided by Tom Cronin and Helen Sestak whom we thank for their generous support. We also thank William Grundy for access to Uranian satellite SpeX spectra collected by his team between 2001 and 2006. Additionally, we thank the people of Hawaii for allowing us to use Mauna Kea for our observations, as well as the IRTF telescope operators and staff for providing observer support. We would also like to thank Sean Lindsay for his assistance with our modifications to the band analysis program SARA. Two anonymous reviewers provided insightful feedback on an earlier version of this paper.

## References

Acuna, M.H., Connerney, J.E.P., Ness, N.F., 1988. Implications of the $GSFCQ_3$ model for trapped particle motion. J. Geophys. Res. 93, 5505-5512.

Beddingfield, C.B., Burr, D.M., Emery, J.P., 2015. Fault geometries on Uranus' satellite Miranda: Implications for internal structure and heat flow. Icarus 247, 35-52, http://dx.doi.org/10.1016/j.icarus.2014.09.048.

Bell III, J.F., McCord, T.B., 1991. A search for spectral units on the Uranian satellites using color ratio images. Lunar Planet. Sci. XXI: $473 - 489$ (abstract).

Bernstein, M.P., Cruikshank, D.P., Sandford, S.A., 2005. Near-infrared laboratory spectra of solid $H_2O/CO_2$ and $CH_3OH/CO_2$ ice mixtures. Icarus 179, 527–534.

Bohren, C.F., Huffman, D.R., 1983. Absorption and scattering of light by small particles. John Wiley & Sons, Inc., Weinheim.

Borucki, J.G., Khare, B., Cruikshank, D.P., 2002. A new energy source for organic synthesis in Europa's surface ice. J. Geophys. Res. 107, $24\text{-}1 - 24\text{-}5$, DOI: 10.1029/2002JE001841.

Brown, R.H., Clark, R.N., 1984. Surface of Miranda: Identification of water ice. Icarus 58, 288-292.

Brown R.H., Cruikshank, D.P., 1983. The Uranian satellites: Surface compositions and opposition brightness surges. Icarus 55, 83-92.

Buratti, B.J., Mosher, J.A., 1991. Comparative global albedo and color maps of the Uranian satellites. Icarus 90, 1-13.

Cassidy, T.A., Coll, P., Raulin, F., Carlson, R.W., Johnson, R.E., Loeffler, M.J., Hand, K.P., Baragiola, R.A., 2010. Radiolysis and photolysis of icy satellite surfaces: Experiments and theory. Space Sci. Rev. 153, 299-315, DOI 10.1007/s11214-009-9625-3.

Cassidy, T.A., Paranicas, C.P., Shirley, J.H., Dalton, J.B., Teolis, B.D., Johnson, R.E., Kamp, L., Hnedrix, A.R., 2013. Magnetospheric ion sputtering and water ice grain size at Europa. Planet. Space Sci. 77, 64-73.

Chakarov, D.V., Gleeson, M.A., Kasemo, B., 2001. Photoreactions of water and carbon at 90 K. J. Chem. Phys. 115, 9477-9483, http://dx.doi.org/10.1063/1.1414375.

Cheng, A.F., Krimigis, S.M., Mauk, B.H., Keath, E.P., Maclennan, C.G., Lanzerotti, L.J., Paonessa, M.T., Armstrong, T.P., 1987. Energetic ion and electron phase space densities in the magnetosphere of Uranus. J. Geophys. Res. 92, 15315-15328.

Clark, R.N., Lucey, P.G., 1984. Spectral properties of ice-particulate mixtures and implications for remote sensing 1. Intimate mixtures. J. Geophys. Res. 89, 6341-6348.

Clark, R.N. et al. 2005. Compositional maps of Saturn's moon Phoebe from imaging spectroscopy. Nature 435, 66–69.

Clark R.N., Curchin J.M., Jaumann R., Cruikshank D.P., Brown R.H., Hoefen T.M., Stephan K., Moore, J.M., Buratti B.J., Baines K.H., Nicholson P.D., Nelson R.M., 2008. Compositional mapping of Saturn's satellite Dione with Cassini VIMS and implications of dark material in the Saturn system. Icarus 193, 372–386.

Clark, R.N., Carlson, R., Grundy, W.M., Noll, K., 2013. Observed ices in the Solar System. In: Gudipati, M.S., Castillo-Rogez, J. (Eds), The Science of Solar System Ices. Springer, New York, pp 3-46, DOI: 10.1007/978-1-4614-3076-6.

Connerney, J.E.P., Acuna, M.H., Ness, N.F., 1987. The Magnetic Field of Uranus. J. Geophys. Res. 92, 15329-15336.

Croft, S.K., Soderblom, L.A., 1991. Geology of the uranian satellites. In: Bergstralh, J.T., Miner, E.D., Matthews, M.S. (Eds.), Uranus. Univ. of Arizona Press, Tucson, pp. 561–628.

Cruikshank, D.P., Roush, T.L., Owen, T.C., Geballe, T.R., de Bergh, C., Schmitt, B., Brown, R.H., Bartholomew, M.J., 1993. Ices on the surface of Triton. Science 261, 742–745.

Cruikshank, D.P. et al., 2010. Carbon dioxide on the satellites of Saturn: Results from the Cassini VIMS investigation and revisions to the VIMS wavelength scale. Icarus 206, 561-572, doi:10.1016/j.icarus.2009.07.012.

Cushing, M.C., Vacca, W.D., Rayner, J.T., 2004. Spextool: A Spectral Extraction Package for SpeX, a $0.8 - 5.5$ Micron Cross-Dispersed Spectrograph. Astron. Soc. of the Pacific 116, 362-376.

Delitsky, M.L., Lane, A.L., 1998. Ice chemistry on the Galilean satellites. J. Geophys. Res. 103, 31391–31403.

Emery, J.P., Burr, D.M., Cruikshank, D.P., Brown R.H., Dalton, J.B., 2005. Near-infrared (0.8–4.0 μm) spectroscopy of Mimas, Enceladus, Tethys, and Rhea. Astro. & Astrophys. 435. 353-362, DOI: 10.1051/0004-6361:20042482.

Emery, J.P., Cruikshank, D.P., Van Cleve, J., 2006. Thermal emission spectroscopy (5.2 – 38 μm) of three Trojan asteroids with the Spitzer Space Telescope: Detection of fine-grained silicates. Icarus 182, 496-512, doi:10.1016/j.icarus.2006.01.011.

Fazio G.G. et al., 2004. The Infrared Array Camera (IRAC) for the Spitzer Space Telescope. Astrophys. J. Supple. Series 154, 10-17.

Gerakines, P.A., Moore, M.H., 2001. Carbon suboxide in astrophysical ice analogs. Icarus 154, 372–380

Gerakines, P.A., Bray, J.J., Davis, A., Richey, C.R., 2005. The strengths of near-infrared absorption features relevant to interstellar and planetary ices. Astrophys. J. 620, 1140–1150.

Gomis, O., Strazzulla, G., 2005. $CO_2$ production by ion irradiation of $H_2O$ ice on top of carbonaceous materials and its relevance to the Galilean satellites. Icarus 177, 570–576.

Grundy, W.M., Young, L.A., Young, E.F., 2003. Discovery of $CO_2$ ice and leading-trailing spectral asymmetry on the Uranian satellite Ariel. Icarus 162, 222-229.

Grundy, W.M., Young, L.A., Spencer, J.R., Johnson, R.E., Young, E.F., Buie, M.W., 2006. Distributions of $H_2O$ and $CO_2$ ices on Ariel, Umbriel, Titania, and Oberon from IRTF/SpeX observations. Icarus 184, 543-555.

Grundy, W.M., Young, L.A., Stansberry, J.A., Buie, M.W., Olkin, C.B., Young E.F., 2010. Near-infrared spectral monitoring of Triton with IRTF/SpeX II: Spatial distribution and evolution of ices. Icarus 205, 594-604.

Hansen, G.B., 1997. The infrared absorption spectrum of carbon dioxide ice from 1.8 to 333 microns. J. Geophys. Res. 102, 21569–21587.

Hansen, G.B., 2005. Ultraviolet to near-infrared absorption spectrum of carbon dioxide ice from 0.174 to 1.8 μm. J. Geophys. Res. 110. E11003.1-18.

Hansen, G.B., McCord, T.B., 2008. Widespread $CO_2$ and other non-ice compounds on the anti-Jovian and trailing sides of Europa from Galileo/NIMS observations. Geophys. Res. Let. 35, LO1202, doi:10.1029/2007GL031748.

Hapke, B., 2002. Theory of reflectance and emittance spectroscopy. Cambridge Univer. Press, New York.

Helfenstein, P., Veverka, J., 1988. Early resurfacing of Umbriel: Evidence from Voyager 2 photometry. Lunar Planet. Sci. XIX: 477 – 478 (abstract).

Helfenstein, P., Thomas, P.C., Veverka, J., 1989. Evidence from Voyager II photometry for early resurfacing of Umbriel. Nature 338, 324-326.

Helfenstein, P., Hillier, J., Weitz, C., Veverka, J., 1991. Oberon Color Photometry from Voyager and Its Geological Implications. Icarus 90, 14-29.

Hibbitts, C.A., McCord, T.B., Hansen, G.B., 2000. Distributions of $CO_2$ and $SO_2$ on the surface of Callisto. J. Geophys. Res. 105, 22541–22557.

Hibbitts, C.A., Klemaszewski, J., McCord, T.B., Hansen, G.B., Greeley, R., 2002. $CO_2$-rich impact craters on Callisto, J. Geophys. Res., 107, 5084, doi:10.1029/2000JE001412.

Hibbitts, C.A., Pappalardo, R.T., Hansen, G.B., McCord, T.B., 2003. Carbon dioxide on Ganymede. J. Geophys. Res. 108. 1956.1-22

Hudson, R.L., Moore, M.H., 2001. Radiation chemical alterations in Solar System ices: An overview. J. Geophys. Res. 106, 33275–33284.

Jamieson, C.S., Mebel, A.M., Kaiser, R.I., 2006. Understanding the kinetics and dynamics of radiation-induced reaction pathways in carbon monoxide ice at 10 K. Astrophys. J. Sup. Series 163, 184-206.

Jankowski, D.G., Squyres, S.W., 1988. Solid-State Ice Volcanism on the Satellites of Uranus. Science 241, 1322-1325.

Jarmillo-Botero, A., Cheng, Q.A.M., Goddard, W.A., 2012. Hypervelocity Impact Effect of Molecules from Enceladus' Plume and Titan's Upper Atmosphere on NASA's Cassini Spectrometer from Reactive Dynamics Simulation. Phys. Rev. Let. 109, 213201-1 – 213201-5.

Johnson, R.E., Lanzerotti, L.J., Brown, W.L., 1984. Sputtering processes: Erosion and chemical change. Adv. Space Rev. 4, 41-51.

Kargel, J.S., 1994. Cryovolcanism on the icy satellites. Earth Moon Planets 67, 101-113.

Karkoschka, E., 2001. Comprehensive photometry of the rings and 16 satellites of Uranus with the Hubble Space Telescope. Icarus 151, 51-68, doi:10.1006/icar.2001.6596.

Kim, Y.S., Kaiser, R.I., 2012. Electron irradiation of Kuiper Belt surface ices: Ternary $N_2$-$CH_4$-CO mixtures as a case study. Astrophys. J. 758, 37-42. doi:10.1088/0004-637X/758/1/37.

Krimigis, S.M., Armstrong, T.P., Axford, W.I., Cheng, A.F., Gloeckler, G., Hamilton, D.C., Keath, E.P., Lanzerotti L.J., Mauk, B.H., 1986. The magnetosphere of Uranus: Hot plasma and radiation environment. Science 233, 97-102.

Lane, A.L., Nelson, R.M., Matson, D.L., 1981. Evidence for sulphur implantation in Europa's UV absorption band. Nature 292, 38–39.

Lanzerotti, L.J., Brown, W.L., Maclennan, C.G., Cheng, A.F., Krimigis, S.M., Johnson, R.E., 1987. Effects of charged particles on the surfaces of the satellites of Uranus. J. Geophys. Res. 92, 14949-14957.

Lebofsky, L.A., 1975, Stability of frosts in the Solar System. Icarus 25, 205-217.

Lindsay, S.S., Marchis, F., Emery, J.P., Enriquez, J.E., Assafin, M., 2015. Composition, mineralogy, and porosity of multiple asteroid systems from visible and near-infrared spectral data. Icarus 247, 53-70. http://dx.doi.org/10.1016/j.icarus.2014.08.040

Loeffler, M.J., Baratta, G.A., Palumbo, M.E., Strazzulla, G., Baragiola, R.A., 2005. $CO_2$ synthesis in solid CO by Lyman-$\alpha$ photons and 200 keV protons. Astro. & Astrophys. 435, 587-594, DOI: 10.1051/0004-6361:20042256.

Mastrapa, R.M., Bernstein, M.P., Sandford, S.A., Roush, T.L., Cruikshank, D.P., Dalle Ore, C.M., 2008. Optical constants of amorphous and crystalline H2O-ice in the near infrared from 1.1 to 2.6 μm. Icarus 197, 307-320.

Mastrapa, R.M., Sandford, S.A., Roush, T.L., Cruikshank, D.P., Dalle Ore, C.M., 2009. Optical constants of amorphous and crystalline H2O-ice: $2.5 - 22$ μm $(4000 - 455$ $cm^{-1})$ optical constant of $H_2O$-ice. Astro. J. 701, 1347-1356.

McCord, T.B. et al., 1997. Organics and other molecules in the surfaces of Callisto and Ganymede. Science 278, 271–275.

McCord, T.B. et al., 1998. Nonwater-ice constituents in the surface material of the icy Galilean satellites from the Galileo near-infrared mapping spectrometer investigation. J. Geophys. Res. 103, 8603–8626

Mennella, V., Palumbo, M.E., Baratta, G.A., 2004. Formation of CO and $CO_2$ molecules by ion irradiation of water ice-covered hydrogenated carbon grains. Astrophys. J. 615, 1073–1080.

Mennella, V., Baratta, G.A., Palumbo, M.E., Bergin, E.A., 2006. Synthesis of CO and $CO_2$ molecules by UV irradiation of water ice-covered hydrogenated carbon grains. Astro. J. 643, 923-931.

Ness, N.F., Acuna, M.H., Behannon, K.W., Burlaga, L.F., Connerney, J.E.P., Lepping, R.P., Neubauer, F.M., 1986. Magnetic Fields at Uranus. Science 233, 85-89.

Noll, K.S., Weaver, H.A., Gonnella, A.M., 1995. The albedo spectrum of Europa from 2200 to 3300 A˚. J. Geophys. Res. 100, 19057–19059.

Noll, K.S., Roush, T.L., Cruikshank, D.P., Johnson, R.E., Pendleton, Y.J., 1997. Detection of ozone on Saturn's satellites Rhea and Dione. Nature 388, 45–47.

Ockert, M.E., Nelson, R.M., Lane, A.L., Matson, D.L., 1987. Europa's ultraviolet absorption band (260 to 320 nm) – temporal and spatial evidence from IUE. Icarus 70, 499–505.

Palmer, E.E., Brown, R.H., 2011. Production and detection of carbon dioxide on Iapetus. Icarus 212, 807-818.

Pappalardo, R.T., Reynolds, S.J., Greeley, R., 1997. Extensional tilt blocks on Miranda: Evidence for an upwelling origin of Arden Corona. J. Geophys. Res.: Planets 102 (E6), 13369–13379.

Pinilla-Alonso, N., Roush, T.L., Marzo, G.A., Cruikshank, D.P., Dalle Ore, C.M., 2011. Iapetus surface variability revealed from statistical clustering of a VIMS mosaic: The distribution of $CO_2$. Icarus 215, 75-82.

Raut, U., Fulvio, D., Loeffler, M.J., Baragiola, R.A., 2012. Radiation synthesis of carbon dioxide in ice-coated carbon: Implications for interstellar grains and icy moons. Astrophys. J. 752, 159 (8pp), doi:10.1088/0004-637X/752/2/159.

Rayner, J.T., Toomey, D.W., Onaka, P.M., Denault, A.J., Stahlberger, W.E., Watanabe, D.Y., Wang, S.I., 1998. SpeX: A medium-resolution IR spectrograph for IRTF. Proc. SPIE 3354, 468–479.

Rayner, J.T., Toomey, D.W., Onaka, P.M., Denault, A.J., Stahlberger, W.E., Vacca, W.D., Cushing, M.C., Wang, S., 2003. SpeX: A medium-resolution 0.8-5.5 micron spectrograph and imager for the NASA Infrared Telescope Facility. Astron. Soc. of the Pacific 115, 362.

Rayner, J.T., Cushing, M.C., Vacca, W.D., 2009. The Infrared Telescope Facility (IRTF) Spectral Library: Cool stars. Astrophys. J. Sup. Series 185, 289-432. doi:10.1088/0067-0049/185/2/289

Reach, W.T. et al., 2005. Absolute Calibration of the Infrared Array Camera on the Spitzer Space Telescope. Astron. Soc. of the Pacific 117, 978-990.

Schenk, P.M., 1991. Fluid volcanism on Miranda and Ariel: Flow morphology and composition. J. Geophys. Res. 96, 1887-1906.

Schmidt, J.A., Johnson, M.S., Schinke, R., 2013. Carbon dioxide photolysis from 150 to 210 nm: Singlet and triplet channel dynamics, UV spectrum, and isotope effects. Proceed. Nation. Acad. Sci. 110, 17691-17696. www.pnas.org/cgi/doi/10.1073/pnas.1213083110

Sedlacko, T., Balog, R., Lafosse, A., Stano, M., Matejcik, S., Azria, R., Illenberger, E., 2005. Reactions in condensed formic acid (HCOOH) induced by low energy (< 20 eV) electrons. Phys. Chem. 7, 1277-1282, DOI:10.1039/b419104h.

Shoemaker, E.M., Wolfe, R.A., 1982. Cratering timescales for the Galilean satellites, in: Morrison, D. (Ed.), Satellites of Jupiter, Univ. of Arizona Press, Tucson, pp. 277–339.

Smith, B.A. et al., 1986. Voyager 2 in the Uranian System: Imaging Science Results. Science 233, 43-64.

Smith, E.V.P., Gottlieb, D.M., 1974. Solar Flux and its Variations. Space Sci. Rev., 16, 771-802.

Soifer, B.T., Neugebauer, G., Matthews, K., 1981. Near-infrared spectrophotometry of the satellites and rings of Uranus. Icarus 45, 612-617.

Spiegel, M.R., 1992. Theory and Problems of Probability and Statistics. McGraw-Hill, New York, pp. 117-118.

Spinks, J.W.T., Woods, R.J., 1990. An Introduction to Radiation Chemistry. John Wiley, New York, pp. 574.

Stone, E.C., Miller, E.D., 1986. The Voyager-2 encounter with the Uranian system. Science 233, 39-43.

Stone, E.C., Cooper, J.T., Cummings, A.C., McDonald, F.B., Trainor, J.H., Lal, N., McGuire, R., Chenette, D.L., 1986. Energetic Charged Particles in the Uranian Magnetosphere. Science 233, 93-97.

Stryk, T., Stooke, P.J., 2008. Voyager 2 images of Uranian satellites: Reprocessing and new interpretations. Lunar Planet. Sci. XXXIX, 1362 (abstract).

Tamayo, D., Burns, J.A., Hamilton, D.P., 2013. Chaotic dust dynamics and implications for the hemispherical color asymmetries of the Uranian satellites. Icarus 226, 655-662, http://dx.doi.org/10.1016/j.icarus.2013.06.018.

Tittemore, W.C., Wisdom, J., 1990. Tidal Evolution of the Uranian Satellites. Icarus 85, 394-443.

Vacca, W.D., Cushing, M.C., Rayner, J.T., 2003. A method of correcting near-infrared spectra for telluric absorption. Astron. Soc. of the Pacific 115, 389-409.

Waite, J.H. et al., 2006. Cassini Ion and Neutral Mass Spectrometer: Enceladus plume composition and structure. Science 311, 1419-1422, DOI: 10.1126/science.1121290.

Werner, M.W. et al., 2004. The Spitzer Space Telescope Mission. Astrophys. J. Supple. Series, 154, 1-9.

Zahnle, K., Schenk, P., Sobieszczyk, S., Dones, L., Levison, H.F., 2001. Differential cratering of synchronously rotating satellites by ecliptic comets. Icarus 153, 111–129.

Zahnle, K., Schenk, P., Levison, H., Dones, L., 2003. Cratering rates in the outer Solar System. Icarus 163, 263–289.

**Table 1: Classical Uranian satellites**

| Satellite | Orbital Radius (km) | Orbital Radius ($R_{Uranus}$) | Orbital Period (days) | Radius (km) | Density (g cm$^{-3}$) | *Geometric Albedo (~0.957 μm) |
|---|---|---|---|---|---|---|
| Miranda | 129,900 | 5.12 | 1.41 | 236 | 1.21 | 0.45 ± 0.02 |
| Ariel | 190,900 | 7.53 | 2.52 | 579 | 1.59 | 0.56 ± 0.02 |
| Umbriel | 266,000 | 10.49 | 4.14 | 585 | 1.46 | 0.26 ± 0.01 |
| Titania | 436,300 | 17.20 | 8.71 | 789 | 1.66 | 0.39 ± 0.02 |
| Oberon | 583,500 | 23.01 | 13.46 | 762 | 1.56 | 0.33 ± 0.01 |

*Geometric albedos digitally extracted from Figure 7 in Karkoschka (2001).*

**Table 2: IRTF/SpeX Observations**

| Satellite | Satellite Long. (°) | Satellite Lat. (°) | Hemisphere Observed[†] | UT Date | UT Time (mid-expos) | $t_{int}$ (min) | Spectral Resolution ($\lambda/\Delta\lambda$) | Observing PI |
|---|---|---|---|---|---|---|---|---|
| Ariel | 53.6 | -16.0 | Leading | 8/9/2003 | 12:15 | 156 | ~1600 − 1700 | Grundy |
| | 79.8 | -19.4 | Leading | 7/17/2002 | 13:25 | 108 | ~1300 − 1400 | Grundy |
| | 87.8 | 24.0 | Leading | 9/5/2013 | 11:10 | 92 | ~650 − 750 | Cartwright |
| | 93.5 | -18.1 | Leading | 10/4/2003 | 5:45 | 108 | ~1600 − 1700 | Grundy |
| | 159.9 | -11.1 | Interm. (L) | 7/15/2004 | 12:00 | 112 | ~1600 − 1700 | Grundy |
| | 200.0 | -15.9 | Interm. (T) | 8/5/2003 | 12:00 | 84 | ~1600 − 1700 | Grundy |
| | 219.8 | -17.2 | Interm. (T) | 9/7/2003 | 9:35 | 90 | ~1600 − 1700 | Grundy |
| | 233.8 | -23.1 | Trailing | 7/5/2001 | 14:10 | 50 | ~1300 − 1400 | Grundy |
| | 257.6 | -29.5 | Trailing | 9/6/2000 | 7:35 | 76 | ~650 − 750 | Rivkin |
| | 278.3 | 24.8 | Trailing | 8/7/2013 | 13:20 | 44 | ~650 − 750 | Cartwright |
| | 294.8 | -19.3 | Trailing | 7/16/2002 | 13:10 | 140 | ~1300 − 1400 | Grundy |
| | 304.8 | -23.2 | Trailing | 7/8/2001 | 14:40 | 48 | ~1300 − 1400 | Grundy |
| | 316.6 | -18.2 | Interm. (T) | 10/8/2003 | 7:55 | 132 | ~1600 − 1700 | Grundy |
| Umbriel | 38.4 | 20.3 | Interm. (L) | 8/13/2012 | 13:55 | 60 | ~650 − 750 | Cartwright |
| | 75.2 | -29.3 | Leading | 9/6/2000 | 10:00 | 80 | ~650 − 750 | Rivkin |
| | 92.1 | -11.1 | Leading | 7/16/2004 | 14:25 | 74 | ~1600 − 1700 | Grundy |
| | 131.3 | 17.5 | Leading | 11/1/2012 | 9:30 | 78 | ~650 − 750 | Cartwright |

| | | | | | | | |
|---|---|---|---|---|---|---|---|
| | 216.2 | -10.9 | Interm. (T) | 7/5/2004 | 14:10 | 80 | ~1600 – 1700 | Grundy |
| | 219.8 | -23.0 | Interm. (T) | 7/7/2001 | 14:10 | 52 | ~1300 – 1400 | Grundy |
| | 261.3 | -9.4 | Trailing | 9/18/2005 | 9:21 | 196 | ~1600 – 1700 | Grundy |
| | 264.0 | 23.7 | Trailing | 9/5/2013 | 13:55 | 52 | ~650 – 750 | Cartwright |
| | 317.6 | -11.4 | Interm. (T) | 7/27/2004 | 11:30 | 184 | ~1600 – 1700 | Grundy |
| Titania | 13.6 | 19.0 | Interm. (L) | 9/21/2012 | 11:25 | 32 | ~650 – 750 | Cartwright |
| | 86.5 | 23.6 | Leading | 9/6/2013 | 10:50 | 60 | ~650 – 750 | Cartwright |
| | 98.0 | -18.1 | Leading | 10/8/2003 | 5:45 | 64 | ~1600 – 1700 | Grundy |
| | 160.0 | 18.2 | Interm. (L) | 10/12/2012 | 10:15 | 60 | ~650 – 750 | Cartwright |
| | 213.9 | -11.1 | Interm. (T) | 7/15/2004 | 14:25 | 64 | ~1600 – 1700 | Grundy |
| | 237.0 | -23.0 | Trailing | 7/6/2001 | 13:25 | 56 | ~1300 – 1400 | Grundy |
| | 258.9 | -29.3 | Trailing | 9/5/2000 | 9:55 | 60 | ~650 – 750 | Rivkin |
| | 277.8 | -23.0 | Trailing | 7/7/2001 | 13:10 | 36 | ~1300 – 1400 | Grundy |
| | 299.6 | -10.2 | Trailing | 10/13/2005 | 9:05 | 106 | ~1600 – 1700 | Grundy |
| | 342.7 | -29.4 | Interm. (T) | 9/7/2000 | 10:35 | 44 | ~650 – 750 | Rivkin |
| Oberon | 1.0 | -29.4 | Interm. (L) | 9/7/2000 | 8:30 | 32 | ~650 – 750 | Rivkin |
| | 64.8 | 18.1 | Leading | 10/12/2012 | 11:05 | 98 | ~650 – 750 | Cartwright |
| | 85.9 | -3.7 | Leading | 8/4/2006 | 13:22 | 92 | ~1600 – 1700 | Grundy |
| | 91.0 | 23.7 | Leading | 9/1/2013 | 11:40 | 56 | ~650 – 750 | Cartwright |
| | 110.6 | -10.2 | Leading | 10/13/2005 | 6:30 | 124 | ~1600 – 1700 | Grundy |
| | 164.0 | -23.0 | Interm. (L) | 7/6/2001 | 14:25 | 32 | ~1300 – 1400 | Grundy |
| | 216.2 | -23.1 | Interm. (T) | 7/8/2001 | 13:10 | 48 | ~1300 – 1400 | Grundy |
| | 223.7 | 18.9 | Trailing | 9/21/2012 | 10:20 | 64 | ~650 – 750 | Cartwright |
| | 236.1 | 17.4 | Trailing | 11/1/2012 | 6:55 | 40 | ~650 – 750 | Cartwright |
| | 266.5 | 20.2 | Trailing | 8/13/2012 | 15:15 | 28 | ~650 – 750 | Cartwright |
| | 307.2 | -29.3 | Trailing | 9/5/2000 | 8:10 | 60 | ~650 – 750 | Rivkin |

[†]'Leading' and 'trailing' table entries represent spectra collected over these moons' leading (046 – 135°) and trailing 226 – 315°) quadrants, respectively. 'Interm. (L)' and 'Interm. (T)' represent spectra collected over Uranus-facing (316 – 045°) and anti-Uranus (136 – 225°) quadrants, respectively.

**Table 3: Local standard and solar analog stars**

| Observing PI | UT date | Solar analogs |
|---|---|---|
| Rivkin | 9/5/2000 | SAO 164, 16 Cygni A |
| | 9/6/2000 | SAO 164, 16 Cygni Q |
| | 9/7/2000 | Sao 164 |
| Cartwright | 8/13/2012 | SAO 109348, SA 115-271 |
| | 9/21/2012 | SAO 109182, SAO 110201, SA 115-271 |
| | 10/12/2012 | SAO 109182, SA 115-271, Hyades 64 |
| | 11/1/2012 | SAO 109182, SAO 126133, Hyades 64 |
| | 8/7/2013 | SAO 109426, SAO 109450 |
| | 9/1/2013 | SAO 109426, SAO 109450, SAO 109348 |
| | 9/5/2013 | SAO 109426, SAO 109450, SAO 109348 |
| | 9/6/2013 | SAO 109426, SAO 109450, SAO 109348 |

## Table 4: Spitzer/IRAC Observations

| Satellite | Satellite Long. (°) | Satellite Lat. (°) | Program 71 Observation[†] | UT Date | UT Time (mid-expos) | $t_{int}$ per channel (sec) | Phase Angle | Heliocentric Distance | Observer Distance |
|---|---|---|---|---|---|---|---|---|---|
| Ariel | 54.0 | -18.4 | Titania-T | 12/3/2003 | 11:08 | 80 | 2.84 | 20.0372 | 20.1741 |
| | 87.8 | -6.7 | Ariel-L | 6/10/2005 | 4:04 | 80 | 2.90 | 20.0660 | 20.0770 |
| | 92.8 | -10.7 | Umbriel-L | 6/29/2004 | 21:55 | 80 | 2.57 | 20.0485 | 19.5529 |
| | 93.9 | -6.7 | Umbriel-T | 6/15/2005 | 6:03 | 80 | 2.90 | 20.0661 | 19.9914 |
| | 127.4 | -14.6 | Oberon-L | 11/23/2004 | 8:16 | 80 | 2.86 | 20.0556 | 19.8728 |
| | 255.5 | -18.4 | Oberon-T | 12/2/2003 | 8:28 | 80 | 2.84 | 20.0361 | 20.1543 |
| | 276.5 | -14.6 | Ariel-T | 11/26/2004 | 21:48 | 80 | 2.88 | 20.0566 | 19.9335 |
| Umbriel | 34.7 | -10.5 | Titania-L | 11/29/2005 | 9:13 | 80 | 2.81 | 20.0748 | 19.7977 |
| | 92.8 | -10.7 | Umbriel-L | 6/29/2004 | 21:55 | 80 | 2.57 | 20.0488 | 19.5532 |
| | 149.9 | -18.3 | Oberon-T | 12/2/2003 | 8:28 | 80 | 2.84 | 20.0349 | 20.1532 |
| | 151.8 | -14.5 | Ariel-T | 11/26/2004 | 21:48 | 80 | 2.88 | 20.0550 | 19.9319 |
| | 246.5 | -18.2 | Titania-T | 12/3/2003 | 11:08 | 80 | 2.84 | 20.0358 | 20.1727 |
| | 276.7 | -6.7 | Umbriel-T | 6/15/2005 | 6:03 | 80 | 2.90 | 20.0663 | 19.9916 |
| Titania | 83.1 | -10.5 | Titania-L | 11/29/2005 | 9:13 | 80 | 2.81 | 20.0737 | 19.7966 |
| | 216.7 | -18.3 | Oberon-T | 12/2/2003 | 8:28 | 80 | 2.84 | 20.0341 | 20.1524 |
| | 219.5 | -14.5 | Oberon-L | 11/23/2004 | 8:16 | 80 | 2.86 | 20.0541 | 19.8713 |
| | 262.6 | -18.3 | Titania-T | 12/3/2003 | 11:08 | 80 | 2.84 | 20.0361 | 20.1729 |
| | 285.1 | -10.7 | Umbriel-L | 6/29/2004 | 21:55 | 80 | 2.57 | 20.0492 | 19.5537 |
| Oberon | 88.3 | -14.6 | Oberon-L | 11/23/2004 | 8:16 | 80 | 2.86 | 20.0564 | 19.8736 |
| | 133.5 | -10.7 | Umbriel-L | 6/29/2004 | 21:55 | 80 | 2.57 | 20.0460 | 19.5504 |
| | 141.9 | -6.7 | Umbriel-T | 6/15/2005 | 6:03 | 80 | 2.90 | 20.0632 | 19.9884 |
| | 261.9 | -18.3 | Oberon-T | 12/2/2003 | 8:28 | 80 | 2.84 | 20.0358 | 20.1541 |
| | 290.3 | -10.6 | Titania-L | 11/29/2005 | 9:13 | 80 | 2.81 | 20.0747 | 19.7975 |
| | 291.6 | -18.3 | Titania-T | 12/3/2003 | 11:08 | 80 | 2.84 | 20.0378 | 20.1747 |

[†]*The names of the Spitzer/IRAC observations of the Uranian satellites collected during Program 71. Spitzer Heritage Archive: http://irsa.ipac.caltech.edu/data/SPITZER/docs/spitzerdataarchives/sha/*

## Table 5: Summary of IRAC fluxes and geometric albedos

| Target | IRAC Channel | Observation | Satellite Long. (°) | Satellite Lat. (°) | Flux (mJy) | ΔFlux (mJy) | Geometric Albedo | ΔGeometric Albedo |
|---|---|---|---|---|---|---|---|---|
| Ariel | Ch.1 | Titania_T | 54.0 | -18.4 | 0.595 | 0.015 | 0.191 | 0.005 |
| | | Ariel_L | 87.8 | -6.7 | 0.594 | 0.033 | 0.189 | 0.010 |
| | | Umbriel_L | 92.8 | -10.7 | 0.621 | 0.019 | 0.185 | 0.006 |
| | | Umbriel_T | 93.9 | -6.7 | 0.568 | 0.008 | 0.188 | 0.006 |
| | | Oberon_L | 127.4 | -14.6 | 0.597 | 0.028 | 0.186 | 0.009 |
| | | Oberon_T | 255.5 | -18.4 | 0.721 | 0.014 | 0.230 | 0.004 |
| | | Ariel_T | 276.5 | -14.6 | 0.744 | 0.016 | 0.233 | 0.005 |
| | Ch.2 | Titania_T | 54.0 | -18.4 | 0.212 | 0.022 | 0.104 | 0.011 |
| | | Ariel_L | 87.8 | -6.7 | 0.195 | 0.026 | 0.095 | 0.012 |
| | | Umbriel_L | 92.8 | -10.7 | 0.187 | 0.017 | 0.085 | 0.008 |
| | | Umbriel_T | 93.9 | -6.7 | 0.185 | 0.018 | 0.089 | 0.008 |
| | | Oberon_L | 127.4 | -14.6 | 0.187 | 0.025 | 0.089 | 0.012 |
| | | Oberon_T | 255.5 | -18.4 | 0.261 | 0.014 | 0.127 | 0.007 |
| | | Ariel_T | 276.5 | -14.6 | 0.254 | 0.018 | 0.121 | 0.008 |
| | Ch.3 | Titania_T | 54.0 | -18.4 | *0.027* | - | - | - |
| | | Ariel_L | 87.8 | -6.7 | 0.042 | 0.015 | 0.033 | 0.012 |
| | | Umbriel_L | 92.8 | -10.7 | 0.094 | 0.018 | 0.070 | 0.011 |
| | | Umbriel_T | 93.9 | -6.7 | 0.103 | 0.020 | 0.081 | 0.016 |
| | | Oberon_L | 127.4 | -14.6 | 0.108 | 0.036 | 0.084 | 0.028 |
| | | Oberon_T | 255.5 | -18.4 | 0.097 | 0.029 | 0.078 | 0.023 |
| | | Ariel_T | 276.5 | -14.6 | 0.127 | 0.025 | 0.099 | 0.020 |
| | Ch.4 | Titania_T | 54.0 | -18.4 | 0.333 | 0.111 | 0.485 | 0.161 |
| | | Ariel_L | 87.8 | -6.7 | 0.153 | 0.046 | 0.221 | 0.066 |
| | | Umbriel_L | 92.8 | -10.7 | 0.118 | 0.034 | 0.160 | 0.046 |
| | | Umbriel_T | 93.9 | -6.7 | 0.114 | 0.019 | 0.164 | 0.028 |
| | | Oberon_L | 127.4 | -14.6 | 0.157 | 0.054 | 0.222 | 0.076 |
| | | Oberon_T | 255.5 | -18.4 | 0.332 | 0.091 | 0.482 | 0.131 |
| | | Ariel_T | 276.5 | -14.6 | 0.168 | 0.033 | 0.239 | 0.047 |
| Umbriel | Ch.1 | Titania_L | 34.7 | -10.5 | 0.542 | 0.010 | 0.155 | 0.003 |
| | | Umbriel_L | 82.8 | -10.7 | 0.590 | 0.027 | 0.163 | 0.008 |
| | | Oberon_T | 149.9 | -18.3 | 0.502 | 0.011 | 0.149 | 0.003 |
| | | Ariel_T | 151.8 | -14.5 | 0.501 | 0.014 | 0.146 | 0.004 |
| | | Titania_T | 246.5 | -18.2 | 0.536 | 0.005 | 0.159 | 0.002 |
| | | Umbriel_T | 276.7 | -6.7 | 0.535 | 0.006 | 0.157 | 0.002 |
| | Ch.2 | Titania_L | 34.7 | -10.5 | 0.198 | 0.020 | 0.087 | 0.009 |
| | | Umbriel_L | 82.8 | -10.7 | 0.174 | 0.011 | 0.073 | 0.005 |
| | | Oberon_T | 149.9 | -18.3 | 0.171 | 0.014 | 0.077 | 0.006 |
| | | Ariel_T | 151.8 | -14.5 | 0.167 | 0.023 | 0.074 | 0.010 |
| | | Titania_T | 246.5 | -18.2 | 0.177 | 0.009 | 0.080 | 0.004 |
| | | Umbriel_T | 276.7 | -6.7 | 0.166 | 0.008 | 0.074 | 0.004 |
| | Ch.3 | Titania_L | 34.7 | -10.5 | 0.101 | 0.043 | 0.072 | 0.030 |
| | | Umbriel_L | 82.8 | -10.7 | 0.084 | 0.016 | 0.058 | 0.011 |
| | | Oberon_T | 149.9 | -18.3 | 0.074 | 0.027 | 0.055 | 0.020 |
| | | Ariel_T | 151.8 | -14.5 | *0.054* | - | - | - |
| | | Titania_T | 246.5 | -18.2 | 0.069 | 0.060 | 0.051 | 0.044 |
| | | Umbriel_T | 276.7 | -6.7 | 0.113 | 0.019 | 0.083 | 0.014 |
| | Ch.4 | Titania_L | 34.7 | -10.5 | 0.115 | 0.022 | 0.150 | 0.028 |
| | | Umbriel_L | 82.8 | -10.7 | 0.102 | 0.033 | 0.128 | 0.041 |
| | | Oberon_T | 149.9 | -18.3 | 0.295 | 0.117 | 0.397 | 0.157 |
| | | Ariel_T | 151.8 | -14.5 | 0.192 | 0.051 | 0.254 | 0.068 |
| | | Titania_T | 246.5 | -18.2 | 0.291 | 0.049 | 0.393 | 0.067 |
| | | Umbriel_T | 276.7 | -6.7 | 0.136 | 0.032 | 0.181 | 0.043 |
| Titania | Ch.1 | Titania_L | 83.1 | -10.5 | 1.001 | 0.008 | 0.160 | 0.001 |
| | | Oberon_T | 216.7 | -18.3 | 0.956 | 0.010 | 0.157 | 0.002 |
| | | Oberon_L | 219.5 | -14.5 | 0.982 | 0.004 | 0.158 | 0.001 |
| | | Titania_T | 262.6 | -18.3 | 0.982 | 0.011 | 0.162 | 0.002 |
| | | Umbriel_L | 285.1 | -10.7 | 1.063 | 0.005 | 0.163 | 0.001 |
| | Ch.2 | Titania_L | 83.1 | -10.5 | 0.261 | 0.009 | 0.063 | 0.002 |
| | | Oberon_T | 216.7 | -18.3 | 0.263 | 0.008 | 0.066 | 0.002 |
| | | Oberon_L | 219.5 | -14.5 | 0.263 | 0.004 | 0.064 | 0.001 |
| | | Titania_T | 262.6 | -18.3 | 0.265 | 0.004 | 0.067 | 0.001 |
| | | Umbriel_L | 285.1 | -10.7 | 0.279 | 0.006 | 0.065 | 0.001 |
| | Ch.3 | Titania_L | 83.1 | -10.5 | 0.112 | 0.016 | 0.044 | 0.006 |
| | | Oberon_T | 216.7 | -18.3 | 0.091 | 0.046 | 0.037 | 0.019 |
| | | Oberon_L | 219.5 | -14.5 | 0.106 | 0.036 | 0.042 | 0.015 |
| | | Titania_T | 262.6 | -18.3 | 0.077 | 0.026 | 0.032 | 0.011 |
| | | Umbriel_L | 285.1 | -10.7 | 0.098 | 0.022 | 0.038 | 0.008 |
| | Ch.4 | Titania_L | 83.1 | -10.5 | 0.130 | 0.033 | 0.094 | 0.024 |
| | | Oberon_T | 216.7 | -18.3 | 0.305 | 0.053 | 0.228 | 0.039 |
| | | Oberon_L | 219.5 | -14.5 | 0.100 | 0.019 | 0.073 | 0.014 |
| | | Titania_T | 262.6 | -18.3 | 0.225 | 0.047 | 0.168 | 0.035 |
| | | Umbriel_L | 285.1 | -10.7 | 0.143 | 0.023 | 0.100 | 0.016 |
| Oberon | Ch.1 | Oberon_L | 88.3 | -14.6 | 0.967 | 0.006 | 0.167 | 0.001 |
| | | Umbriel_L | 133.5 | -10.7 | 1.024 | 0.005 | 0.169 | 0.001 |
| | | Umbriel_T | 141.9 | -6.7 | 0.939 | 0.005 | 0.164 | 0.001 |
| | | Oberon_T | 261.9 | -18.3 | 0.968 | 0.006 | 0.167 | 0.002 |
| | | Titania_L | 290.3 | -10.6 | 0.983 | 0.005 | 0.168 | 0.001 |
| | | Titania_T | 291.6 | -18.3 | 0.955 | 0.007 | 0.169 | 0.002 |
| | Ch.2 | Oberon_L | 88.3 | -14.6 | 0.255 | 0.006 | 0.067 | 0.002 |
| | | Umbriel_L | 133.5 | -10.7 | 0.269 | 0.004 | 0.067 | 0.001 |
| | | Umbriel_T | 141.9 | -6.7 | 0.255 | 0.003 | 0.068 | 0.001 |
| | | Oberon_T | 261.9 | -18.3 | 0.276 | 0.005 | 0.074 | 0.001 |
| | | Titania_L | 290.3 | -10.6 | 0.285 | 0.003 | 0.074 | 0.001 |
| | | Titania_T | 291.6 | -18.3 | 0.293 | 0.019 | 0.079 | 0.005 |
| | Ch.3 | Oberon_L | 88.3 | -14.6 | 0.117 | 0.020 | 0.050 | 0.009 |
| | | Umbriel_L | 133.5 | -10.7 | 0.111 | 0.021 | 0.045 | 0.008 |
| | | Umbriel_T | 141.9 | -6.7 | 0.123 | 0.017 | 0.054 | 0.008 |
| | | Oberon_T | 261.9 | -18.3 | *0.046* | - | - | - |
| | | Titania_L | 290.3 | -10.6 | 0.155 | 0.029 | 0.066 | 0.012 |

|     | Titania_T | 291.6 | -18.3 | 0.120 | 0.032 | 0.053 | 0.014 |
|-----|-----------|-------|-------|-------|-------|-------|-------|
| Ch.4 | Oberon_L | 88.3 | -14.6 | 0.101 | 0.043 | 0.079 | 0.034 |
|     | Umbriel_L | 133.5 | -10.7 | 0.114 | 0.020 | 0.085 | 0.015 |
|     | Umbriel_T | 141.9 | -6.7 | *0.067* | - | - | - |
|     | Oberon_T | 261.9 | -18.3 | 0.285 | 0.068 | 0.229 | 0.054 |
|     | Titania_L | 290.3 | -10.6 | *0.104* | - | - | - |
|     | Titania_T | 291.6 | -18.3 | 0.176 | 0.043 | 0.141 | 0.034 |

*Italicized values represent the 3σ upper limits of the background flux.*

**Table 6: Mean leading and trailing IRAC geometric albedos**

| Target | Hemisphere | IRAC Channel | Number of Observations | Mean Geometric Albedo | ΔMean Geometric Albedo |
|--------|------------|--------------|------------------------|-----------------------|------------------------|
| Ariel | Leading | 1 | 5 | 0.188 | 0.003 |
|       |         | 2 | 5 | 0.092 | 0.005 |
|       |         | 3 | 4 | 0.067 | 0.009 |
|       |         | 4 | 5 | 0.250 | 0.027 |
|       | Trailing | 1 | 2 | 0.232 | 0.003 |
|       |         | 2 | 2 | 0.124 | 0.005 |
|       |         | 3 | 2 | 0.089 | 0.015 |
|       |         | 4 | 2 | 0.361 | 0.048 |
| Umbriel | Leading | 1 | 4 | 0.153 | 0.002 |
|       |         | 2 | 4 | 0.078 | 0.004 |
|       |         | 3 | 4 | 0.052 | 0.011 |
|       |         | 4 | 4 | 0.232 | 0.045 |
|       | Trailing | 1 | 2 | 0.158 | 0.001 |
|       |         | 2 | 2 | 0.077 | 0.003 |
|       |         | 3 | 2 | 0.067 | 0.023 |
|       |         | 4 | 2 | 0.287 | 0.040 |
| Titania | Leading | 1 | 1 | 0.160 | 0.001 |
|       |         | 2 | 1 | 0.063 | 0.002 |
|       |         | 3 | 1 | 0.044 | 0.006 |
|       |         | 4 | 1 | 0.094 | 0.024 |
|       | Trailing | 1 | 4 | 0.160 | 0.001 |
|       |         | 2 | 4 | 0.066 | 0.001 |
|       |         | 3 | 4 | 0.037 | 0.007 |
|       |         | 4 | 4 | 0.142 | 0.019 |
| Oberon | Leading | 1 | 3 | 0.167 | 0.001 |
|       |         | 2 | 3 | 0.067 | 0.001 |
|       |         | 3 | 3 | 0.050 | 0.005 |
|       |         | 4 | 2 | 0.082 | 0.019 |
|       | Trailing | 1 | 3 | 0.168 | 0.001 |

| | | | |
|---|---|---|---|
| 2 | 3 | 0.076 | 0.002 |
| 3 | 2 | 0.060 | 0.009 |
| 4 | 2 | 0.185 | 0.032 |

**Table 7: Wavelength position of $CO_2$ ice bands and adjacent continua**

| Band Name | Combination/Overtone Designation | Band Center (µm) | Band Width (µm) | Continua (µm) |
|---|---|---|---|---|
| $CO_2$ band 1 | $2v_1 + v_3$ | 1.966 | 1.962 - 1.969 | 1.957 - 1.962, 1.969 - 1.974 |
| $CO_2$ band 2 | $v_1 + 2v_2 + v_3$ | 2.012 | 2.008 - 2.015 | 2.002 - 2.008, 2.015 - 2.020 |
| $CO_2$ band 3 | $4v_2 + v_3$ | 2.070 | 2.068 - 2.072 | 2.062 - 2.068, 2.072 - 2.078 |

**Table 8: Summary of CO₂ ice band parameter analysis**

| | | | CO$_2$ Band Area ($10^{-4}$ μm) | | | | | | | | Gauss Fit CO$_2$ Band Centers (μm) | | | | | |
| --- | --- | --- | --- | --- | --- | --- | --- | --- | --- | --- | --- | --- | --- | --- | --- | --- |
| Satellite | Long. (°) | Lat. (°) | Band 1 | ΔArea | Band 2 | ΔArea | Band 3 | ΔArea | Total Area | Total Δarea | Band 1 | Δcenter | Band 2 | Δcenter | Band 3 | Δcenter |
| Ariel | 53.6 | -16.0 | 1.36 | 0.64 | 1.42 | 1.19 | 0.23 | 0.59 | 3.02 | 1.47 | - | - | - | - | - | - |
| | 79.8 | -19.4 | 1.34 | 0.44 | -1.23 | 0.56 | 0.81 | 0.26 | 0.92 | 0.76 | - | - | - | - | - | - |
| | 87.8 | 24.0 | 1.26 | 0.31 | 0.43 | 0.44 | 0.49 | 0.23 | 2.17 | 0.59 | - | - | - | - | - | - |
| | 93.5 | -18.1 | 0.87 | 0.56 | 1.11 | 0.72 | -0.24 | 0.51 | 1.74 | 1.05 | - | - | - | - | - | - |
| | 159.9 | -11.1 | 2.30 | 0.53 | 2.91 | 0.78 | 1.08 | 0.45 | 6.29 | 1.04 | 1.965 | 0.001 | 2.012 | 0.001 | 2.070 | 0.001 |
| | 200.0 | -15.9 | 4.19 | 0.67 | 5.94 | 0.80 | 2.89 | 0.45 | 13.02 | 1.14 | 1.966 | 0.001 | 2.011 | 0.001 | 2.070 | 0.001 |
| | 219.8 | -17.2 | 5.77 | 0.29 | 6.71 | 0.68 | 2.52 | 0.29 | 14.99 | 0.80 | 1.966 | 0.001 | 2.011 | 0.001 | 2.070 | 0.001 |
| | 233.8 | -23.1 | 4.09 | 0.78 | 6.31 | 0.85 | 2.99 | 0.45 | 13.40 | 1.24 | 1.962 | 0.001 | 2.009 | 0.001 | 2.070 | 0.001 |
| | 257.6 | -29.5 | 3.33 | 0.50 | 6.47 | 0.49 | 3.36 | 0.27 | 13.16 | 0.75 | 1.966 | 0.001 | 2.012 | 0.001 | 2.072 | 0.001 |
| | 278.3 | 24.8 | 6.06 | 0.83 | 6.48 | 1.02 | 2.83 | 0.60 | 15.37 | 1.45 | 1.966 | 0.001 | 2.011 | 0.001 | 2.070 | 0.001 |
| | 294.8 | -19.3 | 6.26 | 0.40 | 5.56 | 0.47 | 3.08 | 0.25 | 14.90 | 0.66 | 1.966 | 0.001 | 2.012 | 0.001 | 2.070 | 0.001 |
| | 304.8 | -23.2 | 5.12 | 0.76 | 6.29 | 0.96 | 3.35 | 0.61 | 14.76 | 1.36 | 1.965 | 0.001 | 2.012 | 0.001 | 2.070 | 0.001 |
| | 316.6 | -18.2 | 4.52 | 0.58 | 6.72 | 0.82 | 2.32 | 0.46 | 13.56 | 1.11 | 1.966 | 0.001 | 2.012 | 0.001 | 2.070 | 0.001 |
| Umbriel | 38.4 | 20.3 | -0.74 | 0.61 | -0.66 | 0.72 | 0.14 | 0.37 | -1.26 | 1.01 | - | - | - | - | - | - |
| | 75.2 | -29.3 | -0.04 | 0.75 | -3.07 | 0.78 | -0.37 | 0.50 | -3.48 | 1.19 | - | - | - | - | - | - |
| | 92.1 | -11.1 | 2.59 | 0.99 | 0.74 | 1.41 | 0.83 | 0.63 | 4.16 | 1.83 | - | - | - | - | - | - |
| | 131.3 | 17.5 | 0.23 | 0.37 | 1.18 | 0.44 | 0.07 | 0.22 | 1.48 | 0.61 | - | - | - | - | - | - |
| | 216.2 | -10.9 | 1.09 | 0.92 | 1.54 | 1.22 | 2.56 | 0.74 | 5.19 | 1.69 | - | - | - | - | - | - |
| | 219.8 | -23.0 | 0.65 | 0.92 | 1.42 | 1.10 | 1.11 | 1.03 | 3.18 | 1.77 | - | - | - | - | - | - |
| | 261.3 | -9.4 | 2.11 | 0.52 | 3.13 | 0.52 | 1.67 | 0.29 | 6.92 | 0.79 | 1.963 | 0.001 | 2.011 | 0.001 | 2.070 | 0.001 |
| | 264.0 | 23.7 | 1.73 | 0.42 | 4.27 | 0.49 | 1.39 | 0.25 | 7.39 | 0.69 | 1.963 | 0.001 | 2.015 | 0.001 | 2.070 | 0.001 |
| | 317.6 | -11.4 | 0.95 | 0.51 | 2.12 | 0.72 | 1.40 | 0.36 | 4.47 | 0.95 | 1.968 | 0.001 | 2.013 | 0.001 | 2.068 | 0.001 |
| Titania | 13.6 | 19.0 | 0.01 | 0.25 | 0.20 | 0.28 | 0.48 | 0.16 | 0.70 | 0.41 | - | - | - | - | - | - |
| | 86.5 | 23.6 | 0.40 | 0.21 | 1.24 | 0.27 | -0.02 | 0.15 | 1.62 | 0.37 | - | - | - | - | - | - |
| | 98.0 | -18.1 | 0.95 | 0.89 | 0.88 | 1.50 | 0.53 | 0.66 | 2.36 | 1.87 | - | - | - | - | - | - |
| | 160.0 | 18.2 | -0.65 | 0.51 | 0.03 | 0.70 | -0.12 | 0.39 | -0.74 | 0.95 | - | - | - | - | - | - |

| | | | | | | | | | | | | | | | |
|---|---|---|---|---|---|---|---|---|---|---|---|---|---|---|---|
| 213.9 | -11.1 | 1.55 | 0.48 | 2.54 | 0.61 | 0.53 | 0.35 | 4.62 | 0.85 | - | - | - | - | - | - |
| 237.0 | -23.0 | 2.05 | 1.57 | 1.98 | 1.87 | 1.09 | 1.11 | 5.12 | 2.68 | - | - | - | - | - | - |
| 258.9 | -29.3 | 1.77 | 0.38 | 2.32 | 0.50 | 0.79 | 0.27 | 4.88 | 0.68 | - | - | - | - | - | - |
| 277.8 | -23.0 | 0.53 | 0.46 | 2.94 | 0.71 | -0.11 | 0.89 | 3.36 | 1.22 | - | - | - | - | - | - |
| 299.6 | -10.2 | 1.68 | 0.38 | 1.40 | 0.54 | 1.30 | 0.27 | 4.37 | 0.71 | - | - | - | - | - | - |
| 342.7 | -29.4 | 0.90 | 0.44 | 0.43 | 0.59 | 0.07 | 0.28 | 1.41 | 0.78 | - | - | - | - | - | - |

| Oberon | 1.0 | -29.4 | 1.83 | 0.49 | -0.30 | 0.53 | 0.60 | 0.30 | 2.13 | 0.78 | - | - | - | - | - | - |
|---|---|---|---|---|---|---|---|---|---|---|---|---|---|---|---|---|
| | 64.8 | 18.1 | -0.38 | 0.42 | -0.19 | 0.49 | 0.22 | 0.27 | -0.35 | 0.70 | - | - | - | - | - | - |
| | 85.9 | -3.7 | -3.75 | 0.95 | -2.83 | 1.27 | -0.31 | 0.64 | -6.89 | 1.71 | - | - | - | - | - | - |
| | 91.0 | 23.7 | 0.31 | 0.75 | 0.40 | 0.79 | 0.23 | 0.62 | 0.94 | 1.26 | - | - | - | - | - | - |
| | 110.6 | -10.2 | 0.08 | 0.38 | 0.48 | 0.47 | -0.50 | 0.32 | 0.06 | 0.68 | - | - | - | - | - | - |
| | 164.0 | -23.0 | 0.44 | 1.79 | 0.14 | 2.17 | 0.34 | 1.21 | 0.91 | 3.06 | - | - | - | - | - | - |
| | 216.2 | -23.1 | -0.35 | 0.50 | -0.68 | 0.72 | -0.45 | 0.57 | -1.48 | 1.04 | - | - | - | - | - | - |
| | 223.7 | 18.9 | 0.61 | 0.28 | 0.97 | 0.36 | 0.23 | 0.16 | 1.81 | 0.49 | - | - | - | - | - | - |
| | 236.1 | 17.4 | 0.64 | 0.25 | 1.22 | 0.32 | 0.35 | 0.15 | 2.21 | 0.43 | - | - | - | - | - | - |
| | 266.5 | 20.2 | 1.62 | 0.72 | 1.15 | 0.86 | 0.74 | 0.51 | 3.51 | 1.23 | - | - | - | - | - | - |
| | 307.2 | -29.3 | 0.84 | 0.38 | 0.79 | 0.27 | 0.68 | 0.22 | 2.31 | 0.52 | - | - | - | - | - | - |

**Table 9: F-test analysis on Leading and Trailing Hemispheres**

| Satellite | Spectral Band Area | One Tailed F-test Ratio | Sample Size (n) | Mean Model Deg. Freedom (n - 1) | Sinusoidal Model Deg. Freedom (n - 3) | Probability ($p$) | Reject Null Hypothesis? |
|---|---|---|---|---|---|---|---|
| Ariel | Summed $CO_2$ | 152 | 13 | 12 | 10 | 1.10E-09 | Yes |
| | 1.52 μm $H_2O$ | 129 | 13 | 12 | 10 | 2.50E-09 | Yes |
| | 2.02 μm $H_2O$ | 50.6 | 13 | 12 | 10 | 2.50E-07 | Yes |
| Umbriel | Summed $CO_2$ | 5.62 | 9 | 8 | 6 | 0.025 | Yes |
| | 1.52 μm $H_2O$ | 4.86 | 9 | 8 | 6 | 0.035 | Yes |
| | 2.02 μm $H_2O$ | 82.6 | 9 | 8 | 6 | 1.50E-05 | Yes |
| Titania | Summed $CO_2$ | 6.91 | 10 | 9 | 7 | 9.00E-03 | Yes |
| | 1.52 μm $H_2O$ | 11.7 | 10 | 9 | 7 | 1.90E-03 | Yes |
| | 2.02 μm $H_2O$ | 23.8 | 10 | 9 | 7 | 2.00E-03 | Yes |
| Oberon | Summed $CO_2$ | 6.81 | 11 | 10 | 8 | 6.00E-03 | Yes |
| | 1.52 μm $H_2O$ | 7.31 | 11 | 10 | 8 | 4.80E-03 | Yes |
| | 2.02 μm $H_2O$ | 3.86 | 11 | 10 | 8 | 0.034 | Yes |

**Table 10: F-test analysis on Leading and Trailing Quadrants**

| Satellite | Spectral Band Area | One Tailed F-test Ratio | Sample Size (n) | Mean Model Deg. Freedom (n - 1) | Sinusoidal Model Deg. Freedom (n - 3) | Probability ($p$) | Reject Null Hypothesis? |
|---|---|---|---|---|---|---|---|
| Ariel | Summed $CO_2$ | 247 | 9 | 8 | 6 | 5.50E-07 | Yes |
| | 1.52 μm $H_2O$ | 462 | 9 | 8 | 6 | 8.50E-08 | Yes |
| | 2.02 μm $H_2O$ | 194 | 9 | 8 | 6 | 1.10E-06 | Yes |
| Umbriel | Summed $CO_2$ | 1.42 | 5 | 4 | 2 | 0.45 | No |
| | 1.52 μm $H_2O$ | 16.4 | 5 | 4 | 2 | 0.06 | No |
| | 2.02 μm $H_2O$ | 77.3 | 5 | 4 | 2 | 0.01 | Yes |
| Titania | Summed $CO_2$ | 41.8 | 6 | 5 | 3 | 0.01 | Yes |
| | 1.52 μm $H_2O$ | 44.0 | 6 | 5 | 3 | 5.20E-03 | Yes |
| | 2.02 μm $H_2O$ | 65.3 | 6 | 5 | 3 | 2.90E-03 | Yes |
| Oberon | Summed $CO_2$ | 5.63 | 7 | 6 | 4 | 0.06 | No |
| | 1.52 μm $H_2O$ | 4.43 | 7 | 6 | 4 | 0.09 | No |
| | 2.02 μm $H_2O$ | 0.374 | 7 | 6 | 4 | 0.86 | No |

**Table 11: Summary of $H_2O$ ice band parameter analysis**

| Target | Long. (°) | Lat. (°) | Integrated $H_2O$ Band Area ($10^{-2}$ μm) | | | | | | $H_2O$ Band Depth (μm) | | | | |
|---|---|---|---|---|---|---|---|---|---|---|---|---|---|
| | | | 1.52 μm Band | ΔArea | 2.02 μm Band | ΔArea | 1.52 μm / 2.02 μm Band Ratio | Δband Ratio | 1.52 μm Band | ΔDepth | 2.02 μm Band | ΔDepth |
| Ariel | 53.6 | -16.0 | 4.73 | 0.02 | 8.66 | 0.05 | 0.546 | 0.002 | 0.274 | 0.002 | 0.456 | 0.005 |
| | 79.8 | -19.4 | 4.99 | 0.01 | 9.20 | 0.04 | 0.542 | 0.001 | 0.292 | 0.002 | 0.481 | 0.003 |
| | 87.8 | 24.0 | 5.01 | 0.01 | 9.09 | 0.02 | 0.551 | 0.001 | 0.293 | 0.002 | 0.477 | 0.003 |
| | 93.5 | -18.1 | 4.83 | 0.02 | 8.95 | 0.03 | 0.539 | 0.001 | 0.286 | 0.003 | 0.482 | 0.005 |
| | 159.9 | -11.1 | 3.23 | 0.02 | 7.84 | 0.03 | 0.412 | 0.002 | 0.203 | 0.004 | 0.425 | 0.005 |
| | 200.0 | -15.9 | 2.98 | 0.02 | 7.08 | 0.12 | 0.421 | 0.005 | 0.182 | 0.003 | 0.370 | 0.006 |
| | 219.8 | -17.2 | 2.67 | 0.01 | 6.50 | 0.02 | 0.410 | 0.001 | 0.164 | 0.002 | 0.349 | 0.003 |
| | 233.8 | -23.1 | 3.15 | 0.03 | 6.25 | 0.12 | 0.503 | 0.005 | 0.200 | 0.003 | 0.401 | 0.006 |
| | 257.6 | -29.5 | 3.25 | 0.03 | 6.83 | 0.04 | 0.476 | 0.002 | 0.191 | 0.005 | 0.367 | 0.004 |
| | 278.3 | 24.8 | 2.95 | 0.02 | 6.74 | 0.04 | 0.437 | 0.002 | 0.183 | 0.004 | 0.371 | 0.007 |
| | 294.8 | -19.3 | 2.90 | 0.02 | 6.84 | 0.04 | 0.423 | 0.002 | 0.178 | 0.002 | 0.392 | 0.003 |
| | 304.8 | -23.2 | 3.02 | 0.03 | 6.74 | 0.18 | 0.449 | 0.007 | 0.185 | 0.003 | 0.391 | 0.006 |
| | 316.6 | -18.2 | 3.21 | 0.02 | 6.71 | 0.03 | 0.478 | 0.002 | 0.191 | 0.003 | 0.378 | 0.006 |
| Umbriel | 38.4 | 20.3 | 1.74 | 0.02 | 4.33 | 0.03 | 0.402 | 0.002 | 0.116 | 0.004 | 0.263 | 0.005 |
| | 75.2 | -29.3 | 1.87 | 0.03 | 4.38 | 0.08 | 0.427 | 0.004 | 0.122 | 0.007 | 0.274 | 0.007 |
| | 92.1 | -11.1 | 1.48 | 0.04 | 4.30 | 0.06 | 0.346 | 0.005 | 0.097 | 0.005 | 0.284 | 0.012 |
| | 131.3 | 17.5 | 1.57 | 0.01 | 4.37 | 0.02 | 0.359 | 0.001 | 0.101 | 0.002 | 0.263 | 0.004 |
| | 216.2 | -10.9 | 1.42 | 0.04 | 4.06 | 0.05 | 0.349 | 0.004 | 0.081 | 0.005 | 0.252 | 0.010 |
| | 219.8 | -23.0 | 1.68 | 0.03 | 3.81 | 0.09 | 0.442 | 0.005 | 0.110 | 0.004 | 0.278 | 0.008 |
| | 261.3 | -9.4 | 0.87 | 0.03 | 3.64 | 0.18 | 0.238 | 0.010 | 0.053 | 0.004 | 0.218 | 0.005 |
| | 264.0 | 23.7 | 1.26 | 0.01 | 3.90 | 0.02 | 0.323 | 0.002 | 0.090 | 0.003 | 0.228 | 0.004 |
| | 317.6 | -11.4 | 1.04 | 0.02 | 3.89 | 0.04 | 0.269 | 0.003 | 0.069 | 0.004 | 0.230 | 0.006 |
| Titania | 13.6 | 19.0 | 3.60 | 0.01 | 7.85 | 0.01 | 0.459 | 0.001 | 0.217 | 0.001 | 0.437 | 0.002 |
| | 86.5 | 23.6 | 4.43 | 0.01 | 8.72 | 0.01 | 0.507 | 0.000 | 0.264 | 0.001 | 0.467 | 0.002 |
| | 98.0 | -18.1 | 4.44 | 0.03 | 8.70 | 0.04 | 0.511 | 0.002 | 0.270 | 0.004 | 0.477 | 0.007 |
| | 160.0 | 18.2 | 3.57 | 0.01 | 7.67 | 0.02 | 0.466 | 0.001 | 0.219 | 0.003 | 0.430 | 0.005 |
| | 213.9 | -11.1 | 3.31 | 0.02 | 7.79 | 0.03 | 0.425 | 0.002 | 0.203 | 0.003 | 0.429 | 0.004 |
| | 237.0 | -23.0 | 4.44 | 0.06 | 5.96 | 1.37 | 0.745 | 0.056 | 0.259 | 0.006 | 0.494 | 0.010 |
| | 258.9 | -29.3 | 3.46 | 0.02 | 7.68 | 0.03 | 0.451 | 0.001 | 0.219 | 0.003 | 0.421 | 0.003 |
| | 277.8 | -23.0 | 3.70 | 0.02 | 8.01 | 0.04 | 0.462 | 0.002 | 0.229 | 0.002 | 0.441 | 0.004 |
| | 299.6 | -10.2 | 3.35 | 0.02 | 7.57 | 0.02 | 0.442 | 0.001 | 0.208 | 0.003 | 0.419 | 0.003 |
| | 342.7 | -29.4 | 3.83 | 0.02 | 7.87 | 0.05 | 0.486 | 0.002 | 0.230 | 0.002 | 0.439 | 0.004 |

| Oberon | 1.0 | -29.4 | 2.56 | 0.02 | 6.35 | 0.17 | 0.404 | 0.007 | 0.162 | 0.002 | 0.369 | 0.006 |
| | 64.8 | 18.1 | 2.63 | 0.01 | 6.18 | 0.02 | 0.425 | 0.001 | 0.170 | 0.002 | 0.350 | 0.003 |
| | 85.9 | -3.7 | 2.73 | 0.05 | 6.03 | 0.16 | 0.453 | 0.007 | 0.177 | 0.006 | 0.366 | 0.008 |
| | 91.0 | 23.7 | 2.90 | 0.03 | 6.11 | 0.04 | 0.474 | 0.003 | 0.181 | 0.008 | 0.339 | 0.007 |
| | 110.6 | -10.2 | 2.16 | 0.02 | 5.41 | 0.02 | 0.400 | 0.001 | 0.137 | 0.003 | 0.306 | 0.004 |
| | 164.0 | -23.0 | 3.11 | 0.06 | 2.99 | 1.46 | 1.038 | 0.085 | 0.210 | 0.009 | 0.399 | 0.015 |
| | 216.2 | -23.1 | 2.75 | 0.02 | 6.12 | 0.06 | 0.449 | 0.003 | 0.173 | 0.002 | 0.375 | 0.004 |
| | 223.7 | 18.9 | 2.21 | 0.01 | 5.77 | 0.01 | 0.382 | 0.001 | 0.141 | 0.002 | 0.327 | 0.003 |
| | 236.1 | 17.4 | 2.31 | 0.01 | 5.94 | 0.01 | 0.389 | 0.001 | 0.151 | 0.001 | 0.339 | 0.002 |
| | 266.5 | 20.2 | 2.46 | 0.02 | 5.43 | 0.04 | 0.453 | 0.002 | 0.153 | 0.005 | 0.309 | 0.007 |
| | 307.2 | -29.3 | 2.49 | 0.01 | 5.92 | 0.02 | 0.421 | 0.001 | 0.162 | 0.001 | 0.330 | 0.002 |

**Table 12: H₂O ice band area ratios**

| Satellite Hemisphere | Mean 1.52 μm Band Area | ΔArea | Leading / Trailing Ratio | ΔRatio | Mean 2.02 μm Band Area | ΔArea | Leading / Trailing Ratio | ΔRatio | 1.52 μm / 2.02 μm Band Ratio | ΔRatio |
|---|---|---|---|---|---|---|---|---|---|---|
| Ariel L | 4.56 | 0.01 | | | 8.47 | 0.02 | | | 0.538 | 0.001 |
| Ariel T | 3.02 | 0.01 | 1.51 | 0.002 | 6.71 | 0.03 | 1.26 | 0.004 | 0.449 | 0.002 |
| Umbriel L | 1.67 | 0.02 | | | 4.35 | 0.03 | | | 0.384 | 0.004 |
| Umbriel T | 1.25 | 0.01 | 1.33 | 0.010 | 3.86 | 0.04 | 1.13 | 0.011 | 0.325 | 0.005 |
| Titania L | 4.01 | 0.01 | | | 8.24 | 0.01 | | | 0.487 | 0.001 |
| Titania T | 3.53 | 0.01 | 1.14 | 0.003 | 7.79 | 0.02 | 1.06 | 0.002 | 0.453 | 0.001 |
| Oberon L | 2.60 | 0.01 | | | 6.02 | 0.05 | | | 0.432 | 0.004 |
| Oberon T | 2.44 | 0.01 | 1.06 | 0.005 | 5.84 | 0.01 | 1.03 | 0.008 | 0.419 | 0.001 |

## Appendix A

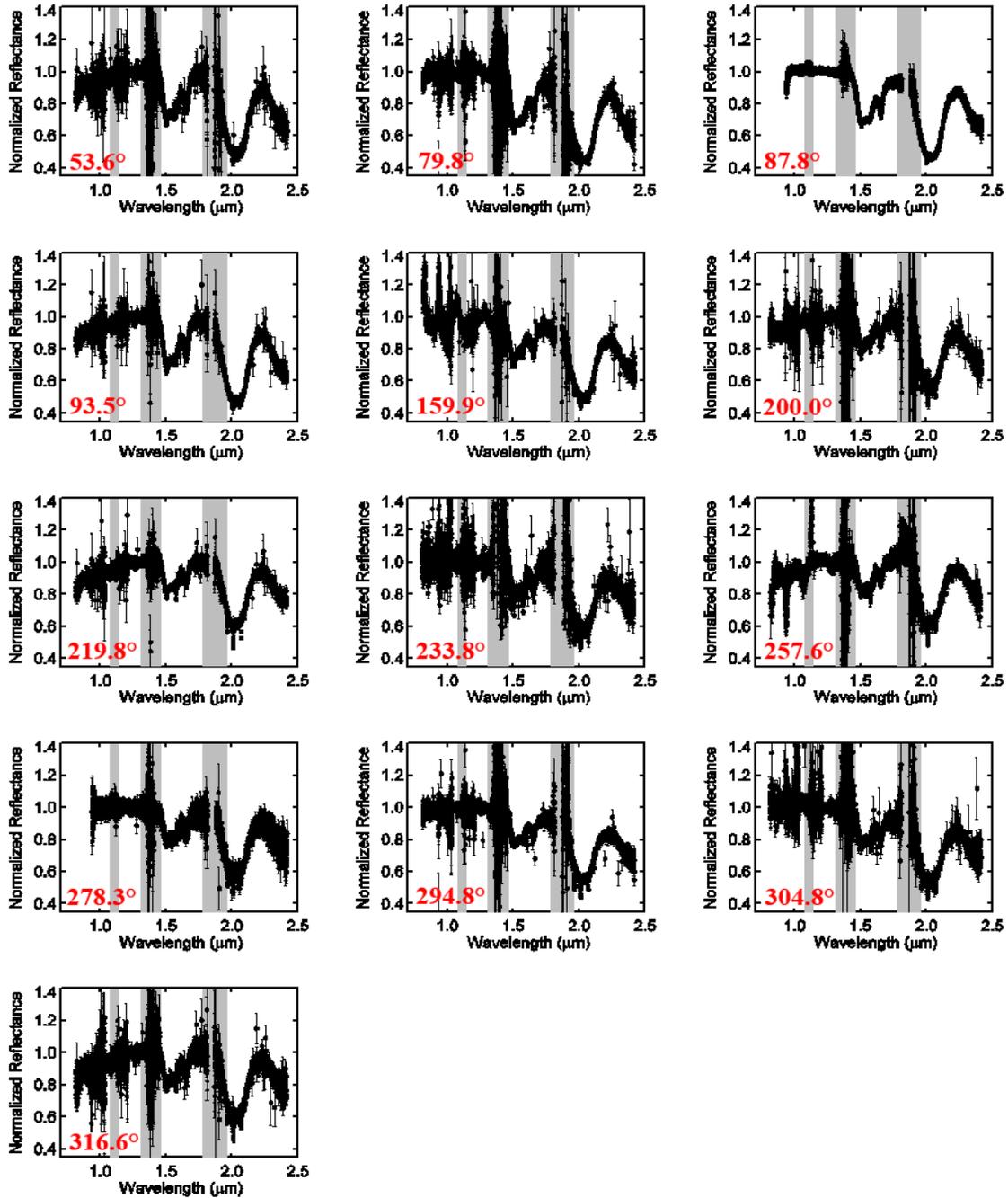

**Figure A1:** 13 SpeX spectra of Ariel gathered by three different teams (Rivkin, Cartwright, Grundy et al., 2003, 2006), organized by increasing mid-observation satellite longitude (listed in top right-hand corner of each plot). Each spectrum has been normalized to its mean reflectance between 1.2 and 1.3 μm. The wavelength ranges of strong telluric bands are indicated by the gray-toned regions (1.08 −1.14 μm, 1.31 − 1.46 μm, and 1.78 − 1.96 μm). See Section 2.1 for data reduction details.

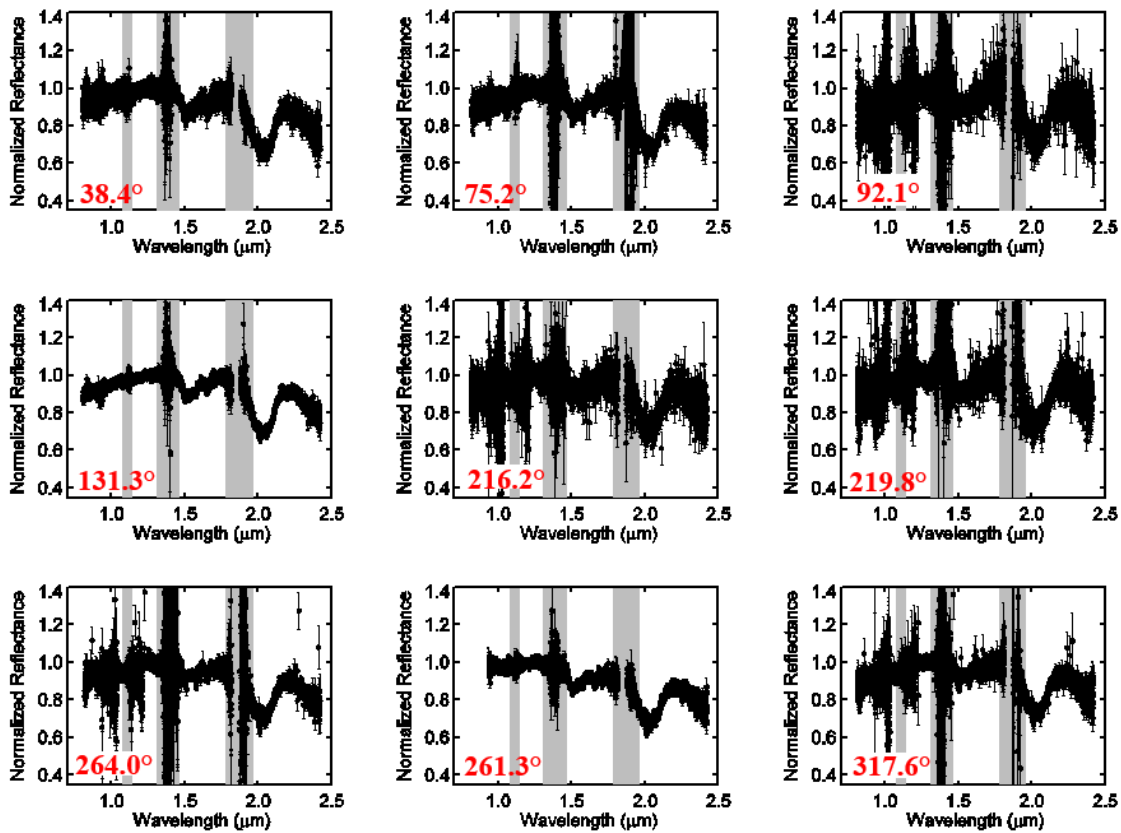

**Figure A2:** 9 SpeX spectra of Umbriel gathered by three different teams (Rivkin, Cartwright, Grundy et al., 2003, 2006), organized by increasing mid-observation satellite longitude (listed in top right-hand corner of each plot). Each spectrum has been normalized to its mean reflectance between 1.2 and 1.3 µm. The wavelength ranges of strong telluric bands are indicated by the gray-toned regions (1.08 −1.14 µm, 1.31 – 1.46 µm, and 1.78 – 1.96 µm). See Section 2.1 for data reduction details.

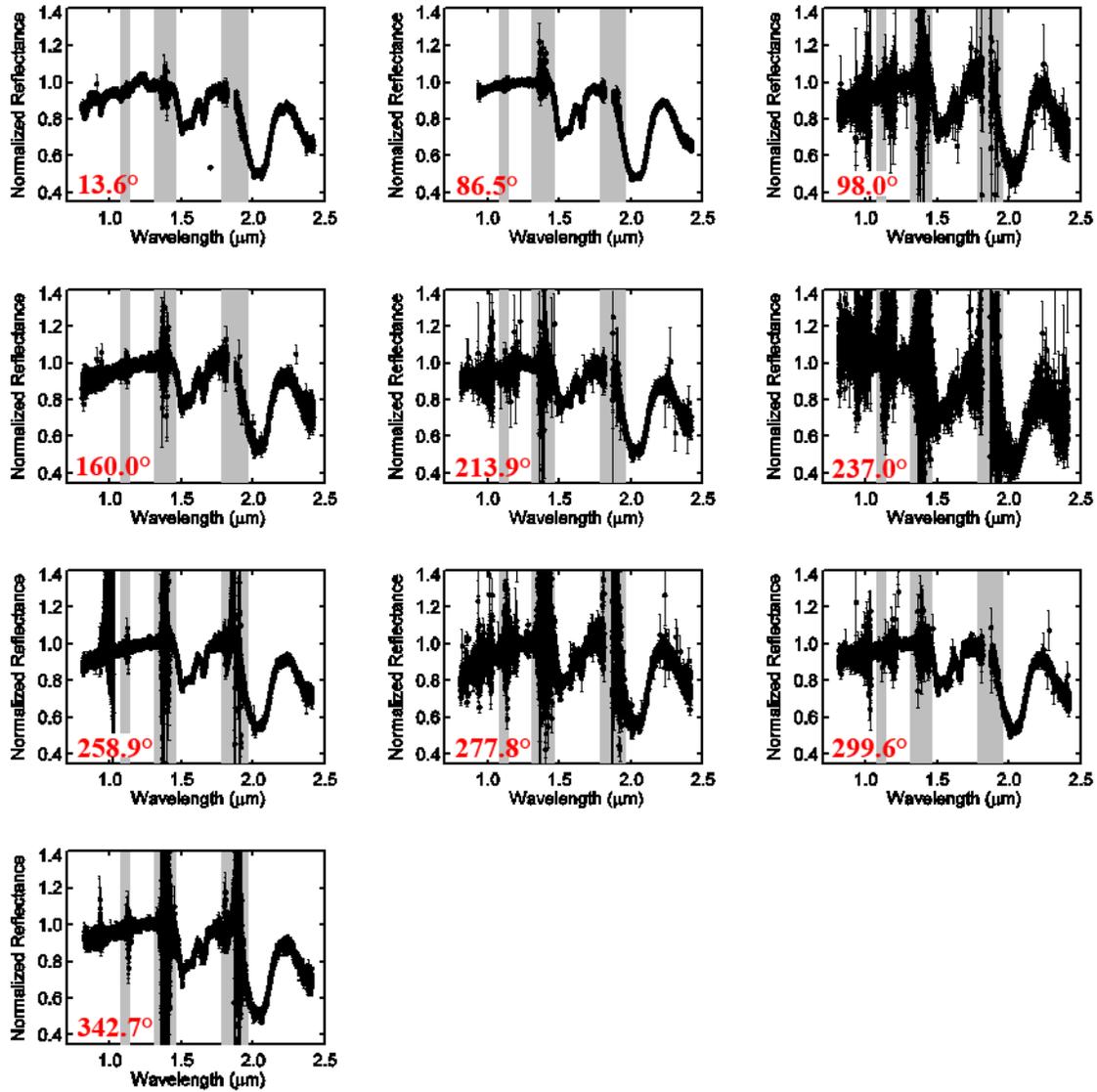

**Figure A3:** 10 SpeX spectra of Titania gathered by three different teams (Rivkin, Cartwright, Grundy et al., 2003, 2006), organized by increasing mid-observation satellite longitude (listed in top right-hand corner of each plot). Each spectrum has been normalized to its mean reflectance between 1.2 and 1.3 µm. The wavelength ranges of strong telluric bands are indicated by the gray-toned regions (1.08 –1.14 µm, 1.31 – 1.46 µm, and 1.78 – 1.96 µm). See Section 2.1 for data reduction details.

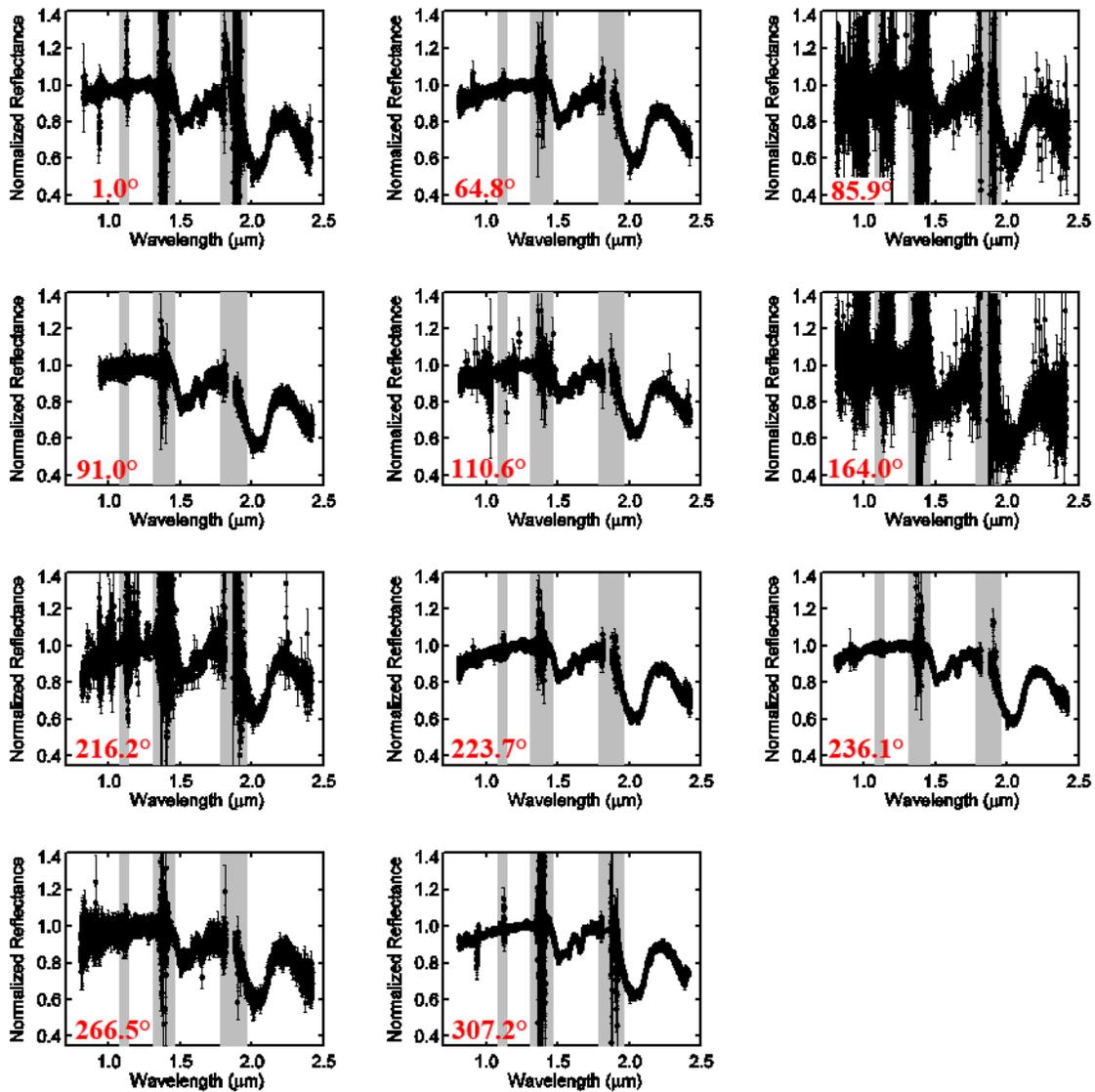

**Figure A4:** 11 SpeX spectra of Oberon gathered by three different teams (Rivkin, Cartwright, Grundy et al., 2003, 2006), organized by increasing mid-observation satellite longitude (listed in top right-hand corner of each plot). Each spectrum has been normalized to its mean reflectance between 1.2 and 1.3 μm. The wavelength ranges of strong telluric bands are indicated by the gray-toned regions (1.08 − 1.14 μm, 1.31 − 1.46 μm, and 1.78 − 1.96 μm). See Section 2.1 for data reduction details.

## Appendix B

### B1 Assessment of $CO_2$ band areas

Our band parameter code indicates that $CO_2$ ice is present in spectra collected over the trailing hemispheres of Ariel, Umbriel, Titania, and Oberon. Telluric $CO_2$ and higher noise levels in spectra with low S/N could be contaminating the measured $CO_2$ bands. In order to test for possible contamination of our $CO_2$ band area measurements, we compared them to the $CO_2$ band areas in a wide range of synthetic spectra generated using a Hapke-Mie numerical model. This hybrid numerical model utilizes Mie theory (*e.g.*, Bohren and Huffman, 1983) to calculate the single scattering albedos for all species before passing these ratios to a Hapke equation that models scattering within a regolith (*e.g.*, Hapke, 2002). This hybrid approach allows us to generate reliable synthetic spectra over a wider range of grain sizes than using Hapke-based codes alone (Emery et al. 2006).

### B1.1 Mean trailing hemisphere spectra and best fit spectral models

In order to increase the S/N of the $CO_2$ bands, we averaged multiple trailing hemisphere spectra together, generating one spectrum to represent the trailing hemisphere of each moon (Figure B1). We used individual spectra with visibly identifiable $CO_2$ bands on Ariel (eight total spectra) and Umbriel (three spectra), along with the trailing hemisphere spectra with non-negative band areas on Titania and Oberon (three and four total spectra, respectively).

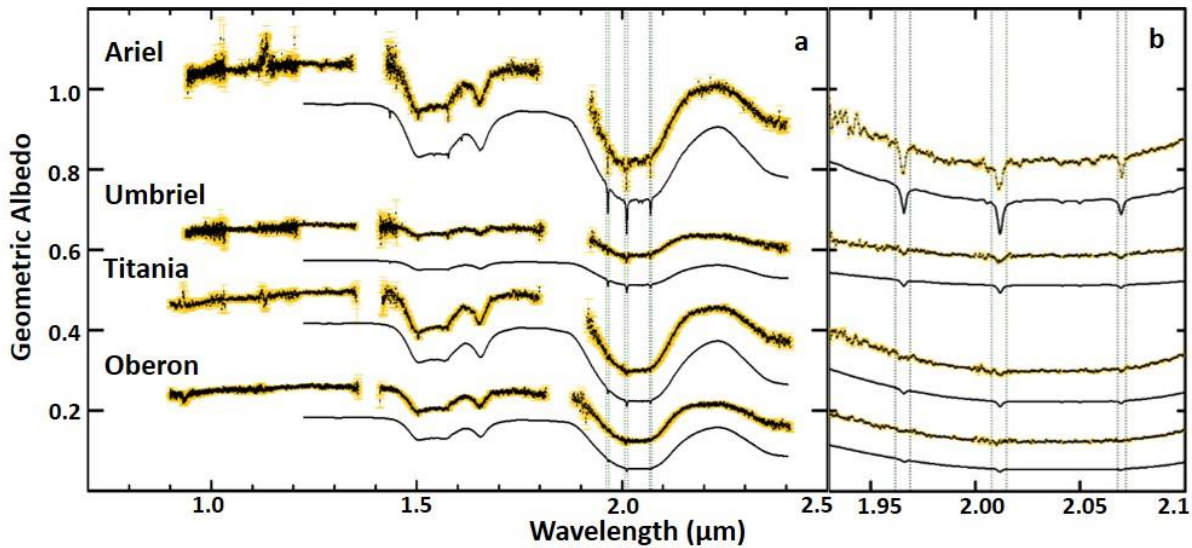

**Figure B1:** (**a**) Black spectra with orange error bars are the mean trailing hemisphere spectra of Ariel, Umbriel, Titania, and Oberon scaled to I band geometric albedos (Karkoschka 2001), offset by +0.47, +0.36, +0.07, -0.09, respectively. The black spectra (directly below each mean spectrum) are best fit models. Green dashed lines indicate positions and widths of $CO_2$ ice bands 1, 2, and 3. (**b**) Close up of the same image shown in (a), focused on the three $CO_2$ bands.

We then generated best fit models for these mean trailing hemisphere spectra using the Hapke-Mie model described above. The best fit spectral models (Figure B1) are a combination of intimately mixed $H_2O$ ice and amorphous carbon and areally mixed $CO_2$ ice (Table B1). The relatively low albedos of the Uranian satellites, coupled with weak 1.52 and 2.02 μm $H_2O$ ice bands (and the absence of shorter wavelength $H_2O$ ice bands) suggests that their surfaces are covered by an intimate mixture of small $H_2O$ ice grains and dark, presumably carbonaceous grains (*e.g.*, Brown and Cruikshank, 1983). The absence of the ~2.134 μm $CO_2$ forbidden transition feature from our spectra, suggests that the detected $CO_2$ is predominantly segregated from other species (*i.e.*, areally mixed). These best fit models are sufficient for characterization of the $CO_2$ bands present in the SpeX spectra; however, they only provide approximations of the $H_2O$ ice band shapes and the spectral continuum levels. Future work by our team will investigate the shape of the $H_2O$ ice bands in greater detail (Cartwright et al. [In prep]).

*B1.2 Analysis of relative $CO_2$ band areas for mean spectra and best fit models*

Gerakines et al. (2005) characterized the relative band strengths of pure $CO_2$ ice combination and overtone bands between 1.9 and 2.1 μm, finding that $CO_2$ band 2 (centered near 2.012 μm) is the strongest band, followed by band 1 (centered near 1.966 μm), and finally band 3 (centered near 2.070 μm) is the weakest of the three. Therefore, if the observed spectral features on the Uranian satellites are dominated by pure $CO_2$ ice, as suggested by the absence of the ~2.134 μm $CO_2$ forbidden transition, they should display the same relative relationships. Consequently, the relative band area contributions of $CO_2$ bands 1, 2, and 3 can be used to test for the dominant $CO_2$ mixing regime (*i.e.*, intimately mixed or segregated).

To this end, we measured $CO_2$ band areas in the mean trailing hemisphere spectra and in the best fit models. The models provide a 'control' measurement for this analysis because we know the contribution of $CO_2$ ice in each. We used two types of best fit models: noise-free versions shown in Figure B1, and noise-added versions (example shown in Figure B2). Given that the detected $CO_2$ bands are weakest in our Oberon spectra, and therefore represent the $CO_2$ bands most likely to be obscured by noise, we chose to replicate the point-to-point variation of the mean Oberon spectrum (between 1.9 and 2.1 μm) in all of our noise-added models.

Using the procedure described in section 4.1, we measured the areas of the $CO_2$ bands in the mean trailing hemisphere spectra and their best fit models. Next, we normalized individual band areas by the total area of the three bands to determine their relative contributions (for example, $CO_2$ band 1 / ($CO_2$ band 1 + $CO_2$ band 2 + $CO_2$ band 3)). The relative $CO_2$ band ratios for the mean Ariel and Oberon spectra (Table B2) are within 2% and 3% of the noise-free models (Table B3) for these two moons, respectively. The ratios between the $CO_2$ bands in the mean spectra of Umbriel and Titania vary by up to 8% for $CO_2$ bands 1 and 2, but only 2% for $CO_2$ band 3. The $CO_2$ band area ratios for each mean spectrum display similar relationships with the band ratios in their corresponding best fit model. For example, $CO_2$ band 2 is strongest in the mean spectra of Ariel, Umbriel, and Oberon (but not Titania), and $CO_2$ band 3 is the weakest in all four spectra.

*B1.3 Analysis of relative $CO_2$ band areas for synthetic spectra*

To further test the dominant $CO_2$ mixing regime present in our spectra, we generated a range of noise-free and noise-added spectral models (1, 10, and 100 μm grain sizes), and investigated their relative $CO_2$ band area contributions (Table B4). We focus primarily on the noise-added synthetic spectra to highlight how point-to-point variations in spectra can alter $CO_2$ band strengths, in particular when only minor amounts of $CO_2$ are present (analogous to detected $CO_2$ levels on Titania and Oberon).

We first investigated the relative band strengths of pure $CO_2$ models using three different grain sizes (1, 10, and 100 μm): $CO_2$ band 1 (34 − 36% of total area), $CO_2$ band 2 (44 − 48%), and $CO_2$ band 3 $(18 - 21\%)$. The $CO_2$ band ratios for the mean spectra of Ariel $(34 \pm 2\%, 45 \pm 2\%, 21 \pm 1\%$, respectively) and Oberon $(35 \pm 11\%, 44 \pm 13\%, 21 \pm 7\%$, respectively) are nearly identical to those of pure $CO_2$ ice. Although the band area ratios for Umbriel $(26 \pm 4\%, 51 \pm 6\%, 23 \pm 3\%$, respectively) and Titania $(41 \pm 8\%, 38 \pm 10\%, 21 \pm 5\%$, respectively) display some variation compared to the relative band ratios for the pure $CO_2$ models, they are $\leq 2\sigma$ removed from the pure $CO_2$ models.

To test whether our band analysis code might incorrectly identify $CO_2$ in spectra of pure $H_2O$ ice, we measured the '$CO_2$ band areas' in both noise-free and noise-added pure $H_2O$ ice models (1, 10, and 100 µm grain sizes). We found non-zero summed '$CO_2$ band areas' in the pure $H_2O$ ice models, but they are at least a factor of ~3 lower than the band areas in all four mean Uranian satellite spectra (Table B4). We also measured the '$CO_2$ bands areas' of models composed of mixtures of different $H_2O$ ice grain sizes (Table B5). We again found that the measured '$CO_2$ bands' are much lower than the mean Uranian satellite spectra.

We also generated a wide range of models using areal and particulate mixtures of different $H_2O$ and $CO_2$ grain sizes (along with amorphous C in our particulate mixtures) (Table B5). At the lowest $CO_2$ abundance levels we investigated (1% and 9% for areal and particulate mixtures, respectively), the measured band areas and relative $CO_2$ band ratios are similar to the pure $H_2O$ ice models described above. At higher $CO_2$ levels ($\geq 5\%$), the measured band areas and relative band area ratios for the areal mixtures of $H_2O$ and $CO_2$ are consistent with the best fit models for the mean Uranian satellite spectra. Much more $CO_2$ ice ($\geq 49\%$) is required before the $CO_2$ band areas in the particulate mixture models can approximate the best fit models. Additionally, the relative band area ratios of the particulate models are not consistent with the mean spectra or their best fit models at any of the tested $CO_2$ abundance levels. The particulate mixture shown in Figure B2 (10 µm grains of $H_2O$ (9%), $CO_2$ (89%), and amorphous C (2%)) has a comparable summed $CO_2$ band area to the mean Ariel spectrum and its best fit model (Figure B2). It is apparent that the $H_2O$ ice bands in the particulate mixture are much weaker than those in the mean Ariel spectrum.

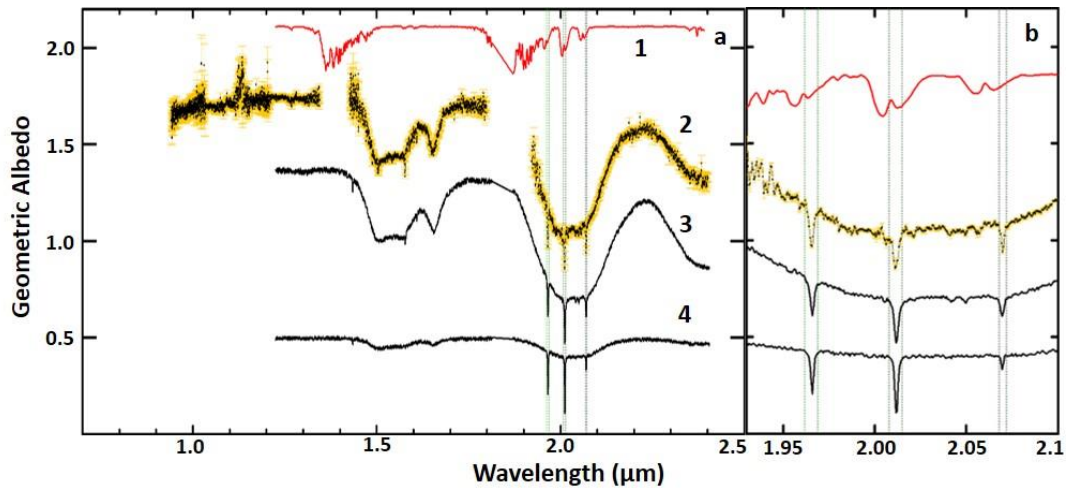

**Figure B2:** (**a**) Atmospheric transmission spectrum taken at the IRTF (red, 1), mean trailing hemisphere spectrum of Ariel (black with orange error bars, 2), best fit model for the mean Ariel spectrum (black, 3), synthetic spectrum composed of a particulate mixture of $CO_2$ ice (89%), $H_2O$ (9%), and amorphous C (2%) (black, 4). Green dashed lines indicate positions and widths of $CO_2$ ice bands 1, 2, and 3. (**b**) Close up of the same image shown in (a), focused on the three $CO_2$ bands. The shape and relative strength of the three $CO_2$ bands in the mean trailing hemisphere spectrum are most similar to the $CO_2$ bands in the best fit model that includes segregated $CO_2$ ice (black, 3).

**Table B1: Best fit synthetic spectra**

| | | | | | | | |
|---|---|---|---|---|---|---|---|
| | Particulate Mixture Model Components | | | | | | |
| Satellite | Component 1 | Mix (%) | Component 2 | Mix (%) | Component 3 | Mix (%) | Component 4 | Mix (%) |

| Satellite | Component 1 | Mix (%) | Component 2 | Mix (%) | Component 3 | Mix (%) | Component 4 | Mix (%) |
|---|---|---|---|---|---|---|---|---|
| Ariel | 50 $\mu$m $H_2O$ | 50.00 | 10 $\mu$m $H_2O$ | 47.00 | 0.3 $\mu$m $H_2O$ | 0.80 | 12.5 $\mu$m amorph C | 2.20 |
| Umbriel | 50 $\mu$m $H_2O$ | 25.05 | 10 $\mu$m $H_2O$ | 34.70 | 0.2 $\mu$m $H_2O$ | 0.25 | 5 $\mu$m amorph C | 40.00 |
| Titania | 50 $\mu$m $H_2O$ | 47.00 | 10 $\mu$m $H_2O$ | 38.70 | 0.20 $\mu$m $H_2O$ | 0.30 | 10 $\mu$m amorph C | 14.00 |
| Oberon | 50 $\mu$m $H_2O$ | 21.68 | 10 $\mu$m $H_2O$ | 52.90 | 0.2 $\mu$m $H_2O$ | 0.32 | 7 $\mu$m amorph C | 25.10 |

| Areal (Final) Mixture Model Components | | | | | | |
|---|---|---|---|---|---|---|
| Satellite | Component 1 | Mix (%) | Component 2 | Mix (%) | Component 3 | Mix (%) |
| Ariel | particulate model described above | 73.0 | 10 $\mu$m $CO_2$ | 22.5 | 50 $\mu$m $CO_2$ | 4.5 |
| Umbriel | particulate model described above | 92.0 | 5 $\mu$m $CO_2$ | 5.2 | 10 $\mu$m $CO_2$ | 2.8 |
| Titania | particulate model described above | 95.0 | 5 $\mu$m $CO_2$ | 1.0 | 10 $\mu$m $CO_2$ | 4.0 |
| Oberon | particulate model described above | 97.0 | 5 $\mu$m $CO_2$ | 1.5 | 10 $\mu$m $CO_2$ | 1.5 |

**Table B2: $CO_2$ band areas for mean trailing hemisphere spectra**

| | $CO_2$ Band Area ($10^{-4}$ $\mu$m) | | | | | | | | Relative $CO_2$ Band Ratios | | | | | |
|---|---|---|---|---|---|---|---|---|---|---|---|---|---|---|
| Satellite | Band 1 | $\Delta$Area | Band 2 | $\Delta$Area | Band 3 | $\Delta$Area | Total Area | Total $\Delta$area | Band 1 (%) | $\Delta$Band 1 (%) | Band 2 (%) | $\Delta$Band 2 (%) | Band 3 (%) | $\Delta$Band 3 (%) |
| Ariel | 4.65 | 0.22 | 6.20 | 0.27 | 2.90 | 0.15 | 13.75 | 0.38 | 34 | 2 | 45 | 2 | 21 | 1 |
| Umbriel | 1.61 | 0.25 | 3.18 | 0.31 | 1.42 | 0.15 | 6.21 | 0.43 | 26 | 4 | 51 | 6 | 23 | 3 |
| Titania | 1.43 | 0.23 | 1.33 | 0.31 | 0.71 | 0.15 | 3.47 | 0.41 | 41 | 8 | 38 | 10 | 20 | 5 |
| Oberon | 0.84 | 0.23 | 1.07 | 0.25 | 0.50 | 0.15 | 2.41 | 0.37 | 35 | 11 | 44 | 13 | 21 | 7 |

**Table B3: $CO_2$ band areas for best fit synthetic spectra**

| | | $CO_2$ Band Area ($10^{-4}$ $\mu$m) | | | Relative $CO_2$ Band Ratios | | |
|---|---|---|---|---|---|---|---|
| Satellite | Noise Added ? | Band 1 | Band 2 | Band 3 | Total Area | Band 1 (%) | Band 2 (%) | Band 3 (%) |
| Ariel | Yes | 4.81 | 6.57 | 2.62 | 14.00 | 34 | 47 | 19 |
| | No | 4.64 | 6.56 | 2.68 | 13.88 | 34 | 47 | 19 |
| Umbriel | Yes | 1.79 | 2.65 | 1.23 | 5.67 | 32 | 47 | 22 |
| | No | 1.91 | 2.66 | 1.02 | 5.59 | 34 | 48 | 18 |
| Titania | Yes | 0.86 | 1.69 | 0.54 | 3.09 | 28 | 55 | 18 |
| | No | 1.18 | 1.69 | 0.68 | 3.55 | 33 | 48 | 19 |
| Oberon | Yes | 0.84 | 1.05 | 0.45 | 2.33 | 36 | 45 | 18 |
| | No | 0.73 | 1.04 | 0.42 | 2.18 | 33 | 48 | 19 |

**Table B4: $CO_2$ band areas for pure synthetic spectra**

| Pure Model Component 1 | Noise Added? | $CO_2$ Band Area ($10^{-4}$ µm) | | | | Relative $CO_2$ Band Ratios | | |
|---|---|---|---|---|---|---|---|---|
| | | Band 1 | Band 2 | Band 3 | Total Area | Band 1 (%) | Band 2 (%) | Band 3 (%) |
| 1 µm $H_2O$ | No | 0.035 | 0.072 | 0.045 | 0.152 | 23 | 47 | 30 |
| 10 µm $H_2O$ | No | 0.051 | 0.091 | 0.057 | 0.200 | 26 | 46 | 29 |
| 100 µm $H_2O$ | No | 0.195 | 0.173 | 0.114 | 0.482 | 40 | 36 | 24 |
| 1 µm $H_2O$ | Yes | -0.035 | -0.148 | -0.013 | -0.195 | 18 | 76 | 7 |
| 10 µm $H_2O$ | Yes | -0.132 | 0.036 | 0.071 | -0.025 | 535 | -146 | -289 |
| 100 µm $H_2O$ | Yes | 0.292 | 0.352 | 0.035 | 0.680 | 43 | 52 | 5 |
| 1 um $CO_2$ | No | 5.15 | 7.29 | 2.73 | 15.17 | 34 | 48 | 18 |
| 10 um $CO_2$ | No | 9.37 | 12.34 | 4.71 | 26.42 | 35 | 47 | 18 |
| 100 um $CO_2$ | No | 20.53 | 25.04 | 11.59 | 57.16 | 36 | 44 | 20 |
| 1 um $CO_2$ | Yes | 5.11 | 7.14 | 2.77 | 15.03 | 34 | 48 | 18 |
| 10 um $CO_2$ | Yes | 9.17 | 12.21 | 4.76 | 26.14 | 35 | 47 | 18 |
| 100 um $CO_2$ | Yes | 20.41 | 25.06 | 11.73 | 57.20 | 36 | 44 | 21 |

**Table B5: $CO_2$ Band areas for mixed synthetic spectra**

| Compositional Model Parameters | | | | | | | | | $CO_2$ Band Area ($10^{-4}$ µm) | | | | Relative $CO_2$ Band Ratios | | |
|---|---|---|---|---|---|---|---|---|---|---|---|---|---|---|---|
| Mixing Regime | Component 1 | Mix (%) | Component 2 | Mix (%) | Component 3 | Mix (%) | Noise Added? | Band 1 | Band 2 | Band 3 | Total Area | Band 1 (%) | Band 2 (%) | Band 3 (%) | |
| Areal | 1 µm $H_2O$ | 20 | 10 µm $H_2O$ | 80 | - | - | Yes | 0.388 | 0.250 | 0.097 | 0.736 | 53 | 34 | 13 | |
| Areal | 1 µm $H_2O$ | 50 | 10 µm $H_2O$ | 50 | - | - | Yes | -0.100 | 0.202 | 0.019 | 0.121 | -83 | 167 | 15 | |

| | | | | | | | | | | | | | | |
|---|---|---|---|---|---|---|---|---|---|---|---|---|---|---|
| Areal | 1 μm $H_2O$ | 80 | 10 μm $H_2O$ | 20 | - | - | Yes | -0.113 | 0.309 | 0.038 | 0.234 | -48 | 132 | 16 |
| Areal | 1 μm $H_2O$ | 20 | 100 μm $H_2O$ | 80 | - | - | Yes | 0.112 | 0.222 | 0.130 | 0.465 | 24 | 48 | 28 |
| Areal | 1 μm $H_2O$ | 50 | 100 μm $H_2O$ | 50 | - | - | Yes | 0.086 | 0.127 | 0.178 | 0.391 | 22 | 32 | 46 |
| Areal | 1 μm $H_2O$ | 80 | 100 μm $H_2O$ | 20 | - | - | Yes | 0.425 | 0.017 | 0.112 | 0.555 | 77 | 3 | 20 |
| Areal | 10 μm $H_2O$ | 99 | 10 μm $CO_2$ | 1 | - | - | Yes | 0.443 | 0.043 | 0.198 | 0.684 | 65 | 6 | 29 |
| Areal | 10 μm $H_2O$ | 95 | 10 μm $CO_2$ | 5 | - | - | Yes | 0.595 | 1.084 | 0.580 | 2.258 | 26 | 48 | 26 |
| Areal | 10 μm $H_2O$ | 90 | 10 μm $CO_2$ | 10 | - | - | Yes | 1.402 | 1.899 | 0.791 | 4.092 | 34 | 46 | 19 |
| Areal | 10 μm $H_2O$ | 80 | 10 μm $CO_2$ | 20 | - | - | Yes | 2.685 | 3.837 | 1.280 | 7.801 | 34 | 49 | 16 |
| Areal | 10 μm $H_2O$ | 70 | 10 μm $CO_2$ | 30 | - | - | Yes | 3.787 | 5.268 | 2.086 | 11.141 | 34 | 47 | 19 |
| Particle | 10 μm $H_2O$ | 89 | 10 μm $CO_2$ | 9 | 10 μm amorph C | 2 | Yes | 0.090 | 0.262 | 0.184 | 0.536 | 17 | 49 | 34 |
| Particle | 10 μm $H_2O$ | 69 | 10 μm $CO_2$ | 29 | 10 μm amorph C | 2 | Yes | 0.932 | 1.350 | 0.173 | 2.455 | 38 | 55 | 7 |
| Particle | 10 μm $H_2O$ | 49 | 10 μm $CO_2$ | 49 | 10 μm amorph C | 2 | Yes | 0.932 | 2.323 | 0.444 | 4.408 | 21 | 53 | 10 |
| Particle | 10 μm $H_2O$ | 29 | 10 μm $CO_2$ | 69 | 10 μm amorph C | 2 | Yes | 2.556 | 4.252 | 0.988 | 7.795 | 33 | 55 | 13 |
| Particle | 10 μm $H_2O$ | 9 | 10 μm $CO_2$ | 89 | 10 μm amorph C | 2 | Yes | 4.582 | 6.582 | 1.181 | 12.345 | 37 | 53 | 10 |

**Appendix C**

*C1 Impact and tectonically exposed $CO_2$*

　　Native $CO_2$ ice buried beneath a protective regolith could be exposed by impact events and tectonic processes. Each moon has a large number of impact features over their spatially resolved southern hemispheres, and these features are generally brighter than the surrounding terrain (*e.g.,* Smith et al., 1986). The higher albedo of these craters is most likely due to the exposure of fresh material from beneath darker regoliths. Similarly, these moons display evidence for high albedo tectonic landforms. Ariel displays abundant evidence for global tectonism, including long grooves suggestive of extensional fractures (*e.g.,* Jankowsky and Squyres, 1988). Titania and Oberon have localized clusters of chasmata and large scale fractures (Smith et al., 1986), and dark polygonal basins, which might be tectonic or cryovolcanic in origin, have been observed on Umbriel (Helfenstein et al., 1989). Therefore, each moon displays evidence for impact and tectonic resurfacing, and perhaps these processes could have exposed native $CO_2$ deposits.

　　Surface age estimates based on crater counting range from ~few 100 Ma for the younger tectonized terrains on Ariel, to > 4 Ga on the dark and ancient surface of Umbriel (Zahnle et al., 2003). Grundy et al. (2006) estimate that pure $CO_2$ ice might survive on the surfaces of the Uranian moons from ~10 kyr to 1.0 Gyr. The large range in this sublimation timescale results from the assumed values of these satellites' bolometric Bond albedos ($A_B$ $0.5 - 0.7$), which are poorly constrained for the Uranian satellites. Therefore, any native $CO_2$ ice exposed by ancient tectonism and impacts has likely been removed, even from the youngest terrains on Ariel. Additionally, because the Uranian moons are tidally-locked, we would expect to find impact-exposed $CO_2$ deposits primarily on their leading hemispheres, which are preferentially bombarded by heliocentric impactors (*e.g.,* Zahnle et al., 2001).

*C2 Cryovolcanic $CO_2$ deposits*

　　$CO_2$ could be produced in these satellites' interiors and deposited on their surfaces by cryovolcanism. Several studies have suggested that various surface features on each of these moons could represent cryovolcanic landforms (*e.g.,* Smith et al., 1986; Schenk, 1991; Kargel, 1994). Long linear grooves at the center of chasmata floors have been interpreted as linear volcanic vents on Ariel (*e.g.,* Jankowsky and Squyres, 1988), and trough-filling ridges with lobate edges and medial crests on Ariel have been interpreted as extrusive cryovolcanic flow deposits (*e.g.,* Schenk, 1991). Numerous crater floors on Umbriel have albedos that are significantly higher than the surrounding terrains (Helfenstein and Veverka, 1988). In particular, the albedo of floor deposits in Wunda crater (~0.49), located at low southern latitudes near the antapex of motion, is over twice as bright as the surrounding terrain (~0.23). Smooth, low albedo regions on Titania have been interpreted as cryovolcanic deposits (Croft and Soderblom, 1991). Similar to Umbriel, crater floors on Oberon have smooth deposits that have been interpreted to be cryovolcanic in origin, albeit these deposits have a much lower albedo ($< 0.1$) than the surrounding terrains (~0.25), counter to the crater floor deposits on Umbriel. Consequently, each moon displays evidence for potential cryovolcanic structures and flow features, which might include large deposits of $CO_2$ ice.

　　Much like the observed tectonic features, however, most of the putative extrusive cryovolcanic flows and ridges are all fairly old according to crater age estimates (few 100 Ma to 1 Ga, Zahnle et al., 2003), and exposed $CO_2$ should be effectively removed over these timescales. Furthermore, there is no obvious reason to expect a leading/trailing dichotomy in cryovolcanism, and therefore in emplaced $CO_2$, on these satellites. Even if emplaced $CO_2$ coincidently displayed a leading/trailing asymmetry on these satellites, perhaps due to global tidal forces enhanced by paleo-resonances (Tittemore and Wisdom, 1990), we would expect to find more $CO_2$ on the relatively younger surfaces of Titania and Oberon than the more ancient surface of Umbriel, counter to the observed $CO_2$ abundances on these moons. Therefore, $CO_2$-rich, extrusive cryolava deposits do not adequately fit the observed distribution of $CO_2$. However, the smooth crater floor deposits on Umbriel and Oberon appear to be devoid of over-printing impact features (albeit, at low spatial resolutions of ~5 – 6 km/pixel), and might represent young cryovolcanic deposits. In particular, Wunda crater is positioned near the antapex of Umbriel and at low southern latitudes where $CO_2$ bands are deepest. Additional observations at higher resolution than the available Voyager 2 data are needed to investigate the morphology and composition of putative cryovolcanic deposits further.

*C3 Micrometeorite impacts*

Micrometeorite impacts eject and vaporize buried material, mixing regolith materials and exposing fresh materials at the surface to irradiation sources. Furthermore, the kinetic energy imparted into surfaces by micrometeorite impacts can dissociate target molecules, even at relatively modest impact velocities of ~5 km/s (Borucki et al., 2002; Jaramillo-Botero et al., 2012). Thus, micrometeorite impacts might represent an important energy source for dissociating $H_2O$, $CO_2$, and others species on icy moons.

Gravitational focusing by planets increases impactor fluxes on the satellites closest to them (Shoemaker and Wolfe, 1982; Zahnle et al., 2003). Because the orbital velocities of the Uranian satellites (~3 – 7 km s$^{-1}$) are comparable to incoming heliocentric and planetocentric dust (~10 km s$^{-1}$), these particles should preferentially impact the leading hemispheres of the Uranian moons (Tamayo et al., 2013). It therefore seems unlikely that kinetic energy from micrometeorite impacts is a primary contributor to active $CO_2$ ice synthesis given that $CO_2$ is concentrated near the antapexes of these satellites, which are predicted to experience the fewest number of heliocentric impacts.

*C4 Sublimation and cold trapping of $CO_2$*

As discussed in section 5.2, $CO_2$ deposits exposed on the surfaces of Uranian moons should sublimate away over timescales shorter than the age of the Solar System (*e.g.,* Lebofsky, 1975). The large obliquity of the Uranian system leads to large seasonal disparities in heating that should drive $CO_2$ migration to low latitude cold traps (Grundy et al., 2006). Additionally, because maximum surface temperatures on airless bodies like the Uranian satellites are a strong function of rotation rate, $CO_2$ should be most resistant against sublimation at low latitudes on the fast rotating inner moons, Ariel and Umbriel – exactly where it has been detected. The absence of $CO_2$ at high latitudes on the Uranian moons has yet to be confirmed, and subsequent observations of these satellites during northern summer will help constrain the abundance of $CO_2$ at high latitudes.

However, it seems unlikely that sublimation alone controls the observed hemispherical asymmetries in $CO_2$ band areas on the Uranian satellites. The time span of maximum subsolar heating should be the same on both their leading and trailing hemispheres, which over time, would drive sublimation on both hemispheres and homogenize the distribution of $CO_2$. It therefore seems unlikely that sublimation alone is responsible for the observed distribution of $CO_2$. As the Uranian system moves toward northern summer over the next two decades, subsequent observations of the Uranian satellites should provide insight into sublimation-driven seasonal migration of $CO_2$.

*C5 UV photolysis*

UV photons interact with icy surfaces, driving molecular dissociation of constituents and the synthesis of new species. Numerous laboratory studies have demonstrated that the absorption of UV photons ($\lesssim 227$ nm, $h\nu \gtrsim 5.5$ eV) by $H_2O$ ice mixed with carbonaceous materials can synthesize $CO_2$ ice (*e.g.,* Chakarov et al., 2001; Hudson and Moore, 2001; Loeffler et al., 2005; Mennella et al., 2006). Other laboratory experiments have demonstrated that $CO_2$ ice is readily dissociated by UV photon irradiation ($\lesssim 227$ nm, $h\nu \gtrsim 5.5$ eV) (*e.g*., Gerakines et al., 2001; Hudson and Moore, 2001; Schmidt et al., 2013).

An important consideration for investigating the efficiency of UV photolysis of $CO_2$ is the differences in penetration depths for chemistry-inducing UV photons ($\lambda \lesssim 227$ nm) and the reflected NIR photons ($\lambda \sim 0.8 - 9.5$ μm) sensed by SpeX and IRAC. As described in section 5.1, NIR photons penetrate ~100 μm into crystalline $H_2O$ ice over the wavelength range relevant for detecting $CO_2$ ice combination and overtone bands ($\lambda \sim 1.9 - 2.1$ μm), but only penetrate ~10 μm into $H_2O$ over the wavelength range relevant for detecting the $CO_2$ $\nu_3$ band. Most of the UV photons that can utilize the highest probability pathway for dissociating $CO_2$ ($\lambda \lesssim 160$ nm, $h\nu \gtrsim 7.7$ eV) only interact with the top ~few microns of crystalline $H_2O$ ice, while UV photons that utilize the much lower probability forbidden transition to dissociate $CO_2$ ($\lambda \lesssim 227$ nm, $h\nu \gtrsim 5.5$ eV) can interact with species in the top ~few meters of crystalline $H_2O$ ice. Therefore, UV photons that can readily dissociate $CO_2$ ($\lambda \lesssim 160$ nm) almost entirely affect the layer sensed by IRAC (~ 10 μm) and do not reach depths sensed by SpeX (~ 100 μm). This overlap between far-UV photons and IRAC penetration depths is consistent with our best fit IRAC models, which are dominated by small $H_2O$ ice grains with only minor amounts of small grained $CO_2$ ($\lesssim 5\%$).

UV photons interact equally with both the leading and trailing hemispheres of synchronously-locked satellites. Hence, it seems unlikely that UV photolysis can account for the observed hemispherical asymmetries in $CO_2$ abundance on these satellites. Furthermore, UV photons should interact equally with each satellite, which makes the observed planetocentric trend in $CO_2$ abundance difficult to explain via UV photolysis. While UV photons plausibly contribute to the production and destruction of $CO_2$ on the Uranian moons, as hypothesized for the detected $CO_2$ on Iapetus (Palmer and Brown, 2011), it is unlikely to represent the primary controlling mechanism on the distribution of $CO_2$ ice in the Uranian system.

*C6 Ion and electron radiolysis*

Charged particle bombardment of icy surfaces leads to the destruction of constituent species, the synthesis of new species, the ejection of material, and amorphization of $H_2O$ ice (Spinks and Wood, 1990). Synthesis of $CO_2$ via ion irradiation of carbon-rich and $H_2O$ ice mixtures at cryogenic temperatures has been demonstrated to occur rapidly in laboratory settings, using a wide range of C and O rich constituents, substrate temperatures, and ion energies (*e.g.,* Gerakines et al., 2001; Hudson and Moore, 2001; Mennella et al., 2004; Gomis and Strazzulla, 2005; Raut et al., 2012). Similarly, experimental work investigating electron irradiation of icy mixtures has demonstrated that electrons can readily synthesize $CO_2$ under a wide range of conditions and electron energy levels (*e.g.,* Sedlacko et al., 2005; Jamieson et al., 2006; Kim and Kaiser, 2012).

Penetration depths of heavy ions and protons are significant larger than the few μm depths of far-UV photons. Hudson and Moore (2001) state that a 1 MeV proton, along with secondary electrons, can chemically alter the top 20 μm of an icy surface. Delitsky and Lane (1998) designate the top ~100 μm of icy surfaces as the "plasma deposition zone" where heavy ions and protons initiate chemical changes in the near-surface species. According to these authors, energetic electrons can penetrate much deeper than ions, on the order of ~5 mm. Penetration depths of NIR photons sensed by IRAC over the wavelengths covered by the $CO_2$ $v_3$ band (~10 μm) clearly overlap with energetic ions and protons as well as electrons. However, only electrons have a high probability of penetrating to the depths probed by SpeX over the wavelength range covered by $CO_2$ bands 1 – 3 (~0.1 mm).

Similar to UV photons, solar wind particles ($\gtrsim$ 400 km s$^{-1}$) and cosmic rays ($\gtrsim$ 1% speed of light) interact with both the leading and trailing hemispheres of the synchronously locked Uranian moons (~3 – 7 km s$^{-1}$ orbital velocities). Consequently, these radiation sources would tend to generate (and destroy) $CO_2$ molecules ubiquitously, without pronounced hemispherical asymmetries. In a simplified Offset Tilted Dipole (OTD) model of magnetospheric interactions with the surfaces of Uranus' tidally-locked moons, embedded charged particles should preferentially bombard their trailing hemispheres as the co-rotating field lines sweep over the satellites, similar to the Jovian and Saturnian systems. The observed distribution of $CO_2$ supports this simple model of magnetospheric-satellite surface interactions in the Uranian system.

*C7 Charged particle sputtering*

Impacts by charged particles, micrometeorites, and high energy photons can eject material from surfaces (e.g., Johnson et al., 1984). In the Uranian system, charged particle sputtering is estimated to be much lower than at Saturn because of lower ion fluxes in the Uranian satellite orbital plane, and the lower sputtering efficiency of protons compared to heavy ions (Lanzerotti et al., 1987). Grundy et al. (2006) estimate that charged particle sputtering rates for the more volatile $CO_2$ should be 50 – 100 times higher than $H_2O$. These authors determined that to remove a global surface layer of $CO_2$ (~0.5 cm thick) using charged particle sputtering would take on the order of a few 100 Ma for the inner moons and longer than the age of the Solar System for the outer moons. Moreover, our $CO_2$ band parameter analysis indicates that the largest $CO_2$ band areas occur in spectra collected near the antapexes of these satellites, which likely represent the longitudes that experience the highest rates of charged particle sputtering. Therefore, charged particle sputtering does not appear to represent an important control on the distribution of $CO_2$.

**Appendix D**

*D1 Uranian magnetic field and plasma environment*

The magnetic dipole of Uranus is significantly offset from its rotational axis (58.6°), and its magnetic center is offset by ~0.3 Uranian radii ($R_U$) from the planetary center (Ness et al., 1986). The magnetopause is at a distance of ~18 $R_U$ (~450,000 km), encompassing the orbits of Ariel, Umbriel, and Titania, but not entirely that of Oberon, which orbits upstream of the magnetopause on Uranus' Sun-facing hemisphere (< 50% of its orbital period). The rotation period of Uranus' magnetic field (~17.2 hours) is much shorter than the orbital periods of the large Uranian moons (~2.5 – 13.5 Earth days, Table 1). Due to this differential velocity, charged particles embedded in the Uranian magnetosphere bombard the trailing hemispheres of the satellites as the magnetic field lines sweep past them. While higher order multipole models are needed to adequately describe the structure of Uranus' magnetic field (Connerney et al., 1987), higher order terms die out faster than the dipole term and are negligible outwards of ~5 $U_R$ and a simple OTD model provides a reasonable approximation at the radial distances of the classical satellites (Acuna et al., 1988). Subsequent discussion of the Uranian plasma environment relies on this simpler OTD model.

Analysis of data collected by the Low Energy Charge Particle (LECP) instrument onboard Voyager 2 demonstrated that the magnetosphere of Uranus is populated by protons and electrons (sourced by Uranus' ionosphere) with little evidence of heavier ions (< 1% of total ion population) (Krimigis et al., 1986). Voyager 2's Cosmic Ray System (CRS) measured a steep increase in the electron and proton counts interior to the orbit of Titania (Stone et al., 1986). Localized reductions in high energy (> 1 MeV) electron and proton counts coincide with the orbits of Miranda, Ariel, and Umbriel, indicating that these satellites are bombarded by high energy particles that intersect their orbits (Stone et al., 1986). The lack of heavy ions in the Uranian magnetosphere is likely due in part to the large offset between the orbital plane of the satellites and the magnetic equator, which results in heavy ions and (pre-cursor) neutrals occupying distinctly different volumes of the magnetosphere (Cheng et al., 1987). Consequently, implantation of heavy ions is unlikely to be an important driver of chemistry on the Uranian satellites.

Due to the large offset of the magnetic equator, the satellite orbital plane intersects a wide range of magnetic latitudes and L shells (measure of the magnetic field distance in Uranian radii, with lower L shell values typically corresponding to higher charged particle fluxes within radiation belts) (Stone et al., 1986). The net result is that electron and proton fluxes in the satellite orbital plane are low compared to the magnetic equator (Cheng et al., 1987). Laboratory experiments, informed by LECP measurements of proton and electron fluxes at each satellite, demonstrate that cryogenic $H_2O$ ice is readily modified. Consequently, electrons and protons embedded in Uranus' magnetosphere should chemically alter the surfaces of these moons over short timescales (Lanzerotti et al., 1987).

The large offset of Uranus' magnetic axis from its rotational axis makes predictions of the satellite latitudes where charged particle irradiation is greatest much more uncertain than in the Jovian or Saturnian systems. Detailed models simulating the interactions between Uranus' magnetic field and its satellites are therefore required to constrain the locations of peak radiation dose and fluence on these satellites.